\documentclass[letterpaper,titlepage,11pt]{article}
\usepackage{hyperref}
\usepackage{amssymb,amsmath,amsfonts}
\usepackage{epsfig}

\def\clock{{\count0=\time
           \divide\count0 60
           \ifnum\count0<10 0\fi\the\count0
           \multiply\count0 -60 \advance\count0 \time
           :\ifnum\count0<10 0\fi \the\count0
         }}
\newcommand{\timestamp}{{\small\vbox{\hbox{\tt\jobname.tex}
\hbox{\the\day/\the\month/\the\year, \clock}}}}

\newcommand{\be}{\begin{equation}} \newcommand{\ee}{\end{equation}}
\newcommand{\bea}{\begin{eqnarray}} \newcommand{\eea}{\end{eqnarray}}

\newcommand{\CO}{\mathcal{O}}
\newcommand{\CN}{\mathcal{N}}

\newcommand{\CT}{\mathcal{T}}
\newcommand{\CTu}{\CT_i^{(u)}}
\newcommand{\hCTu}{\hat{\CT}_i^{(u)}}

\newcommand{\CM}{\mathcal{M}}

\newcommand{\CK}{\mathcal{K}}
\newcommand{\Lu}{L_i^{(u)}}
 
\newcommand{\id}{\hbox{1\kern-.27em l}}
\newcommand{\sid}{\hbox{\scriptsize1\kern-.27em l}}

\newcommand{\we}{\kern-.1em\wedge\kern-.1em}
\newcommand{\scal}{\kern-.13em\cdot\kern-.13em}

\newcommand{\II}{I\kern-.09em I}

\newcommand{\al}{\alpha}

\newcommand{\R}{\mathbb{R}}
\newcommand{\T}{\mathbb{T}}

\newcommand{\nn}{\nonumber}

\newcommand{\spa}{\ , \ \ }

\newcommand{\gym}{g_{\mathrm{YM}}}
\newcommand{\Ord}{{\cal{O}}}

\newcommand{\vecto}[2]{\left( \begin{array}{c} #1 \\ #2 \end{array}
\right) }
\newcommand{\matrto}[4]{\left( \begin{array}{cc} #1 & #2 \\
#3 & #4 \end{array} \right) }

\newcommand{\rht}{\tilde{\rho}}
\newcommand{\tht}{\tilde{\theta}}

\newcommand{\mt}{\mathfrak{t}}
\newcommand{\ms}{\mathfrak{s}}
\newcommand{\bmt}{\bar{\mathfrak{t}}}
\newcommand{\bms}{\bar{\mathfrak{s}}}
\newcommand{\hmt}{\hat{\mathfrak{t}}}
\newcommand{\hms}{\hat{\mathfrak{s}}}
\newcommand{\mf}{\mathfrak{f}}

\newcommand{\hmf}{\hat{\mathfrak{f}}}
\newcommand{\bc}{\bar{c}}

\numberwithin{equation}{section}

\begin{document}

\begin{titlepage}

\rightline{\vbox{\small\hbox{\tt hep-th/0407094} }}
\vskip 3cm

\centerline{\Large \bf New Phases of Near-Extremal Branes on a Circle}

\vskip 1.6cm
\centerline{\bf Troels Harmark and Niels A. Obers}
\vskip 0.5cm
\centerline{\sl The Niels Bohr Institute}
\centerline{\sl Blegdamsvej 17, 2100 Copenhagen \O, Denmark}

\vskip 0.5cm

\centerline{\small\tt harmark@nbi.dk, obers@nbi.dk}

\vskip 1.6cm

\centerline{\bf Abstract} \vskip 0.2cm \noindent
We study the phases of near-extremal branes on a circle,
by which we mean near-extremal branes of string theory
and M-theory with a circle in their transverse space.
We find a map that takes any static and neutral Kaluza-Klein black hole, i.e.
 any static and neutral black hole on Minkowski-space times a circle
 $\CM^d \times S^1$, and map it to a corresponding solution for
 a near-extremal brane on a circle.
 The map is derived using first
a combined boost and U-duality transformation on the
 Kaluza-Klein black hole,
transforming it to a solution for a non-extremal brane on a circle.
The resulting solution for a near-extremal brane on a circle
is then obtained by taking a certain near-extremal limit.
As a consequence of the map,
 we can transform the neutral non-uniform black string branch into a
 new non-uniform phase of near-extremal branes on a circle.
 Furthermore, we use recently obtained analytical results on small
 black holes in Minkowski-space times a circle to get
new information about the
 localized phase of near-extremal branes on a circle.
 This gives in turn predictions for the thermal behavior of the
 non-gravitational theories dual to these near-extremal branes.
 In particular, we give predictions for the thermodynamics of
 supersymmetric  Yang-Mills theories on a circle,
 and we find a new stable phase of $(2,0)$
 Little String Theory in the canonical ensemble
for temperatures above its Hagedorn temperature.


\end{titlepage}

\pagestyle{empty}
\small
\tableofcontents
\normalsize

\pagestyle{plain}
\setcounter{page}{1}

\section{Introduction}

It has been established in recent years that near-extremal
branes in string theory and M-theory provide
a link between black hole phenomena in gravity and the thermal physics
of non-gravitational theories.
Among the most prominent examples is
the duality between near-extremal D3-branes
and $\CN=4$ supersymmetric Yang-Mills theory
\cite{Maldacena:1997re,Gubser:1998bc,Witten:1998qj,Witten:1998zw}.
More generally, for all the supersymmetric branes in string/M-theory
one has dualities
between the near-extremal limit of the brane
and a non-gravitational theory \cite{Itzhaki:1998dd}.%
\footnote{See also the review \cite{Aharony:1999ti} and references therein.}

In this paper, we use U-duality to find a map from
static and neutral
Kaluza-Klein black holes to near-extremal branes on a transverse
circle.
This gives a precise connection between
phases of static and neutral Kaluza-Klein black holes and the
thermodynamic
behavior of certain non-gravitational theories.
The non-gravitational theories include
($p+1$)-dimensional supersymmetric Yang-Mills theories
with 16 supercharges compactified
on a circle, the uncompactified
($2+1$)-dimensional supersymmetric Yang-Mills theory
with 16 supercharges, and
$(2,0)$ Little String Theory.

Static and neutral Kaluza-Klein black holes is the term we use for
static solutions of pure gravity with an event horizon
and asymptoting to $d$-dimensional Minkowski-space times a circle
$\CM^d \times S^1$, i.e. Kaluza-Klein space, with $d \geq 4$.
Kaluza-Klein black holes have been studied from several
points of view.
In \cite{Gregory:1993vy,Gregory:1994bj}
Gregory and Laflamme found that a uniform black string
wrapped on the circle is classically unstable if the
mass is sufficiently small.
A consequence of this is that there exists a non-uniform black
string solution, i.e. a string wrapping the circle but without
translational invariance along the circle.
This has been found in
\cite{Gregory:1988nb,Gubser:2001ac,Wiseman:2002zc,Sorkin:2004qq}.%
\footnote{Part of the motivation to look for the non-uniform
branch has been to reveal the endpoint of the Gregory-Laflamme
instability \cite{Horowitz:2001cz}.}
Furthermore, the black hole branch has been studied
\cite{Harmark:2002tr,Kol:2003if,Sorkin:2003ka,Kudoh:2003ki,Harmark:2003yz,Gorbonos:2004uc}.
In this branch, the black hole
is localized on the circle contrary to the string branches.
Finally, static
solutions including both event horizons and Kaluza-Klein bubbles,
called bubble-black hole sequences, have been constructed
\cite{Emparan:2001wk,Elvang:2002br,Elvang:2004iz}.
All these different types of solutions correspond to points in
the $(\mu,n)$ phase diagram introduced
in \cite{Harmark:2003dg,Harmark:2003eg} (see also \cite{Kol:2003if}),
where $\mu$ is the mass, rescaled to make it dimensionless, and
$n$ is the relative tension, i.e. the ratio of the tension
along the circle-direction and the mass.

The class of supersymmetric branes that we consider in this paper, is
the set of $1/2$ BPS branes of type IIA/B string theory and M-theory.
For these branes we consider the situation in which we have a
circle in the transverse space and a non-zero temperature.
We denote this class as {\sl non-extremal branes on a circle} in the paper.
We consider furthermore the near-extremal limit of the
non-extremal branes on a circle. The near-extremal limit
that we take is defined such
that one keeps the non-trivial physics related to the presence
of the circle.
We denote this class of near-extremal branes as
{\sl near-extremal branes on a circle}.
The dual non-gravitational theories of these branes
are the ones listed above.

In order to set up a theoretical framework in which we can
analyze these two classes of brane solutions, we define
phase diagrams for the non- and near-extremal branes on a circle.
In particular for near-extremal branes on a circle we define the
$(\epsilon,r)$ phase diagram, where $\epsilon$ is the energy,
rescaled to be dimensionless, and the {\sl relative tension}
$r$ which is the ratio between the tension
of the brane along the transverse circle and the energy.
The tension is measured using the general tension formula
found in \cite{Harmark:2004ch}.

Two of the main results of this paper are then:
\begin{itemize}
\item It is shown that one can transform any
static and neutral Kaluza-Klein black hole
to a non-extremal brane on a circle by a combined boost
and U-duality transformation.
\item By taking a particular near-extremal limit of the map to
non-extremal
branes, it is shown that one can transform
any static and neutral Kaluza-Klein black hole
to a near-extremal brane on a circle.
In particular, we find a map relating points in the $(\mu,n)$
phase diagram to points in the $(\epsilon,r)$ phase diagram.
\end{itemize}
This has the consequence that any Kaluza-Klein black hole solution
in $d+1$ dimensions can be mapped to a corresponding solution describing
non- or near-extremal $p$-branes on a circle. Here $d$ and $p$
are related by $D=d+p+1$ with $D=10$ for string theory and
$D=11$ for M-theory.
Therefore, we can
map all the known phases of Kaluza-Klein black holes into phases
of non- or near-extremal branes on a circle.

This map is a development of an earlier result in \cite{Harmark:2002tr}.
In \cite{Harmark:2002tr} it was observed
that for the class of Kaluza-Klein black holes that fall
into the $SO(d-1)$-symmetric ansatz of Ref.~\cite{Harmark:2002tr},
one can map any solution into a corresponding non- and near-extremal
solution.
While this observation was made at the level of the equations of
motion, we show in this paper that it follows more generally from a
combined boost and U-duality transformation, thus revealing
the physical reason behind the existence
of this map.
The boost/U-duality map of this paper is an extension of the
map of \cite{Harmark:2002tr} since it works for
any static and neutral Kaluza-Klein black hole.

If we apply the map to the uniform string branch we recover the
known phase of non- and near-extremal branes smeared on a circle,
which we call the {\it uniform phase}. However, by applying the
map to the non-uniform black string branch we generate a new phase of
non- and near-extremal branes, which we denote as the
{\it non-uniform phase}.%
\footnote{In \cite{Horowitz:2002ym}
a new non-uniform phase of certain near-extremal
branes on a circle was conjectured to exist for small energies.
There does not seem to be any direct connection between the branch
that we find and the one of \cite{Horowitz:2002ym}.
We comment further on  \cite{Horowitz:2002ym}
in the conclusions in Section \ref{s:concl}.}
We use the map to study the thermodynamics of this new phase
in detail, both for the non- and near-extremal $p$-branes on a circle.
In particular, using the results of Ref.~\cite{Sorkin:2004qq}
for $4 \leq d \leq 9$
we obtain the first correction around the point where the non-uniform
phase emanates from the uniform phase. For the particular case
of the M5-brane on a circle we can go even further and use
the numerically obtained $d=5$ non-uniform branch of Wiseman
\cite{Wiseman:2002zc} to obtain the
corresponding near-extremal non-uniform phase.

In particular, the Gregory-Laflamme mass $\mu_{\rm GL}$ of the
uniform black string branch is mapped into a critical mass
$\bar \mu_{c}$ of the uniform non-extremal branch and
a critical energy $\epsilon_{c}$ of the uniform near-extremal branch.
$\bar \mu_{c}$ and $\epsilon_{c}$ mark where
the non-uniform phase connects to the uniform phase for
the non- and near-extremal branes on a circle, respectively.
The existence of this new non-uniform phase suggests that
non- and near-extremal branes smeared on a circle
have a critical mass/energy below which
they are classically unstable.
This seems to provide a counter-example to the
Gubser-Mitra conjecture \cite{Gubser:2000ec,Gubser:2000mm,Reall:2001ag,Gregory:2001bd}.

As a second application of the map we apply it to the small black hole
branch. This generates non- and near-extremal localized on a circle,
and we denote this as the {\it localized phase}.
For this phase we use the analytical results of
Ref.~\cite{Harmark:2003yz} to explicitly compute the first correction
to the
solution for the non- and near-extremal branes localized on a circle.
We also study the thermodynamics in detail, obtaining
the first correction due to the presence of the circle.

The results for the non-uniform and localized phase
 of near-extremal branes are particularly interesting,
since they provide us with new information about the
dual non-gravitational theories at finite temperature.
In particular, we study:
\begin{itemize}
\item The M5-brane on a circle, which is dual to thermal (2,0)
Little String Theory (LST).
\item The D$(p-1)$-brane
on circle, which is dual to thermal $(p+1)$-dimensional
supersymmetric Yang-Mills (SYM) theory on $\R^{p-1} \times S^1$.
\item The M2-brane on a circle, which is dual to
thermal $(2+1)$-dimensional SYM theory on $\R^2$.
\end{itemize}
In these dual non-gravitational theories,
the localized phase corresponds to the low temperature/low
energy regime of the dual theory, whereas the uniform phase corresponds
to the high temperature/high energy regime of the theory.
The non-uniform phase
appears instead for intermediate temperatures/energies.

In particular, by translating the results for the thermodynamics
of the localized and non-uniform phase in terms of the dual
non-gravitational theories we find:
\begin{itemize}
\item The first correction to the thermodynamics for the localized
phase of the SYM theories.
\item A prediction of a new non-uniform phase of the SYM theories
including the first correction around the point where the non-uniform
phase emanates from the uniform phase.
\item Using the numerical data of Wiseman \cite{Wiseman:2002zc}
for the $d=5$ non-uniform branch we numerically
 compute the corresponding thermodynamics in
$(2,0)$ LST. This gives
 a new stable phase of $(2,0)$ LST in the canonical ensemble,
for temperatures above its Hagedorn temperature.
We furthermore compute the first correction to the
thermodynamics in the infrared region, when moving away from
the infrared fixed point, which is superconformal $(2,0)$
theory.
\end{itemize}

The outline of this paper is as follows.
In Section \ref{review} we review the current knowledge on phases
of static and neutral Kaluza-Klein black holes.
This is important since these are the phases that later in the paper
are transformed
to phases of non- and near-extremal branes on a circle.
We review in particular the $(\mu,n)$ phase diagram introduced
in \cite{Harmark:2003dg},
with $\mu$ being the rescaled mass and $n$ the relative tension.
Moreover, we review the ansatz of \cite{Harmark:2002tr} for
Kaluza-Klein black holes with $SO(d-1)$ symmetry.

In Section \ref{phasnonex}
we describe in detail the class of non-extremal branes that we
consider in this paper, i.e. the non-extremal branes on a circle.
We furthermore define a $(\bar{\mu},\bar{n})$ phase diagram
for non-extremal branes on a circle for a given rescaled
charge $q$, where $\bar{\mu}$ and $\bar{n}$ are the rescaled mass
and the relative tension, respectively.

In Section \ref{getnoe}
we first describe the combined boost and U-duality transformation
on a static and neutral Kaluza-Klein black hole
that maps it to a solution for non-extremal branes on a circle.
Then we find the induced map from the $(\mu,n)$ phase
diagram to the $(\bar{\mu},\bar{n})$ phase diagram
for non-extremal branes on a circle, for a given rescaled
charge $q$.
Finally, we notice that by using the ansatz of \cite{Harmark:2002tr} for
Kaluza-Klein black holes with $SO(d-1)$ symmetry, we get
an ansatz for non-extremal branes on a circle that was also
proposed in \cite{Harmark:2002tr}, though
on rather different grounds.

In Section \ref{phasenearex}
we describe the near-extremal limit taken on non-extremal branes
on a circle that defines the class of near-extremal branes,
i.e. near-extremal branes on a circle, that we consider in this paper.
We define the $(\epsilon,r)$ phase diagram, with $\epsilon$ being
the rescaled energy above extremality and $r$ the relative tension.
This is done using the general definition of gravitational tension given in
\cite{Harmark:2004ch}.

In Section \ref{getnee}
we take the near-extremal limit of the map of Section \ref{getnoe} from
the neutral to the non-extremal case, giving a map
from static and neutral Kaluza-Klein black holes
to near-extremal branes on a circle.
We derive the induced map from the $(\mu,n)$ phase diagram
to the $(\epsilon,r)$ phase diagram.
This map has the simple form
\[
\epsilon = \frac{d+n}{2(d-1)} \mu \spa
r = 2 \frac{(d-1)n}{d+n}
\spa
\hat{\mt} =  \mt \sqrt{\mt \ms}
\spa
\hat{\ms} = \frac{\ms}{\sqrt{ \mt \ms}}
 \ ,
\]
where $\mt$, $\ms$ and  $\hmt$, $\hms$ are the rescaled
temperature, entropy of the Kaluza-Klein black hole and
near-extremal brane on a circle, respectively.
We furthermore use the map on the ansatz of \cite{Harmark:2002tr} for
Kaluza-Klein black holes with $SO(d-1)$ symmetry,
getting the ansatz for near-extremal branes on a circle
proposed in \cite{Harmark:2002tr}.

In Section \ref{s:gencon}
we explore the consequences of the map of Section \ref{getnee}
for the general understanding of near-extremal branes on a circle.
We find several features that are analogous to static and neutral
Kaluza-Klein black holes, including physical bounds on the relative
tension $r$, a generalized Smarr formula and from that an intersection rule
for the $(\epsilon,r)$ phase diagram.
We furthermore examine the general conditions
for the pressure on the world-volume of the near-extremal branes
to be positive.

In Section \ref{secloc} we first review the recently obtained analytical
results  \cite{Harmark:2003yz} (see also \cite{Gorbonos:2004uc})
for the corrected metric and thermodynamics of the black hole on
cylinder branch. We then use these together with the U-duality
mapping of Sections \ref{getnoe} and \ref{getnee} to obtain the
corrected metric and thermodynamics of non- and near-extremal
branes localized on a circle. In particular, we obtain the leading
correction to the entropy and free energy of the localized phase
of near-extremal branes. This is applied in Sections \ref{s:LST}
and \ref{secqft} to find the corrected free energy of the dual
non-gravitational theories in the localized phase.

In Section \ref{secnune} we first review the behavior of the
non-uniform string branch near the Gregory-Laflamme point on the
uniform string branch, using the numerical results for $4 \leq d
\leq 9$ obtained by
 Sorkin \cite{Sorkin:2004qq}.%
\footnote{For $d=4$ and $d=5$ these were obtained earlier in
\cite{Gubser:2001ac} and \cite{Wiseman:2002zc}.} These first
order results are then used together with the U-duality map, to
show that there exists a non-uniform phase for non-/near-extremal
branes on a circle, emanating from the uniform phase at a critical
mass/energy that is related to the Gregory-Laflamme mass. In
particular, for the near-extremal case the resulting corrections
to the entropy and free energy are analyzed in detail. Finally, a
possible violation of the Gubser-Mitra conjecture
\cite{Gubser:2000ec,Gubser:2000mm,Reall:2001ag,Gregory:2001bd} is pointed out.

Section \ref{secM5} is devoted to a special study of the
non-uniform phase of near-extremal M5-branes on a circle. This
case is particularly interesting for two reasons. First, the
thermodynamics of the uniform phase of the near-extremal M5-brane
on a circle, which is related to that of the NS5-brane, is of a
rather singular nature. Moreover, we can use the
numerical data that are available for the non-uniform black string
branch with $d=5$ \cite{Wiseman:2002zc} to map these to the
corresponding non-uniform phase of near-extremal M5-branes on a
circle. In this way, we are able to find detailed information on
the phase diagram and thermodynamics of this non-uniform phase.

In Section \ref{s:LST} we apply the results of Section \ref{secM5}
to study the thermal behavior of the non-gravitational dual
\cite{Maldacena:1997cg,Aharony:1998ub} of the near-extremal
M5-brane on a circle, which is $(2,0)$ LST
\cite{Seiberg:1997zk,Berkooz:1997cq,Dijkgraaf:1998ku}. It is shown
that the interpretation of the non-uniform phase
is the existence of a new stable phase%
\footnote{This is especially interesting in view of the earlier
results of \cite{Harmark:2000hw,Berkooz:2000mz,Kutasov:2000jp}
where the string corrections to the NS5-brane supergravity
description were
 considered. It was found
that the leading correction gives rise to a negative specific heat
of the NS5-brane, and that the temperature of the near-extremal
NS5-brane is larger than $\hat T_{\rm hg}$ \cite{Kutasov:2000jp}.}
 of $(2,0)$ LST in the canonical ensemble for temperatures
above its Hagedorn temperature
\cite{Maldacena:1996ya,Maldacena:1997cg}. We  give a quantitative
prediction for the free energy for temperatures near the Hagedorn
temperature. We also use the results obtained in Section
\ref{secloc}
 for the localized phase of near-extremal M5-branes to give a prediction
for the first correction to the thermodynamics of the
superconformal (2,0) theory, as one moves away from the infrared
fixed point. The correction depends on the dimensionless parameter
$\hat T/\hat T_{\rm hg}$.

In Section \ref{secqft} we use the localized and non-uniform phase
 of the near-extremal D0, D1, D2, D3 and M2-brane on a circle,
to obtain non-trivial predictions for the thermodynamics of the
corresponding dual SYM theories
\cite{Maldacena:1997re,Itzhaki:1998dd}. Here, the localized phase
corresponds to the low temperature/low energy regime and the
non-uniform phase emerges from the uniform phase, which
corresponds to the high temperature/high energy regime. For the
near-extremal D($p-1)$-brane on a circle the dual is thermal
($p+1$)-dimensional SYM on $\R^{p-1} \times S^1$, and we give
quantitative predictions for the first correction to the free
energy in both phases. The dimensionless expansion parameter in
the localized phase is $\hat T/\hat T_0$ with
$ \hat T_0 =\sqrt{2\pi} (\lambda \hat L^{3-p} )^{-1/2} \hat L^{-1}$,
with $\lambda$ the 't
Hooft coupling of the gauge theory and $\hat L$ the circumference of the
field theory circle. The critical temperature that characterizes
the emergence of
 the non-uniform phase is
$\hat T_c =  \hat T_0 \hmt_c$, with $\hmt_c$ a numerically
determined constant that depends on $p$. For the near-extremal
M2-brane on a circle the dual is (uncompactified) thermal
($2+1$)-dimensional SYM on $\R^{2}$. Also in this case do we give
quantitative predictions for the corrected free energy in the two
phases. In particular, it is found that the dimensionless
expansion parameter in the localized phase is $\hat T/\hat T_0$,
with $\hat T_0 =\lambda/(2\pi N^{3/2})$ and the critical
temperature of the non-uniform phase is $\hat T_c = 0.97 \hat T_0
$.

We conclude in Section \ref{s:concl} with a discussion of some of
our results and outlook for future developments.

Two appendices are included. In Appendix \ref{appnearex} we give a
direct computation of the energy and tension (using the
formulas of \cite{Hawking:1996fd} and \cite{Harmark:2004ch}
respectively) for the class of near-extremal branes that fall into
the ansatz of \cite{Harmark:2002tr}. This provides an important
consistency check on our near-extremal map. In Appendix
\ref{appcoord} we review the flat space metric of $\CM^d \times
S^1$ in the special coordinates used in \cite{Harmark:2003yz} to
write down the metric of small black holes on the cylinder. This
is relevant for the corrected metric of non- and near-extremal branes
localized on  a circle.

\subsubsection*{Note added}

While this paper was in its final stages of preparation,
the papers \cite{Bostock:2004mg,Aharony:2004ig} appeared,
discussing related matters.

\section{Review of phases of Kaluza-Klein black holes}
\label{review}

In this section we briefly review the main ideas and results
of \cite{Harmark:2002tr,Harmark:2003dg,Harmark:2003eg,Elvang:2004iz}
for use below.

We consider in this section static solutions of the vacuum Einstein
equations (i.e. pure gravity) that have an event horizon, and
that asymptote to
Minkowski-space times a circle $\CM^d \times S^1$, i.e.
Kaluza-Klein space, with $d \geq 4$. We call
these solutions static and neutral Kaluza-Klein black holes.

We write here the metric for $\CM^d \times S^1$ as
\begin{equation}
\label{flatneut}
ds^2 = - dt^2 + dr^2 + r^2 d\Omega_{d-2}^2 + dz^2 \ .
\end{equation}
Here $t$ is the time-coordinate, $r$ is the radius on $\R^{d-1}$
and $z$ is the coordinate of the circle. We define $L$
to be the circumference of the circle.
Then for any given Kaluza-Klein black hole solution with metric
$g_{\mu \nu}$ we can write
\begin{equation}
\label{neutass}
g_{tt} \simeq -1 + \frac{c_t}{r^{d-3}}
\spa
g_{zz} \simeq 1 + \frac{c_z}{r^{d-3}} \ ,
\end{equation}
as the asymptotics
for $r \rightarrow \infty$. In terms of $c_t$, $c_z$ the mass $M$
and the tension $\CT$ along the $z$-direction are given by
\cite{Harmark:2003dg,Kol:2003if}%
\footnote{Here the unit $k$-sphere has volume
$\Omega_k = 2\pi^{(k+1)/2}/ \Gamma((k+1)/2)$.}
\begin{equation}
M = \frac{\Omega_{d-2} L}{16 \pi G_{\rm N}}
\left[ (d-2) c_t - c_z \right]
\spa
\CT = \frac{\Omega_{d-2}}{16 \pi G_{\rm N}}
\left[ c_t - (d-2) c_z \right] \ .
\end{equation}
It is useful to work only with dimensionless quantities when
discussing the solutions since two solutions can only be truly
physically different if they have some dimensionless parameter
(made out of physical observables) that is different for the two
solutions. We use therefore instead the rescaled mass $\mu$ and
the relative tension $n$ \cite{Harmark:2003dg} given by
\begin{equation}
\label{munneut}
\mu = \frac{16 \pi G_{\rm N}}{L^{d-2}} M
= \frac{\Omega_{d-2}}{L^{d-3}} \left[ (d-2) c_t - c_z \right]
\spa
n = \frac{\CT L}{M} = \frac{c_t - (d-2) c_z}{(d-2)c_t - c_z} \ .
\end{equation}

The program set forth in \cite{Harmark:2003dg} is to find all
possible phases of Kaluza-Klein black holes and draw them in a
$(\mu,n)$ phase diagram. An advantage of using $n$ as a parameter
is that it satisfies the bounds \cite{Harmark:2003dg}
\begin{equation}
\label{neutbound}
0 \leq n \leq d-2 \ .
\end{equation}
We also have the bound $\mu \geq 0$. Thus, the $(\mu,n)$
diagram is restricted to these ranges on $n$ and $\mu$ which
helps to decide how many branches of solutions exist.

According to the present knowledge of phases of
static and neutral Kaluza-Klein black holes,
the $(\mu,n)$ phase diagram appears to be divided into two separate regions
\cite{Elvang:2004iz,Elvang:2004ny}:
\begin{itemize}
\item The region $0 \leq n \leq 1/(d-2)$ contains solutions
without Kaluza-Klein bubbles, and the solutions have
a local $SO(d-1)$ symmetry.
These solutions are the subject of \cite{Harmark:2003dg,Harmark:2003eg}
where they are called black holes and strings on cylinders (since
$\CM^d \times S^1$ is the cylinder $\R^{d-1} \times S^1$ for a fixed
time).
Due to the $SO(d-1)$ symmetry
there are only two types of event horizon topologies:
The event horizon topology
is $S^{d-1}$ ($S^{d-2} \times S^1$) for the black hole (string) on a cylinder.
\item The region $1/(d-2) < n \leq d-2$ contains
solutions with Kaluza-Klein bubbles.
This part of the phase diagram is the subject of \cite{Elvang:2004iz}.
We do not consider this class of solutions in this paper.
See the conclusions in Section \ref{s:concl} for further remarks
on these solutions.
\end{itemize}

In the rest of this section we review some additional facts on the region
$0 \leq n \leq 1/(d-2)$ of the $(\mu,n)$ phase diagram
that are important for this paper.
Three branches are known for $0 \leq n \leq 1/(d-2)$:

\begin{itemize}
\item[$\blacksquare$] {\sl The uniform black string branch.}
The uniform black string has the metric
\begin{equation}
ds^2 = - \left( 1 - \frac{r_0^{d-3}}{r^{d-3}} \right) dt^2
+ \left( 1 - \frac{r_0^{d-3}}{r^{d-3}} \right)^{-1} dr^2
+ r^2 d\Omega_{d-2}^2
+ dz^2 \ .
\end{equation}
{}From the metric it is easy to see that uniform black
string has $n=1/(d-2)$.
As discovered by
Gregory and Laflamme \cite{Gregory:1993vy,Gregory:1994bj},
the uniform string branch is classically stable
for $\mu > \mu_{\rm GL}$ and classically unstable for $\mu < \mu_{\rm GL}$
where $\mu_{\rm GL}$ can be obtained numerically for each dimension $d$.
See Table \ref{tabnonuni} in Section \ref{revnune}
for the numerical values of $\mu_{\rm GL}$ for $4 \leq d \leq 9$.
\item[$\blacksquare$] {\sl The non-uniform black string branch.}
This branch was discovered in \cite{Gregory:1988nb,Gubser:2001ac}.
It starts at $\mu = \mu_{\rm GL}$ with $n=1/(d-2)$ in
the uniform string branch.
The approximate behavior near the Gregory-Laflamme point
$\mu=\mu_{\rm GL}$
is studied in \cite{Gubser:2001ac,Wiseman:2002zc,Sorkin:2004qq}.
For $4 \leq d \leq 9$ the results are that
the branch moves away from the Gregory-Laflamme point
$\mu=\mu_{\rm GL}$ with decreasing $n$ and increasing
$\mu$. As shown in \cite{Harmark:2003dg} this means that the uniform
string branch has higher entropy than the non-uniform
string branch for a given mass.
For $d=5$ a large piece of the branch was found
numerically in \cite{Wiseman:2002zc} thus providing detailed knowledge
of the behavior of the branch away from $\mu=\mu_{\rm GL}$.
We give further details on this branch in Section \ref{revnune}.

\item[$\blacksquare$] {\sl The black hole on cylinder branch.}
This branch has been studied analytically in
\cite{Harmark:2002tr,Harmark:2003yz,Kol:2003if,Harmark:2003eg,%
Gorbonos:2004uc} (see also \cite{Harmark:2003dg})
and numerically for $d=4$ in \cite{Sorkin:2003ka}
and for $d=5$ in \cite{Kudoh:2003ki}.
The branch starts in $(\mu,n)=(0,0)$ and then has increasing $n$ and $\mu$.
The first part of the branch has been found analytically in
\cite{Harmark:2003yz,Gorbonos:2004uc}.
We review the computations and results of
Ref.~\cite{Harmark:2003yz} in Section
\ref{smallBH}.
\end{itemize}

\begin{figure}[ht]
\centerline{\epsfig{file=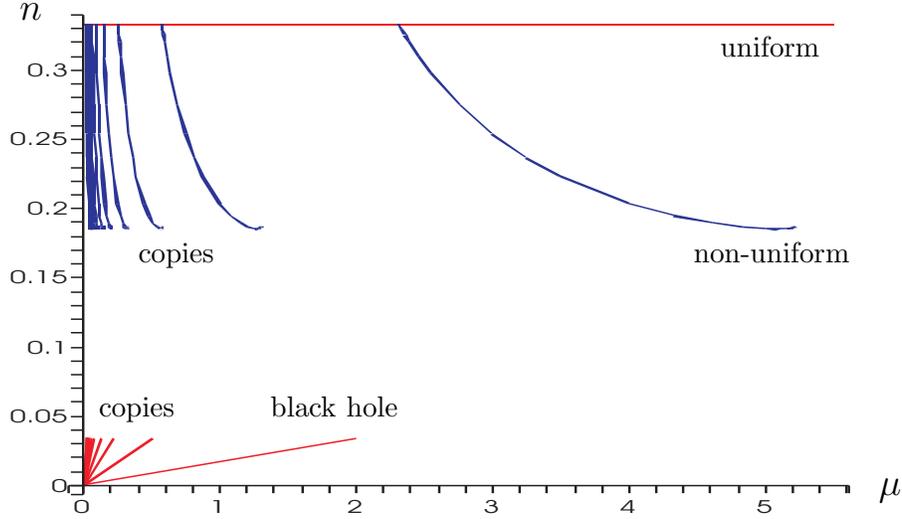,width=13 cm,height=8cm} }
 \caption{Phase diagram for $d=5$ in the region $n \leq 1/3$.}
\label{fig_neut}
\begin{picture}(0,0)(0,0)
\put(370,65){\Large $\mu$}
\put(45,245){\Large $n$}
\put(75,95){copies}
\put(90,153){copies}
\put(140,95){black hole}
\put(300,153){non-uniform}
\put(310,231){uniform}
\end{picture}
\end{figure}

In Figure \ref{fig_neut} we have displayed
the $(\mu,n)$ phase diagram
for $d=5$ for the known solutions with $0 \leq n \leq 1/3$.
The non-uniform black string branch was drawn in
\cite{Harmark:2003dg} using the data of Wiseman
\cite{Wiseman:2002zc}.

\subsubsection*{Thermodynamics}

For a neutral Kaluza-Klein black hole with a single connected
horizon, we can find the temperature $T$ and entropy $S$ directly
from the metric.
It is useful to define the rescaled
temperature $\mt$ and entropy $\ms$ by
\begin{equation}
\label{tsneut}
\mt = L T \spa
\ms = \frac{16 \pi G_{\rm N}}{L^{d-1}} S \ .
\end{equation}
In terms of these quantities,
the Smarr formula for Kaluza-Klein black holes is given by
\cite{Harmark:2003dg,Kol:2003if}
\begin{equation}
\label{neutsmarr} \mt \ms = \frac{d-2-n}{d-1} \mu \ .
\end{equation}
The first law of thermodynamics is
\begin{equation}
\label{neutfirst1}
\delta \mu = \mt\, \delta \ms \ .
\end{equation}
Combining \eqref{neutsmarr} and \eqref{neutfirst1}, we get the useful
relation
\begin{equation}
\label{neutfirst2}
\frac{\delta \log \ms}{\delta \log \mu} = \frac{d-1}{d-2-n} \ ,
\end{equation}
so that, given a curve $n(\mu)$ in
the phase diagram, the entire thermodynamics can be obtained.

As seen in \cite{Elvang:2004iz}, there are also neutral Kaluza-Klein black
hole solution with more than one connected event horizon.
The generalization of the Smarr formula \eqref{neutsmarr}
and first law \eqref{neutfirst1} were found in \cite{Elvang:2004iz}
for the specific class of solutions considered there.

\subsubsection*{The ansatz}

As mentioned above, the solutions with $0 \leq n \leq 1/(d-2)$
have, to our present knowledge, a local $SO(d-1)$ symmetry.
Using this symmetry it has been shown \cite{Wiseman:2002ti,Harmark:2003eg}
that the metric of these solutions can be written in the form
\begin{equation}
\label{ansatz}
ds^2_{d+1} = - f dt^2 + \frac{L^2}{(2\pi)^2} \left[ \frac{A}{f} dR^2
+ \frac{A}{K^{d-2}} dv^2 + K R^2 d\Omega_{d-2}^2 \right] \spa
f = 1 - \frac{R_0^{d-3}}{R^{d-3}} \ ,
\end{equation}
where $R_0$ is a
dimensionless parameter, $R$ and $v$ are dimensionless coordinates
and the metric is determined by the two functions
$A=A(R,v)$ and $K=K(R,v)$. The form \eqref{ansatz}
was originally proposed in Ref.~\cite{Harmark:2002tr}
as an ansatz for the metric of black holes on cylinders.

The properties of the ansatz \eqref{ansatz} were extensively
considered in \cite{Harmark:2002tr}.
It was found that the function $A=A(R,v)$ can be written explicitly
in terms of the function $K(R,v)$ thus reducing the number of free
unknown functions to one. The functions $A(R,v)$ and $K(R,v)$ are
periodic in $v$ with the period $2\pi$.
Note that $R = R_0$ is the location of the horizon in \eqref{ansatz}.

The asymptotic region, i.e. the region far away from the black
hole or string, is located at $R \rightarrow \infty$.
We demand that $r/R \rightarrow L/(2\pi)$ and $z/v \rightarrow L/(2\pi)$
for $R \rightarrow \infty$. This is equivalent to
demanding that $A,K \rightarrow 1$ for $R \rightarrow \infty$.
Since $K(R,v)$ in principle determines the complete metric
\eqref{ansatz}, we can read off the asymptotic quantities $\mu$
and $n$ from the first correction to $K(R,v)$.
We define the parameter $\chi$ from $K(R,v)$
in the limit $R \rightarrow \infty$ by%
\footnote{Note here that with \eqref{defchi} as the behavior of
$K(R,v)$ for $R \rightarrow \infty$ we get that
$A(R,v) = 1 - \chi \frac{R_0^{d-3}}{R^{d-3}} + \CO ( R^{-2(d-3)} )$
from the equations of motion \cite{Harmark:2002tr}.}
\begin{equation}
\label{defchi}
K(R,v) = 1 - \chi \frac{R_0^{d-3}}{R^{d-3}} +
\CO ( R^{-2(d-3)} ) \ .
\end{equation}
Then, using \eqref{munneut},
 we can write the rescaled mass $\mu$ and relative tension $n$
in terms of $R_0$ and $\chi$ as
\begin{equation}
\label{ct1}
\mu = \frac{(d-3)\Omega_{d-2}}{(2\pi)^{d-3}} R_0^{d-3}
\left[ \frac{d-2}{d-3} -  \chi \right] \spa
n =\frac{1-(d-2)(d-3)\chi}{d-2 - (d-3)\chi} \ .
\end{equation}

If we consider instead the solution near the horizon at $R=R_0$,
we can read off the temperature and entropy in a precise manner.
To this end, define
\begin{equation}
\label{defAh}
A_h \equiv A(R,v) |_{R=R_0} \ .
\end{equation}
This is a meaningful definition since,
as shown in \cite{Harmark:2002tr}, $A(R,v)$ is independent of $v$
on the horizon $R=R_0$.
The thermodynamics \cite{Harmark:2002tr,Harmark:2003dg} is then
\begin{equation}
\label{ct2}
\mt = \frac{d-3}{2 \sqrt{A_h}R_0} \spa
\ms = \frac{4\pi\Omega_{d-2}}{(2\pi)^{d-2}} \sqrt{A_h} R_0^{d-2} \ ,
\end{equation}
in terms of the rescaled temperature and entropy defined in \eqref{tsneut}.

\subsubsection*{Copies}

Finally, we note that for any solution in the ansatz
\eqref{ansatz} one can generate
an infinite number of copies.
This was found in \cite{Horowitz:2002dc,Harmark:2003eg}.
We refer to Ref.~\cite{Harmark:2003eg} for the transformation of
the solution and mention here the relation between the physical
quantities of the original solution and those of the copies
\begin{equation}
\label{neutcopies}
\mu' = \frac{\mu}{k^{d-3}} \spa
n' = n \spa
\mt' = k \mt \spa
\ms' = \frac{\ms}{k^{d-2}} \ ,
\end{equation}
where $k$ is any positive integer.
In Figure \ref{fig_neut} we depicted
the copies of the black hole on cylinder and non-uniform black
string branches for $d=5$.

\section{Defining a phase diagram for non-extremal branes on a circle}
\label{phasnonex}

We define in this section what we precisely mean by a {\sl non-extremal
brane on a circle}. We define furthermore the asymptotic parameters
that we use to categorize the solutions.

\subsection{$1/2$ BPS branes of String/M-theory \label{forma} }

Before defining more precisely the class of non-extremal branes
that we consider, we begin by
specifying what class of
extremal branes they are thermal excitations of.
As we explain in the following,
the class of branes that we consider are thermal excitations of
the $1/2$ BPS branes of Type IIA and IIB string theory and M-theory.

We consider singly-charged dilatonic $p$-branes in a $D$-dimensional space-time
which are solutions to the equations of motion of the action
\begin{equation}
\label{IDact}
I_D = \frac{1}{16 \pi G_D} \int d^D x \sqrt{-g} \left(
R - \frac{1}{2} \partial_\mu \phi \partial^\mu \phi
- \frac{1}{2(p+2)!} e^{a\phi} (F_{(p+2)})^2 \right) \ ,
\end{equation}
with $\phi$ the dilaton field and $F_{(p+2)}$ a $(p+2)$-form
field strength with $F_{(p+2)} = d A_{(p+1)}$ where
$A_{(p+1)}$ is the corresponding $(p+1)$-form gauge field.
The action \eqref{IDact} is for $D=10$ the bosonic part of the low energy
action of Type IIA and IIB string theory (in the Einstein-frame)
when only one of the gauge-fields is present,
and for $D=11$ it is the low energy action
of M-theory, for suitable choices of $a$ and $p$.
For use below we write the Einstein equations that follow from \eqref{IDact}
\begin{equation}
\label{e:einst}
R_{\mu \nu} - \frac{1}{2} g_{\mu \nu} R
= 8 \pi G_D T^{\rm mat}_{\mu \nu}
= 8 \pi G_D ( T^{\rm dil}_{\mu \nu} + T^{\rm el}_{\mu \nu}) \ ,
\end{equation}
\begin{equation}
\begin{array}{c}
\label{Tmat}
\displaystyle 8 \pi G_D T^{\rm dil}_{\mu \nu}
= - \frac{1}{4} g_{\mu \nu} \partial^\rho \phi \partial_\rho \phi
+ \frac{1}{2} \partial_\mu \phi \partial_\nu \phi \ ,
\\ \displaystyle
8 \pi G_D T^{\rm el}_{\mu \nu} =
- \frac{1}{2} g_{\mu \nu} \frac{1}{2(p+2)!} e^{a\phi} (F_{(p+2)})^2
+ \frac{1}{2(p+1)!} e^{a\phi} F_\mu^{\ \rho_1 \cdots \rho_{p+1}}
F_{\nu \rho_1 \cdots \rho_{p+1}} \ ,
\end{array}
\end{equation}
where $T^{\rm mat}_{\mu \nu}$ is the ``matter'' part of the action
\eqref{IDact} consisting of the dilaton part
$T^{\rm dil}_{\mu \nu}$ and the ``electric''
part $T^{\rm el}_{\mu \nu}$ due to the $(p+1)$-form gauge field .

We write in the following $d = D - p - 1$
as the number of directions transverse to the $p$-brane.

The extremal $1/2$ BPS $p$-brane solutions in string theory and M-theory
are then given by
\begin{equation}
\label{extr1}
ds^2 = H^{-\frac{d-2}{D-2}} \left[ - dt^2 + \sum_{i=1}^p (du^i)^2
+ H ds_d^2 \right] \ ,
\end{equation}
\begin{equation}
\label{extr2}
e^{2\phi} = H^a \spa
A_{(p+1)}= ( H^{-1} - 1 ) dt \wedge du^1 \wedge \cdots \wedge
du^p \ ,
\end{equation}
with $\nabla^2 H = 0$ (away from the sources)
where $ds_d^2$ is the (flat) metric and $\nabla^2$ the Laplacian for
the $d$-dimensional transverse space. These solutions correspond
to $1/2$ BPS extremal $p$-branes of String/M-theory when $D=10,11$ and
\begin{equation}
\label{asq}
a^2 = 4 - 2 \frac{(p+1)(d-2)}{D-2} \ .
\end{equation}
Thus with \eqref{asq} obeyed we get for $D=10$ the D-branes,
NS5-brane and the F-string of Type IIA and IIB string theory, and
for $D=11$ the M2-brane and M5-brane of M-theory. Note that the
string theory solutions are written here in the Einstein frame.

We see that for the extremal solution \eqref{extr1}-\eqref{extr2}
we can in particular consider the extremal $p$-branes to have
$\R^{d-1} \times S^1$ as the transverse space.
The thermal excitations of this class of extremal branes
are precisely the non-extremal branes that we consider below.

\subsection{Measuring asymptotic quantities \label{secmeas}}

As mentioned above, the class of non-extremal branes we
consider are thermal excitations of the extremal $1/2$ BPS branes
in Type IIA/IIB String theory and M-theory with
transverse space $\R^{d-1} \times S^1$.
Therefore, since these non-extremal branes have a transverse circle
we refer to them as {\sl non-extremal branes on a circle} in this paper.

Clearly, for a given $p$-brane solution of this type
we have that the solution asymptotes to
$\CM^{D-1} \times S^1 = \CM^{p+1} \times \R^{d-1} \times S^1$
far away from the brane,
where the $p$-brane world-volume is along the $\CM^{p+1}$ part.
To be more specific, write the
metric of $\CM^{p+1} \times \R^{d-1} \times S^1$ as
\begin{equation}
\label{flatt}
ds^2 = - dt^2 + \sum_{i=1}^p (du^i)^2
+ dr^2 + r^2 d\Omega_{d-2}^2 + dz^2 \ .
\end{equation}
Here $t$ parameterizes the time, $u^i$, $i=1,..,p$, parameterize
$\R^p$ (the spatial part of the world-volume), $r$ is the radius on $\R^{d-1}$ and $z$ is a periodic coordinate
with period $L$ that parameterizes $S^1$.
The $p$-brane metric thus asymptotes to the metric \eqref{flatt}
far away from the brane.

In the discussion below it is useful, however,
to consider each spatial world-volume direction $u^i$ of the $p$-brane
to be compactified on a circle of length $\Lu$. The solution then
asymptotes to $\CM^{d} \times \T^p \times S^1$ where $\T^p$ is
a rectangular torus with volume $V_p = \prod_{i=1}^p \Lu$.

For a given $p$-brane solution we consider the asymptotic region
defined by $r \rightarrow \infty$. For $r \rightarrow \infty$ we
have to leading order
\begin{equation}
\label{gttne}
g_{tt} \simeq -1 + \frac{\bc_{t}}{r^{d-3}}
\spa g_{zz} \simeq 1 + \frac{\bc_{z}}{r^{d-3}}
\spa g_{ii} \simeq 1 + \frac{\bc_{u}}{r^{d-3}} \spa
i = 1 \ldots p \ ,
\end{equation}
for the metric and
\begin{equation}
\label{asymAphi}
A_{(p+1)} \simeq \frac{\bc_{A}}{r^{d-3}} dt \wedge du^1 \wedge \cdots \wedge du^p
\spa
\phi \simeq \frac{\bc_{\phi}}{r^{d-3}} \ ,
\end{equation}
for the gauge field and dilaton.
By writing the leading part of the gauge field $A_{(p+1)}$
as in \eqref{asymAphi} we have specified that it is only charged along
the $p$-plane spanned by the $u^1,...,u^p$ directions.
We now explain how to measure the various asymptotic quantities
in terms of $\bc_t$, $\bc_{z}$, $\bc_{u}$, $\bc_{A}$ and $\bc_{\phi}$.

It is clear that we can use the general formulas
of \cite{Harmark:2004ch} (see also \cite{Townsend:2001rg,Traschen:2001pb})
to find the total mass $\bar{M}$, the total tension $\CTu$ along
the $i$'th world-volume direction and the tension
$\bar{\CT}$ along the $z$-direction.
This gives
\begin{equation}
\label{barM}
\bar M = \frac{V_p L \Omega_{d-2}}{16 \pi G_D}
\left[ (d-2) \bc_{t} - \bc_{z} - p \bc_{u} \right]
\spa
\bar{\CT} = \frac{V_p \Omega_{d-2}}{16 \pi G_D}
\left[ \bc_{t} - (d-2) \bc_{z} - p \bc_{u} \right] \ ,
\end{equation}
\begin{equation}
\label{CTa}
\Lu \CTu = \frac{V_p L \Omega_{d-2}}{16 \pi G_D}
\left[ \bc_{t} - \bc_{z} - (D-4) \bc_{u} \right] \ .
\end{equation}
Moreover, the charge $Q = \frac{V_p}{16\pi G_D} \int e^{a\phi} * \! F$
is measured to be
\begin{equation}
\label{theQ}
Q = \frac{V_p L \Omega_{d-2}}{16 \pi G_D} (d-3) | \bc_{A} | \ .
\end{equation}
We can furthermore split up the mass in a gravitational,
a dilatonic and an electric part as
$\bar{M} = M^{\rm gr} + M^{\rm dil} + M^{\rm el}$,
following \eqref{e:einst}-\eqref{Tmat}.
Similarly, we can split up
the tension along the world-volume as $\CTu = (\CTu)^{\rm gr}
+(\CTu)^{\rm dil} +(\CTu)^{\rm el}$.
We now explain how to measure $M^{\rm el}$ and $(\CTu)^{\rm el}$.

It is useful in the following to apply the ``principle of equivalent sources''
and think of the total energy-momentum tensor
$T_{\mu \nu} = T^{\rm gr}_{\mu \nu} +
T^{\rm dil}_{\mu \nu} + T^{\rm el}_{\mu \nu}$
as the sum of a gravitational, a dilatonic and an electric part.
Then we can write $M^{\rm gr} = \int T^{\rm gr}_{00}$,
$M^{\rm dil} = \int T^{\rm dil}_{00}$
and $M^{\rm el} = \int T^{\rm el}_{00}$, and moreover
$\Lu (\CTu)^{\rm gr} = - \int T^{\rm gr}_{ii}$,
$\Lu (\CTu)^{\rm dil} = - \int T^{\rm dil}_{ii}$
and $\Lu (\CTu)^{\rm el} = - \int T^{\rm el}_{ii}$.

If we consider the electric part of the energy-momentum tensor
$T^{\rm el}_{\mu \nu}$ in \eqref{Tmat}, we require
boost invariance on the world-volume. This means
that $T^{\rm el}_{00} = - T^{\rm el}_{ii}$ for all
$i=1,...,p$. We see that this fixes
\begin{equation}
\label{cond1}
M^{\rm el} = \Lu (\CTu)^{\rm el} \ .
\end{equation}
We require furthermore that the gravitational parts of
the energy and tensions do not affect the world-volume directions,
i.e. $\bc^{\rm gr}_{u} = 0$. Since from \cite{Harmark:2004ch}
we have that $\nabla^2 g_{ii} = - 16 \pi G_D ( T_{ii}
- \frac{1}{D-2} T^\rho{}_\rho)$ we see that this implies
$T^{\rm gr}_{ii} = \frac{1}{D-2} (T^{\rm gr})^\rho{}_\rho$.
Since furthermore one can see from \eqref{Tmat} that
$T^{\rm dil}_{ii} = \frac{1}{D-2} (T^{\rm dil})^\rho{}_\rho$,
we have that $(T^{\rm gr} + T^{\rm dil})_{ii}
= \frac{1}{D-2} (T^{\rm gr}+T^{\rm dil})^\rho{}_\rho$.
This gives the condition
\begin{equation}
\label{cond2}
\Lu ( \CTu - (\CTu)^{\rm el} ) = \frac{1}{d-1} ( \bar{M} -
M_{\rm el} + L \bar{\CT} ) \ .
\end{equation}
We get therefore
\begin{equation}
\label{Mmat}
M_{\rm el} = \frac{1}{d-2} \left[ (d-1) \Lu \CTu - \bar{M} - \bar{\CT}
\right] \ ,
\end{equation}
where $i$ can correspond to any of the world-volume directions.
Using now \eqref{barM}-\eqref{CTa} we get
\begin{equation}
M_{\rm el} = -\frac{ V_p L \Omega_{d-2}}{16 \pi G_D}
\frac{(D-2)(d-3)}{d-2} \bc_{u} \ .
\end{equation}

It is important to note that, for the class of non-extremal
$p$-branes that we consider here, $\bc_{u}$ and $\bc_{\phi}$ are
not independent. We have the relation
\begin{equation}
\label{cpcp}
\bc_{u} = - \frac{d-2}{D-2} \frac{2 \bc_{\phi}}{a} \ ,
\end{equation}
where $a$ is the constant defined in \eqref{IDact}
determining the type of brane.
This relation follows from the fact that we are looking at
$p$-branes which are thermally excited supersymmetric
$p$-branes, for which this relation applies.

The relation \eqref{cpcp} is important for the D0-brane
since $\bc_{u}$ cannot be measured in that case.
However, one can measure $\bc_{\phi}$ and use \eqref{cpcp}
to find $\bc_{u}$ and use this in the formulas \eqref{barM}-\eqref{theQ}.

\subsubsection*{Defining dimensionless quantities}

Instead of using $\bar{M}$, $Q$, $\bar{\CT}$ and $\Lu \CTu$, it is
useful to introduce the dimensionless quantities%
\footnote{Note that the definitions of the relative tensions
here are different from \cite{Harmark:2004ch}
in that they include the dilatonic
contribution.}
\begin{equation}
\label{dimqne}
\bar{\mu} = \frac{16\pi G_D}{V_p L^{d-2}} \bar{M} \spa
q = \frac{16\pi G_D}{V_p L^{d-2}} Q \spa
\bar{\mu}_{\rm el} = \frac{16\pi G_D}{V_p L^{d-2}} M^{\rm el} \ ,
\end{equation}
\begin{equation}
\label{dimqne2}
\bar n = \frac{L \bar{\CT}}{\bar{M} - M_{\rm el}} \spa
\bar{n}_i = \frac{\Lu (\CTu - (\CTu)^{\rm el})}{\bar{M} - M_{\rm el}} \ ,
\end{equation}
where $\bar{\mu}$ is the rescaled mass, $q$ the rescaled charge,
$\bar n$ the relative tension along the $z$-direction and
$\bar{n}_i$ the relative tension along the $u^i$ world-volume direction.

We get from \eqref{barM}-\eqref{theQ} the results
\begin{equation}
\label{barmun}
\bar \mu = \frac{\Omega_{d-2}}{L^{d-3}}
\left[ (d-2) \bc_{t} - p \bc_{u} - \bc_{z} \right]
\spa
q = \frac{\Omega_{d-2}}{L^{d-3}} (d-3) | \bc_{A} | \ ,
\end{equation}
\begin{equation}
\label{barn}
\bar n
=\frac{\bc_{t} - (d-2) \bc_{z} - p \bc_{u}}{(d-2) \bc_{t}
- \bc_{z} + \frac{d^2 - 4d - p + 3}{d-2} \bc_{u} } \ ,
\end{equation}
which enable the determination of the three independent physical quantities
$(\bar \mu, \bar n, q)$ from the asymptotics of the metric.

The relative tension $\bar{n}_i$ along each of the world-volume directions is
\begin{equation}
\label{ens}
\bar{n}_i = \frac{1 + \bar n}{d-1} \spa i = 1,\ldots, p \ .
\end{equation}
Eq.~\eqref{ens} shows that the relative tension $\bar{n}_i$ is affected by
the tension in the
transverse direction. The physical reason is that extra tension
is necessary in order to keep
the $p$-brane directions flat while increasing the tension in the
transverse direction. We also see from \eqref{ens} that $\bar n =0$
correctly gives $\bar{n}_i = 1/(d-1)$, reproducing the known result
for $p$-branes \cite{Townsend:2001rg}. On the other hand,
substituting $\bar n = 1/(d-2)$ gives $\bar{n}_i = 1/(d-2)$, which
is the correct result for a $(p+1)$-brane which is uniform
in all directions.
Note that from \eqref{cond1}, \eqref{cond2} and the definitions
\eqref{dimqne}, \eqref{dimqne2} it follows that the total dimensionless
world-volume tension is
\begin{equation}
\label{tauu}
\tau_{(u)} \equiv \frac{16\pi G_D}{V_p L^{d-2}}
\Lu \CTu = \bar{n}_i (\bar \mu - \bar \mu_{\rm el}) + \bar \mu_{\rm el}
= \frac{1+\bar n}{d-1} \bar \mu + \frac{d-2-\bar n}{d-1} \bar \mu_{\rm el} \ ,
\end{equation}
where we used the expression \eqref{ens} for the
relative tension $\bar{n}_i$ along the world-volume direction in terms
of the relative tension in the transverse direction.

\subsubsection*{Bounds on $\bar n$}

We examine now the general bounds \cite{Harmark:2004ch}
\begin{equation}
\label{bounds}
\bar{M}
\geq \frac{1}{D-3} \left( \sum_{i=1}^p \Lu \CTu + L \bar{\CT} \right) \spa
\CTu \geq 0  \spa \bar{\CT} \geq 0  \ ,
\end{equation}
which we require to hold separately  for all three
contributions to the mass and tension.
Firstly, for the electric contribution to the mass and tension it is
easy to see that both bounds are trivially satisfied since
$M_{\rm el}  = \Lu (\CTu)^{\rm el} \geq 0$.
If we consider \eqref{bounds}
for the gravitational plus the dilatonic contributions
together, these conditions are
easily seen to be equivalent to the conditions
\begin{equation}
\label{nbarbound}
0 \leq \bar n \leq d-2 \ ,
\end{equation}
using \eqref{dimqne2} and \eqref{ens}.
The upper bound on $\bar n$ implies the bound
$\tau_{(u)} \leq \bar{\mu}$
for the dimensionless world-volume tension.

\subsubsection*{Thermodynamics}

It will also be useful to define, in analogy with \eqref{tsneut},
the dimensionless temperature and entropy for non-extremal branes
\begin{equation}
\label{nonexts}
\bar{\mt} = L \bar{T}
\spa
\bar{\ms} = \frac{16\pi G_D}{V_p L^{d-1}} \bar{S} \ ,
\end{equation}
where $\bar{T}$ and $\bar{S}$ are the temperature and entropy
of a non-extremal brane on a circle.
Here we assume that we have only one connected event horizon. For
more event horizons, one should consider more temperatures
and entropies.
The first law of thermodynamics, for constant charge $q$,
takes the form
\begin{equation}
\label{nonfirst}
\delta \bar{\mu} = \bar{\mt} \, \delta \bar{\ms} \ .
\end{equation}
The analogue of the neutral Smarr formula \eqref{neutsmarr}
follows easily from the general Smarr formula of \cite{Harmark:2004ch}
and the definitions \eqref{dimqne}, \eqref{dimqne2} and \eqref{nonexts}.
The resulting non-extremal Smarr formula then takes the form
\begin{equation}
\label{nonexsmarr} \bar{\mt} \bar{\ms} = \frac{d-2-\bar n}{d-1}
\left( \bar{\mu} - \bar{\mu}_{\rm el} \right) \ ,
\end{equation}
which is consistent with the bounds on $\bar{n}$ in \eqref{nbarbound}.

\subsubsection*{World-volume tension is free energy}

The dimensionless Helmholtz free energy is defined by
\begin{equation}
\bar{\mf} \equiv \bar \mu - \bmt \bms \ .
\end{equation}
Using the non-extremal Smarr formula \eqref{nonexsmarr} and comparing
to \eqref{tauu} it immediately
follows that
\begin{equation}
\label{tauf}
\tau_{(u)} = \bar{\mf} \ .
\end{equation}
This shows that the world-volume tension is equal to the free energy
for the general class of near-extremal brane that we have defined.
Note in particular that this relation is a direct consequence of
the physical properties that we have imposed on the branes.

\subsubsection*{Categorizing solutions}

The program for non-extremal branes on a circle can now be formulated
as taking any solution of the class defined above
and reading off $\bar{\mu}$, $\bar{n}$ and $q$ using \eqref{barmun},
\eqref{barn}.
We can then consider all possible solutions for a given
charge $q$ and draw a $(\bar{\mu},\bar{n})$ phase diagram
depicting the values of $\bar{\mu}$ and $\bar{n}$ for all
possible solutions.
In this way we can categorize non-extremal branes on a circle
in much the same way as for neutral and static Kaluza-Klein black holes.

Note that we can immediately dismiss certain values of $\bar{\mu}$
and $\bar{n}$ as being realized for physically viable solutions.
We need clearly that $\bar{\mu} \geq q$ due to the fact that
our non-extremal branes are thermal excitations of the supersymmetric
branes with $\bar{\mu} = q$. Moreover, from \eqref{nbarbound}
we have the bounds $0 \leq \bar n \leq d-2$.

The question is now what one can say about the phase structure
for a given charge $q$: How many solutions are known and
how does the $(\bar{\mu},\bar{n})$ diagram look.
In Section \ref{getnoe} we show that we can get at least part of the
$(\bar{\mu},\bar{n})$ phase diagram by U-dualizing
 neutral and static Kaluza-Klein black holes.

\section{Generating non-extremal branes \label{getnoe}}

In this section we demonstrate that any static and neutral Kaluza-Klein
black hole solution can be mapped to a solution for a non-extremal
brane on a circle in Type IIA/B String theory and M-theory
via U-duality.%
\footnote{See e.g. Ref.~\cite{Obers:1998fb} for a review on
U-duality.} That one can ``charge up'' neutral solutions in this
fashion was originally conceived in \cite{Hassan:1992mq} where
U-duality was used to obtain black $p$-branes from neutral black
holes. Using this result we then prove  that there is a map from
the $(\mu,n)$ phase diagram for Kaluza-Klein black holes to the
$(\bar{\mu},\bar{n})$ phase diagram for non-extremal branes on a
circle. Finally, we show that one can use the map on the ansatz
\eqref{ansatz} to get an ansatz for certain phases of non-extremal
branes on a circle.

\subsection{Charging up solutions via U-duality}
\label{secUdual}

Consider a static and neutral
Kaluza-Klein black hole solution, i.e. a static vacuum solution
of $(d+1)$-dimensional General Relativity that asymptotes
to $\CM^d \times S^1$. We can always write the metric of such
a solution in the form
\begin{equation}
\label{startmet}
ds^2_{d+1} = - U dt^2 + \frac{L^2}{(2\pi)^2} V_{ab} dx^a dx^b \ ,
\end{equation}
where $a,b=1,...,d$ and $V_{ab}$ is a symmetric tensor.
Moreover, $U$ and $V_{ab}$ are functions of $x^1,..,x^d$.
In the asymptotic region we have that
$U \rightarrow 1$ and $L^2 (2\pi)^{-2} V_{ab} dx^a dx^b$
describes the cylinder $\R^{d-1} \times S^1$.
The factor $L^2/(2\pi)^2$, where $L$ is the circumference
of the circle in $\CM^d \times S^1$,
has been included in \eqref{startmet}
for later convenience.

Using the $(d+1)$-dimensional metric we can construct the eleven-dimensional
metric
\begin{equation}
\label{mthsol}
ds^2_{11} = - U dt^2 + \frac{L^2}{(2\pi)^2} V_{ab} dx^a dx^b
+ dy^2 + \sum_{i=1}^{9-d} (du^i)^2 \ .
\end{equation}
The metric \eqref{mthsol} is clearly a vacuum solution of M-theory.
Make now a Lorentz-boost along the $y$-axis so that
\begin{equation}
\label{boost}
\vecto{t_{\rm new}}{y_{\rm new}}
= \matrto{\cosh \alpha}{\sinh \alpha}{\sinh \alpha}{\cosh \alpha}
\vecto{t_{\rm old}}{y_{\rm old}} \ .
\end{equation}
This gives the boosted metric
\begin{eqnarray}
ds^2_{11} &=& \left( - U \cosh^2 \alpha + \sinh^2 \alpha \right) dt^2
+ 2 (1-U) \cosh \alpha \sinh \alpha \, dt dy \nn \\ &&
+ \left( - U \sinh^2 \alpha + \cosh^2 \alpha \right) dy^2
+ \frac{L^2}{(2\pi)^2} V_{ab} dx^a dx^b + \sum_{i=1}^{9-d} (du^i)^2 \ .
\end{eqnarray}
Since we have an isometry in the $y$-direction we can now
make an S-duality in the $y$-direction to obtain
a solution of type IIA string theory.
This gives
\begin{equation}
\label{D0a}
ds^2 = - H^{-\frac{1}{2}} U dt^2
+ H^{\frac{1}{2}} \left(  \frac{L^2}{(2\pi)^2} V_{ab} dx^a dx^b
+ \sum_{i=1}^{9-d} (du^i)^2 \right) \ ,
\end{equation}
\begin{equation}
\label{D0b}
H = 1 + \sinh^2 \alpha \, (1 - U) \spa
e^{2\phi} = H^{\frac{3}{2}} \spa
A = \coth \alpha \, ( H^{-1} - 1 ) dt  \ .
\end{equation}
This is a non-extremal D0-brane solution in type IIA string theory.
For $d < 9$ the D0-branes are uniformly smeared along an $\R^{9-d}$
space and
they have transverse space $\R^{d-1} \times S^1$, i.e. they have
a transverse circle.
The solution \eqref{D0a}-\eqref{D0b} is written in the string-frame.

We can now use U-duality to transform the solution
\eqref{D0a}-\eqref{D0b}
into a D$p$-brane solution, an F-string solution
or an NS5-brane solution of
Type IIA/B String theory, or to an M2-brane or M5-brane solution
of M-theory.
In general, we U-dualize
into the class of non-extremal branes on a circle
defined in Section \ref{phasnonex},
which are
non-extremal singly charged $p$-branes
on a circle in $D$ dimensions with $D=d+p+1$
(for String/M-theory we have $D=10$/$D=11$).%
\footnote{One could easily also use the U-duality to get
branes that are smeared uniformly in some directions, but we
choose not to consider this here.}
We can then write all these solutions as solutions to the EOMs
of the action \eqref{IDact}, as explained in Section \ref{forma}.
Thus, by U-duality we get
\begin{equation}
\label{gen1}
ds^2 = H^{-\frac{d-2}{D-2}} \left( - U dt^2 + \sum_{i=1}^p (du^i)^2
+ \frac{L^2}{(2\pi)^2} H \, V_{ab} dx^a dx^b \right) \ ,
\end{equation}
\begin{equation}
\label{gen2}
H = 1 + \sinh^2 \alpha \, (1 - U) \ ,
\end{equation}
\begin{equation}
\label{gen3}
e^{2\phi} = H^a \spa
A_{(p+1)} = \coth \alpha \, (H^{-1} -1) dt \wedge du^1 \wedge \cdots \wedge
du^p \ .
\end{equation}
This solution is a non-extremal $p$-brane with a transverse circle.
It is written in the Einstein frame for the Type IIA/B String theory
solutions.

In conclusion, we have shown by employing a boost and a U-duality
transformation that we can transform any static and neutral
Kaluza-Klein black hole solution \eqref{startmet} into
the solution \eqref{gen1}-\eqref{gen3} describing non-extremal $p$-branes
on a circle.
This is one of the central results of this paper.
Note that the transformation only works if $U$ is non-constant.

We also see that if we have an event horizon
in the metric \eqref{startmet}, $U \rightarrow 0$ near the event horizon.
This translates into an event horizon in the non-extremal
solution \eqref{gen1}-\eqref{gen3}.
Since the harmonic function \eqref{gen2} stays finite and non-zero for
$U \rightarrow 0$, we see from this that the source for the electric
potential $A_{(p+1)}$ in \eqref{gen3}
is hidden behind the event horizon.

Finally, we note that if we have a number of event horizons defined by
$U = 0$, then from \eqref{gen3} we can measure the
{\sl chemical potential} $\nu$ to be
\begin{equation}
\label{nutan}
\nu = - A_{t u^1 \cdots u^p} |_{U=0} = \tanh \alpha \ .
\end{equation}
Thus, for the class of non-extremal solutions \eqref{gen1}-\eqref{gen3},
obtained by a boost/U-duality transformation
on neutral Kaluza-Klein black holes, we have that the chemical
potential is given by \eqref{nutan}.
Therefore, even though we might have several
disconnected event horizons,
we have that $\nu = \tanh \alpha $ on all of them.

In the following we assume that we have at least
one event horizon present.

\subsection{Mapping of phase diagram  \label{secmapnoe}}

Given the above map from neutral Kaluza-Klein black holes
to non-extremal branes on a circle,
we now use the results of Section \ref{phasnonex} to
find the induced map for the physical quantities.
Our goal in the following is: Given $\mu$ and $n$ for the
Kaluza-Klein black hole, find $\bar{\mu}$ and $\bar{n}$
for the corresponding non-extremal brane on a circle for
a given charge $q$.

We begin by expressing the leading-order behavior
of the non-extremal solution \eqref{gen1}-\eqref{gen3} in terms
of that of the neutral solution \eqref{startmet}, for which we have the
asymptotics  \eqref{neutass}. We have
\begin{equation}
\label{Uexpr}
U = -g_{tt} = 1 - \frac{c_t}{r^{d-3}} + \CO ( r^{-2(d-3)} ) \ ,
\end{equation}
and hence
\begin{equation}
H = 1  + \frac{c_t\sinh^2 \al }{r^{d-3}}+ \CO ( r^{-2(d-3)} ) \ .
\end{equation}
Moreover, for the spatial part $V_{ab}$ of the neutral metric, we only
need that in spherically symmetric coordinates at infinity,
$g_{zz} = 1 + c_z/r^{d-3} + \CO ( r^{-2(d-3)} )$.
Substituting this into \eqref{gen1}-\eqref{gen3} and comparing to
\eqref{gttne}, \eqref{asymAphi} then gives the relations
\begin{equation}
\label{barctz}
\bc_{t} = c_t  + \frac{d-2}{D-2} c_t  \sinh^2 \al  \spa
\bc_{z} = c_z + \frac{p+1}{D-2} c_t  \sinh^2 \al \ ,
\end{equation}
\begin{equation}
\label{barcua}
\bc_{u} =  -\frac{d-2}{D-2} c_t  \sinh^2 \al \spa
| \bc_{A} | =c_t\cosh \al \sinh \al \ ,
\end{equation}
and $\bc_{\phi}$ related to $\bc_{u}$ by \eqref{cpcp}.
Using these expressions in \eqref{dimqne}, \eqref{dimqne2} we
can then write the three independent physical parameters
$(\bar \mu,\bar n, q )$ of the non-extremal brane
in terms of the boost parameter $\al$ and the two independent quantities
$c_t$, $c_z$ of the neutral solution. Using \eqref{munneut}, the latter two can  be eliminated
in favor of the physical parameters $(\mu,n)$ of the neutral
solution, yielding the result%
\footnote{Notice that from \eqref{barctz}-\eqref{mapne0} we get that
the electric contribution to the mass is
\begin{equation}
\bar{\mu}_{\rm el} = \nu q \ ,
\end{equation}
where the chemical potential $\nu$ is given in \eqref{nutan}.
This is the well-known relation for non-extremal $p$-branes stating
that the electric contribution to the
mass is equal to the chemical potential times the charge.}
\begin{equation}
\label{mapne0}
\bar \mu =  \mu \left(1 + \frac{d-2-n}{d-1} \sinh^2 \al \right)
 \spa
q = \mu \frac{d-2-n}{d-1} \cosh \al \, \sinh \al \spa \bar n = n \ .
\end{equation}
We can now solve the second relation in \eqref{mapne0} for $\al$,
\begin{equation}
\label{alpha}
\cosh \al (\mu,n,q) = \frac{1 + b + \sqrt{1 + b^2}}{2 \sqrt{b(1 + \sqrt{1+b^2})}}
\spa b \equiv \frac{d-2-n}{2(d-1)} \frac{\mu}{q} \ .
\end{equation}
Substituting  Eq.~\eqref{alpha}
in Eqs.~\eqref{mapne0} we obtain
the map
\begin{equation}
\label{mapp}
\bar \mu = q  +  \frac{(d+n)}{2(d-1)} \mu
+ \frac{\left(\frac{d-2-n}{2(d-1)}\right)^2 \frac{\mu^2}{q}}{
1 + \sqrt{1 + \left( \frac{(d-2-n)\mu}{2(d-1)q}   \right)^2 }}
\spa
\bar n = n \ .
\end{equation}
Thus, given a neutral Kaluza-Klein black hole
with mass $\mu$ and relative tension $n$,
the transformation \eqref{mapp} gives us the mass
$\bar \mu$ and relative tension
$\bar{n}$ for the corresponding non-extremal $p$-brane on a circle
for a given charge $q$. Note that we can also write $\bar \mu =
\mu + \nu (\mu,n,q) q $ where $\nu (\mu,n,q) = \tanh \al(\mu,n,q)$.

As a consequence of the map \eqref{mapp}
we see that from knowing the phase structure of
the $(\mu,n)$ phase diagram for neutral Kaluza-Klein black holes
we get at least part of the phase structure of the $(\bar{\mu},\bar{n})$
phase diagram for non-extremal branes on a circle.

\subsubsection*{Mapping of thermodynamics}

If we assume that the neutral Kaluza-Klein black hole with
metric \eqref{startmet} has a single connected event horizon
defined by $U =0$, we can measure the (rescaled) temperature
$\mt$ and entropy $\ms$.
For the corresponding solution of
non-extremal branes on a circle \eqref{gen1}-\eqref{gen3}
we can then read off the (rescaled) temperature
$\bar{\mt}$ and entropy $\bar{\ms}$.
By using that $H = \coth^2 \alpha$ on the horizon, it is then
easily seen that we get the map
\begin{equation}
\label{mapp2}
\bar{\mt}\bar{\ms} = \mt \ms \spa
\bar{\mt} \cosh \alpha  = \mt \ ,
\end{equation}
where $\al$ is given in \eqref{alpha} in terms of $\mu$, $n$ and $q$.

\subsubsection*{Tension on world-volume}

For use below we also give the expression of the tension
in the spatial world-volume of the brane
\begin{equation}
\label{taup}
\tau_{(u)} =  \mu \left( \frac{1+n}{d-1}
+ \frac{d-2-n}{d-1} \sinh^2 \al \right) \ ,
\end{equation}
which follows from \eqref{tauu} and \eqref{mapne0}.
Using \eqref{alpha} to eliminate $\al$, we get
\begin{equation}
\label{taup2}
\tau_{(u)} =q  +  \frac{(4-d+3n)}{2(d-1)} \mu
+ \frac{\left(\frac{d-2-n}{2(d-1)}\right)^2 \frac{\mu^2}{q}}{
1 + \sqrt{1 + \left( \frac{(d-2-n)\mu}{2(d-1)q}   \right)^2 }} \ ,
\end{equation}
which shows a similar structure as the mass $\bar \mu$ in
\eqref{mapp}.

\subsection{Ansatz for non-extremal branes on a circle}

As reviewed in Section \ref{review} we can write all
Kaluza-Klein black hole solutions
with a local $SO(d-1)$ symmetry, which seems to apply to all
solutions with $n \leq 1/(d-2)$, with the metric in the form
of the ansatz \eqref{ansatz}.
Now, using this $(d+1)$-dimensional metric \eqref{ansatz} we can apply
the same boost/U-duality transformation as for the
general metric \eqref{startmet}. Using
\eqref{gen1}-\eqref{gen3} we get the
following $p$-brane solution in $D=d+p+1$ dimensions
\begin{equation}
\label{pans1}
ds^2 = H^{-\frac{d-2}{D-2}} \left( - f dt^2 + \sum_{i=1}^p (du^i)^2
+ H \frac{L^2}{(2\pi)^2} \left[ \frac{A}{f} dR^2
+ \frac{A}{K^{d-2}} dv^2 + K R^2 d\Omega_{d-2}^2 \right] \right) \ ,
\end{equation}
\begin{equation}
e^{2\phi} = H^a \spa
A_{(p+1)} = \coth \alpha \, (H^{-1} -1) dt \wedge du^1 \wedge \cdots \wedge
du^p \ ,
\end{equation}
\begin{equation}
\label{pans3}
f = 1 - \frac{R_0^{d-3}}{R^{d-3}} \spa
H = 1 + \sinh^2 \alpha \frac{R_0^{d-3}}{R^{d-3}} \ .
\end{equation}
This solution describes non-extremal $p$-branes on a circle.
It is written in the Einstein frame for the Type IIA/B String theory
solutions.

It is interesting to notice that
this correspondence  was already discovered in Ref.~\cite{Harmark:2002tr}.
In Ref.~\cite{Harmark:2002tr} it was shown at the level of
equations of motion that for any solution that can be written in the ansatz
\eqref{ansatz} a
corresponding non-extremal $p$-brane solution
of string theory or M-theory could be obtained.
The above boost/U-duality derivation thus
provides us with a physical understanding of this correspondence.

Regarding the topology of the horizon for the non-extremal solution
\eqref{pans1}-\eqref{pans3} we note that if
the neutral solution \eqref{ansatz} is a black hole (black string)
on a cylinder it has horizon topology
$S^{d-1}$ ($S^{d-2} \times S^1$) and then the topology of the
horizon in the non-extremal $p$-brane solution \eqref{pans1}-\eqref{pans3} is
$\R^{p} \times S^{d-1}$ ($\R^{p} \times S^{d-2} \times S^1$).%
\footnote{In case the world-volume is compactified one should
replace $\R^p$ with $\T^p$.}

The dimensionless
physical quantities $\bar{\mu}$, $\bar{n}$, $\bar{\mt}$, $\bar{\ms}$,
$\nu$ and $q$ can now be found directly for a given solution in
the ansatz \eqref{pans1}-\eqref{pans3}.
Defining $\chi$ and $A_h$ as in \eqref{defchi} and \eqref{defAh},
we find \cite{Harmark:2002tr}
\begin{equation}
\label{nont1}
\bar{\mu} = \frac{ (d-3)\Omega_{d-2}}{ ( 2\pi)^{d-3} } R_0^{d-3}
\left[ \frac{d-2}{d-3} -  \chi + \sinh^2 \alpha \right] \spa
\bar{n} =\frac{1-(d-2)(d-3)\chi}{d-2 - (d-3)\chi} \ ,
\end{equation}
\begin{equation}
\bar{\mt} = \frac{d-3}{2 \sqrt{A_h}R_0  \cosh \alpha} \spa
\bar{\ms} = \frac{4\pi \Omega_{d-2}}{( 2\pi)^{d-2} }
\sqrt{A_h} R_0^{d-2} \cosh \alpha \ ,
\end{equation}
\begin{equation}
\label{nont3}
\nu = \tanh \alpha \spa
q = \frac{(d-3) \Omega_{d-2}}{ ( 2\pi)^{d-3} }
 R_0^{d-3} \cosh \alpha \, \sinh \alpha \ .
\end{equation}
It is easy to see that Eqs.~\eqref{nont1}-\eqref{nont3}
are consistent with the transformation rules \eqref{mapne0} and
\eqref{mapp2} using the thermodynamics \eqref{ct1}, \eqref{ct2} for
the ansatz \eqref{ansatz} for Kaluza-Klein black holes.

As stated in Section \ref{review} we have three phases of Kaluza-Klein
black holes with local $SO(d-1)$ symmetry: The uniform
black string, the non-uniform black string and
the black hole on cylinder branch.
Via the map of Section \ref{secUdual}, all these branches generate
a non-extremal solutions that
fit into the ansatz \eqref{pans1}-\eqref{pans3}.

We treat the non-extremal branch generated from the black hole on cylinder
in Section \ref{secloc} (the localized phase)
and the non-extremal branch generated from the non-uniform black string
in Section \ref{secnune} (the non-uniform phase).
The non-extremal branch generated from the
uniform black string branch is treated below.

\subsubsection*{The uniform phase: Non-extremal branes smeared
on a circle}

We use here the boost/U-duality map discussed above on
the uniform black string branch. This generates
non-extremal branes smeared on a circle. We refer to this
as {\sl the uniform phase} of non-extremal branes on a circle.

The uniform black string branch is obtained in
the ansatz \eqref{ansatz} by putting $A(R,v) = K(R,v) = 1$.
Therefore, we obtain the solution for non-extremal branes smeared on a circle
by setting $A(R,v) = K(R,v) = 1$ in Eqs.~\eqref{pans1}-\eqref{pans3},
yielding
\begin{equation}
\label{uni1}
ds^2 = H^{-\frac{d-2}{D-2}} \left( - f dt^2 + \sum_{i=1}^p (du^i)^2
+ H \frac{L^2}{(2\pi)^2} \left[ \frac{1}{f} dR^2
+ dv^2 + R^2 d\Omega_{d-2}^2 \right] \right) \ ,
\end{equation}
\begin{equation}
e^{2\phi} = H^a \spa
A_{(p+1)} = \coth \alpha \, (H^{-1} -1) dt \wedge du^1 \wedge \cdots \wedge
du^p \ ,
\end{equation}
\begin{equation}
\label{uni3}
f = 1 - \frac{R_0^{d-3}}{R^{d-3}} \spa
H = 1 + \sinh^2 \alpha \frac{R_0^{d-3}}{R^{d-3}} \ .
\end{equation}
This solution is well known, but we write it here explicitly
for the sake of clarity.
Note that comparing with \eqref{flatt} for $R\rightarrow \infty$
we see that $R = 2\pi r/L$ and $v = 2\pi z/L$.
The physical quantities for this solution can easily be found
by setting $\chi = 0$ and $A_h = 1$ in Eqs.~\eqref{nont1}-\eqref{nont3}.

\section{Defining a phase diagram for near-extremal branes on a circle
\label{phasenearex}}

In Section \ref{phasnonex} we defined the class of non-extremal
$p$-branes that we consider in the main part of this paper to be
thermally excited states of the $1/2$ BPS branes of String and M-theory.
In this class of branes, we can define
a near-extremal brane to be a non-extremal brane with infinitesimally
small temperature, or, equivalently, a non-extremal brane with
infinitely high charge.
As we shall see in Section \ref{getnee}, it is possible
for any of the phases of non-extremal
branes on a circle obtained from the map of Section \ref{secUdual}
to take a near-extremal limit and obtain a corresponding
phase of a near-extremal brane on a circle.
Therefore, we define in this section what we mean by
a {\sl near-extremal brane on a circle} and how to measure
the energy and tension for such a brane.
This defines a new two-dimensional phase diagram for near-extremal
branes on a circle, analogous to the two-dimensional $(\mu,n)$ phase
diagram for static and neutral Kaluza-Klein black holes.

\subsubsection*{Near-extremal branes on a circle}

Consider a given non-extremal $p$-brane on a circle, as
defined in Section \ref{phasnonex}. We can write the metric
as
\begin{equation}
\label{gnem}
ds^2 = C^{(f)} \bigg( - C^{(t)} dt^2 + \sum_{i=1}^p ( du^i)^2
+  \sum_{a,b=1}^d C^{(x)}_{ab} dx^a dx^b \bigg) \ ,
\end{equation}
where $t$ is the time-direction, $u^i$ the spatial world-volume
directions, $x^a$ the transverse directions
parameterizing $\R^{d-1} \times S^1$ in the asymptotic region, and
$C^{(f)}$, $C^{(t)}$, $C^{(x)}_{ab}$ are functions of
$x^1,...,x^d$.
For such a solution, we want to take a near-extremal limit
such that the size of the circle has the same scale as the
excitations of the energy above extremality.
This is because we want to keep the non-trivial physics related
to the presence of the circle.
Therefore, for a non-extremal $p$-brane with volume $V_p$, circumference
$L$ and rescaled charge $q$, the near-extremal limit is
\begin{equation}
\label{nelimit}
q \rightarrow \infty \spa L \rightarrow 0 \spa g
\equiv \frac{16\pi G_D}{V_p L^{d-2}}\ \, \mbox{fixed } \spa l
\equiv L \sqrt{q} \ \, \mbox{fixed } \spa
x^a\ \, \mbox{fixed } \ .
\end{equation}
Note that in the metric \eqref{gnem} we keep
$C^{(t)}(x^1,...,x^d)$
and $C^{(x)}_{ab}(x^1,...,x^d)$ fixed in the near-extremal limit \eqref{nelimit}.
As we shall see below, the near-extremal limit
\eqref{nelimit} is defined so that the
energy above extremality $\bar{\mu} - q$
is finite.

In the near-extremal limit \eqref{nelimit} we rescale the metric
in the transverse directions $x^a$. Given a non-extremal brane, this means that
after the near-extremal limit the asymptotics of the solution has
changed, and we find below how the asymptotic region looks
for near-extremal branes on a circle.
For the non-extremal branes we use the asymptotic
coordinate system \eqref{flatt}. Therefore, following \eqref{gnem}
and \eqref{nelimit},
we should rescale $\hat{r} = 2\pi r/L$ and $\hat{z} = 2\pi z/L$.
{}From this we see that the circumference of the circle
transverse to the branes has the length $2\pi$.
We also see that the asymptotic region for the near-extremal branes is
$\hat{r} \rightarrow \infty$.

To understand better the near-extremal limit \eqref{nelimit}
it is important to consider this limit taken on an extremal
brane on a circle, both since that can tell us about the asymptotic
region $\hat{r} \rightarrow \infty$
of general near-extremal branes on a circle and also since it will
be the reference space when measuring asymptotic physical quantities,
as we shall see below.

The solution of extremal branes on a circle is given in
Eqs.~\eqref{extr1}-\eqref{extr2}.
Taking the near-extremal limit \eqref{nelimit} of this solution,
we get
\begin{equation}
\label{nearsole}
ds^2
= \hat{H}^{-\frac{d-2}{D-2}} \left( - dt^2
+ \sum_{i=1}^p (du^i)^2 + \hat{H} \left[  d\hat{r}^2 +
  \hat{r}^2 d\Omega_{d-2}^2  + d\hat{z}^2  \right] \right) \ ,
\end{equation}
\begin{equation}
\label{nsex}
e^{2\phi} = \hat{H}^a
\spa
A_{(p+1)} = \hat H^{-1}  dt \wedge du^1 \wedge \cdots \wedge du^p \ ,
\end{equation}
\begin{equation}
\label{nearsole2}
\hat{H} = \frac{\hat h_d}{\hat{r}^{d-3}} \spa
\hat h_d \equiv \frac{l^2 (2\pi)^{d-5} }{(d-3) \Omega_{d-2}} \ ,
\end{equation}
for $\hat{r} \rightarrow \infty$.
Note that we defined $\hat{H} = H L^2/(2\pi)^2$
and that the solution is written in units of $L/(2\pi)$, which
means that we have rescaled the fields with powers of $L/(2\pi)$.
The solution \eqref{nearsole}-\eqref{nearsole2} is exact
for an extremal brane smeared uniformly on the circle, while there are
for example corrections for an extremal brane localized at
one point of the circle.
In general all extremal branes localized in a point of the $\R^{d-1}$
of the cylinder $\R^{d-1} \times S^1$ are described by
the solution \eqref{nearsole}-\eqref{nsex} with
\begin{equation}
\label{extrH}
\hat{H} = \frac{\hat h_d}{\hat{r}^{d-3}}
\left( 1+ \CO ( e^{-\hat{r}} ) \right) \ ,
\end{equation}
for $\hat{r} \rightarrow \infty$. Thus, the first correction
to the solution \eqref{nearsole}-\eqref{nearsole2} comes in
at order $e^{-\hat{r}}$.
This will be important below since we use the extremal solution
as the reference space for measuring asymptotic physical quantities.

If we consider now a near-extremal brane on a circle,
we have that in the asymptotic region $\hat{r} \rightarrow \infty$,
the solution asymptotes to that given by
Eqs.~\eqref{nearsole}-\eqref{nearsole2},
for a given value of $l$.
We clearly see that the near-extremal solutions asymptote
to a non-flat space-time described by the metric  \eqref{nearsole}.
As opposed to the
extremal case, the first
correction for a near-extremal brane on a circle
to the solution given by Eqs.~\eqref{nearsole}-\eqref{nearsole2}
will in general appear at order $\hat{r}^{-(d-3)}$ relative to the leading
order solution.

\subsubsection*{Measuring energy and tension}

Let a near-extremal brane on a circle be given.
We can then define asymptotic physical quantities
by comparing the solution to the extremal reference background
given by Eqs.~\eqref{nearsole}-\eqref{nsex} and \eqref{extrH}.
Using the definition of energy in \cite{Hawking:1996fd}
and the definition of tension in \cite{Harmark:2004ch}
we have that the energy $E$, the tension $\hat{\CT}$ along $z$
and the tension $\hCTu$ in a world-volume direction $u^i$,
are given by
\begin{equation}
\label{energ}
E = - \frac{2}{(2\pi)^{d-2} g V_p} \int_{S^\infty_t } N_t \left(
{\cal{K}}^{(D-2)}
- {\cal{K}}_0^{(D-2)} \right) \ ,
\end{equation}
\begin{equation}
\label{tens}
\hat{\CT} = - \frac{2}{(2\pi)^{d-2} g V_p}
\int_{\hat{S}^\infty_z } N_z \left( {\cal{K}}^{(D-2)}
- {\cal{K}}_0^{(D-2)} \right)  \ ,
\end{equation}
\begin{equation}
\label{tensl}
 \hCTu = - \frac{2}{(2\pi)^{d-2} g V_p}
\int_{\hat{S}^\infty_u} N_u \left( {\cal{K}}^{(D-2)}
- {\cal{K}}_0^{(D-2)} \right) \ .
\end{equation}
Here $N_t$, $N_z$ and $N_u$ are the lapse functions
in the $t$, $z$ and $u^i$ directions, respectively.
$S^\infty_t$ is a $(D-2)$-dimensional
surface at infinity (i.e. with $\hat{r}\rightarrow \infty$)
transverse to the $t$-direction. $\hat{S}^\infty_z$ and $\hat{S}^\infty_u$
are instead $(D-3)$-dimensional surfaces transverse to the
$z$ and $u^i$ directions, respectively, and both transverse
to the $t$-direction
since we do not want to integrate over the time-direction.
Moreover, ${\cal{K}}^{(D-2)}$ is the extrinsic curvature of the near-extremal
brane solution while ${\cal{K}}_0^{(D-2)}$ is the extrinsic curvature
of the reference space, which is the extremal brane solution
given by \eqref{nearsole}-\eqref{nsex} and \eqref{extrH}.%
\footnote{Note that for the $z$-direction we define ${\cal{K}}^{(D-2)}$
and ${\cal{K}}_0^{(D-2)}$ to be the extrinsic curvatures of the
$(D-2)$-dimensional surface $]t_1,t_2[ \times \hat{S}_z^\infty$,
i.e. including the time-direction $t$. But since the extrinsic curvatures
do not depend on the time-direction $t$ we do not integrate
over the time-direction in Eq.~\eqref{tens}. Similar comments
applies to Eq.~\eqref{tensl}.}
In Appendix \ref{appnearex} we compute $E$, $\hat{\CT}$ and $\hCTu$
using Eqs.~\eqref{energ}-\eqref{tensl} for a particular class of solutions.

We define now the
rescaled energy $\epsilon$, the relative tension $r$
and the relative world-volume tension $r_u$ as%
\footnote{The symbol $r$ is also used for the radial coordinate
in \eqref{flatneut} and \eqref{flatt}.
However, it should be clear from the context
whether $r$ means the radial coordinate or the relative tension.}
\begin{equation}
\label{epsrdef}
\epsilon = g E \spa
r = \frac{2 \pi \hat{\CT}}{E} \spa
r_u = \frac{\Lu \hCTu}{E} \ .
\end{equation}
These quantities are useful since they are dimensionless.

Given any near-extremal brane on a circle, we can read off the two
quantities $(\epsilon,r)$. Therefore, we can make an
$(\epsilon,r)$ phase diagram for all the near-extremal brane on a
circle.%
\footnote{Note that we shall see in Section \ref{secmapne} that
$r_u$ is not an independent physical quantity for the class
of near-extremal branes we consider.}
 The program for near-extremal branes on a circle can now
be formulated as taking any solution of the class defined above,
and depicting the corresponding values for $\epsilon$ and $r$ in a
$(\epsilon,r)$ phase diagram for all possible solutions. This is
analogous to the $(\mu,n)$ phase diagram for neutral and static
Kaluza-Klein black holes reviewed in Section \ref{review}.

In Sections \ref{getnee}-\ref{secnune} we show that many of the features
of the $(\mu,n)$ phase diagram for neutral and static
Kaluza-Klein black holes are carried over
to the near-extremal branes on a circle. In particular, we show
in Section \ref{getnee} that
we have a map that gives a near-extremal brane on a circle from a
neutral and static Kaluza-Klein black hole solution.

For a given near-extremal brane on a circle with a single connected
horizon we can also read off the temperature $\hat{T}$ and
entropy $\hat{S}$.%
\footnote{Note that the entropy $\hat{S}$ is found
as the area of the event horizon divided by
$\frac{1}{2} g V_p (2\pi)^{d-3}$.}
It is useful to define the dimensionless versions of
the temperature and entropy
\begin{equation}
\label{tsneex}
\hat{\mt} = l \, \hat{T} \spa
\hat{\ms} = \frac{g}{l} \hat{S} \ ,
\end{equation}
since $l$ and $g$ in \eqref{nelimit} have dimension length.
The first law of thermodynamics for near-extremal branes then
takes the form
\begin{equation}
\label{firstne}
\delta \epsilon = \hmt \, \delta \hms \ .
\end{equation}

Finally, we note that one can get the physical quantities for
the near-extremal brane directly from the physical quantities
of the non-extremal brane in the near-extremal limit \eqref{nelimit}.
We have
\begin{equation}
\label{rel1}
E = \lim_{L \rightarrow 0} \left( \bar{M} - Q \right)
\spa
\hat{\CT} = \lim_{L \rightarrow 0} \frac{L}{2\pi} \bar{\CT}
\spa
\Lu \hCTu
= \lim_{L \rightarrow 0} \left( \Lu \CTu - Q \right) \ ,
\end{equation}
\begin{equation}
\label{rel2}
\hat{T} = \lim_{L \rightarrow 0} \bar{T}
\spa
\hat{S} = \lim_{L \rightarrow 0} \bar{S} \ .
\end{equation}
We only prove the validity of these
relations for a special class of solutions in this paper, namely for
the near-extremal limit of the non-extremal branes on a circle
that can be written in the ansatz \eqref{pans1}-\eqref{pans3}.%
\footnote{In Appendix \ref{appnearex} we
compute the energy $E$ and tension $\hat{\CT}$
for near-extremal branes on a circle
in the ansatz \eqref{nearsol1}-\eqref{nearsol3}, using the general expression
for energy and tension
\eqref{energ} and \eqref{tens}. The results
match what one gets by taking the near-extremal limit
directly on the non-extremal quantities as prescribed in \eqref{rel1}.}
However, from a physical perspective, these relations are expected
to hold for the complete class of non-extremal branes on a circle
in the near-extremal limit \eqref{nelimit}.

Note that as a consequence of the first relation in \eqref{rel1}
we have that%
\footnote{In the following sections we sometimes write
$\epsilon = \bar{\mu} - q$, by a slight abuse of notation.}
\begin{equation}
\epsilon = \lim_{L \rightarrow 0} ( \bar{\mu} - q) \ .
\end{equation}
It follows from this that the energy
above extremality $\bar{\mu} - q$ is finite
in the near-extremal limit.

\section{Generating near-extremal branes \label{getnee} }

We constructed in Section \ref{getnoe} a map that takes
a given neutral and static Kaluza-Klein black hole and transforms
it into a solution for non-extremal branes on a circle,
using a boost/U-duality transformation.
In this section, we take the near-extremal limit of this
map and thereby obtain a map that instead takes
a given Kaluza-Klein black hole solution to a solution for
near-extremal branes on a circle.
This in turn induces a map from the $(\mu,n)$ phase diagram
for Kaluza-Klein black holes to the $(\epsilon,r)$
phase diagram for near-extremal branes on a circle.
Finally, we show that one can use the map on the ansatz \eqref{ansatz} to
get an ansatz for certain phases of near-extremal branes on
a circle.

\subsection{Near-extremal limit of U-dual brane solution \label{secnelim}}

Let a static and neutral Kaluza-Klein black hole be given.
The metric can be written in the form \eqref{startmet}.
In Section \ref{getnoe} we learned that we can transform
the Kaluza-Klein black hole into the solution \eqref{gen1}-\eqref{gen3}
describing non-extremal $p$-branes on a circle.
We now take the near-extremal limit \eqref{nelimit}
of \eqref{gen1}-\eqref{gen3}.
Note first that we need to write $U$ and $V_{ab}$ as functions
of the dimensionless variables $x^1,...,x^d$. In this way
$U$ and $V_{ab}$ do not change under the near-horizon limit
\eqref{nelimit}.
Note also that $\hat{r} = 2\pi r / L$ and $\hat{z} = 2\pi z / L$
can be used as two of these dimensionless variables.
Then, from \eqref{neutass} we see that
\begin{equation}
\label{Une}
U = 1 - \frac{\hat{c}_t}{\hat{r}^{d-3}} + \CO ( \hat{r}^{-2(d-3)} )
\spa
\hat{c}_t \equiv c_t \frac{L^{d-3}}{(2\pi)^{d-3}} \ ,
\end{equation}
for $\hat{r} \rightarrow \infty$.
Using now \eqref{barmun} and \eqref{barcua} we get that
the charge $q$ can be written
\begin{equation}
\label{qne}
q = \frac{(d-3)\Omega_{d-2}}{(2\pi)^{d-3}} \hat{c}_t
\cosh \alpha \sinh \alpha \ .
\end{equation}
Since $L\sqrt{q}$ should be fixed in the limit $L\rightarrow 0$
we see that $\alpha \rightarrow \infty$ with $L e^\alpha$ being fixed.
Define the rescaled harmonic function
\begin{equation}
\label{defhatH}
\hat{H} \equiv \lim_{L \rightarrow 0} \frac{L^2}{(2\pi)^2} H \ .
\end{equation}
One can check that $\hat{H}$ is finite in the near-extremal limit
\eqref{nelimit}. We can then write the resulting solution for
near-extremal $p$-branes on a circle as
\begin{equation}
\label{gensolnh}
ds^2 = \hat H^{-\frac{d-2}{D-2}} \left( - U
dt^2 + \sum_{i=1}^p (du^i)^2 + \hat H V_{ab} d x^a d x^b
\right) \ ,
\end{equation}
\begin{equation}
\label{gensolnh2}
e^{2\phi} = \hat H^a \spa
A_{(p+1)} = \hat H^{-1}  dt \wedge du^1 \wedge \cdots \wedge du^p \ ,
\end{equation}
where
\begin{equation}
\label{Hhat}
\hat H = \hat{h}_d \frac{1-U}{\hat{c}_t} \spa
\hat h_d \equiv \frac{ l^2 (2\pi)^{d-5} }{(d-3) \Omega_{d-2}} \ .
\end{equation}
Here the fields in \eqref{gensolnh}-\eqref{gensolnh2}
have been written in units of $L/(2\pi)$, i.e.
we have rescaled the fields with the appropriate powers of $L/(2\pi)$
to get a finite solution.

It is important to note that,
using \eqref{Une} and \eqref{Hhat}, we have to leading order
for $\hat{r} \rightarrow \infty$
\begin{equation}
\hat H =\frac{\hat h_d }{ \hat r^{d-3}} + \CO ( \hat{r}^{-2(d-3)} ) \ ,
\end{equation}
which agrees with the leading order harmonic function of the
near-horizon limit of the extremal case, as given in
\eqref{nearsole}-\eqref{nearsole2}.
As a consequence, the near-extremal $p$-brane solutions generated
this way correctly asymptote to the   near-extremal limit of the
extremal $p$-brane on a transverse circle, which is taken as the
reference space when calculating energy and tensions.

In conclusion, we have that for any neutral Kaluza-Klein black hole
with metric \eqref{startmet} we get the
solution \eqref{gensolnh}-\eqref{Hhat} describing
near-extremal $p$-branes on a circle.
This is one of the central results of this paper
since it enables us to find new phases
of near-extremal branes on a circle from the known
phases of neutral Kaluza-Klein black holes.
In particular, in the following we study the
near-extremal $p$-brane solutions that follow from the black hole on
cylinder branch and the non-uniform black string branch.
The resulting solutions are respectively
the phase for near-extremal branes localized on
the circle (see Section \ref{secloc}) and a new non-uniform phase
for near-extremal branes (see Section \ref{secnune}).

\subsection{Mapping of phase diagram \label{secmapne}}

We have shown above that any neutral Kaluza-Klein black hole
can be mapped into a solution for near-extremal branes on a circle.
As a consequence of this, we show in this section
that we can map the physical parameters $\mu$ and $n$ for
the neutral Kaluza-Klein black hole into $\epsilon$ and $r$
for the near-extremal branes on a circle.
We furthermore find a map for the thermodynamics.

To find this map, we first note that we can easily read off
what $\epsilon = \bar{\mu} - q$ becomes in terms of $\mu$
and $n$ using \eqref{mapp}, since $q \rightarrow \infty$
in the near-extremal limit \eqref{nelimit}.
For the tension, on the other hand, we can use that it follows
from \eqref{barM}, \eqref{barctz} and \eqref{barcua} that
\begin{equation}
\bar{\CT} = \frac{V_p \Omega_{d-2}}{16\pi G_D} \left[ c_t
- (d-2) c_z \right] \ ,
\end{equation}
for non-extremal branes on a circle related to neutral Kaluza-Klein
black holes by the map in Section \ref{secUdual}.
Then, using the near-extremal limit \eqref{nelimit} and \eqref{rel1}
we get
\begin{equation}
\hat{\CT} = \frac{\mu n}{2\pi g } \ .
\end{equation}
Collecting the above observations, we
see that
given a neutral
Kaluza-Klein black hole with mass $\mu$ and relative tension $n$
we find that the corresponding  near-extremal
$p$-brane on a circle ($D=d+p+1$) has
energy $\epsilon$ and relative tension $r$ given by
\begin{equation}
\label{nemap}
\epsilon = \frac{d+n}{2(d-1)} \mu \spa
r = 2 \frac{(d-1)n}{d+n} \ .
\end{equation}
We remind the reader that the energy $\epsilon$ and
relative tension  $r$  are defined in \eqref{epsrdef}.

The near-extremal map \eqref{nemap} is another of the central results
of this paper since it
gives a direct way to get the near-extremal $(\epsilon,r)$
phase diagram from the $(\mu,n)$ phase diagram for
neutral Kaluza-Klein black holes.

For a solution with a single connected event horizon (defined
by $U=0$) we have furthermore, that for a neutral
Kaluza-Klein black hole with (rescaled)
temperature $\mt$ and entropy $\ms$
the corresponding near-extremal brane on a circle
has the (rescaled) temperature $\hmt$ and entropy $\hms$ given by
\begin{equation}
\label{nets}
\hat{\mt} \hat{\ms} = \mt \ms \spa
\hat{\mt} =  \mt \sqrt{\mt \ms} \ .
\end{equation}
To show this, one first finds that $\hmt = \bar{\mt} \sqrt{q}$
and $\hms = \bar{\ms} / \sqrt{q}$ using \eqref{nonexts},
\eqref{nelimit}, \eqref{tsneex} and \eqref{rel2}.
Then one finds from \eqref{mapne0} and \eqref{neutsmarr} that
$q/\cosh^2 \alpha \rightarrow \mt \ms$ in the near-extremal limit
\eqref{nelimit}. Using this together with \eqref{mapp2}, one
derives the map \eqref{nets}.

\subsubsection*{World-volume tension}

Besides the mapping \eqref{nemap}, it is also interesting   to compute the
near-extremal limit \eqref{nelimit} of the expression
\eqref{taup2} for the world-volume tension of  non-extremal branes.
In terms of the relative world-volume tension
$r_u$ defined in \eqref{epsrdef} we find that%
\footnote{In terms of the non-extremal quantities, we have
$r_u = \lim_{q \rightarrow \infty} (\tau_{(u)}-q)/(\bar{\mu} -q)$.}
\begin{equation}
\label{ra}
r_u = \frac{4-d +3n}{d+n} \ .
\end{equation}
This is not an independent quantity for the near-extremal brane,
since $r_u$ is related to relative tension in the transverse direction via
\begin{equation}
\label{rar}
r_u = \frac{4-d +2r}{d} \ ,
\end{equation}
which follows  using \eqref{nemap} in \eqref{ra}. This relation is
the analogue of \eqref{ens} for non-extremal branes, which
physically expresses the fact that the gravitational part of
energy and tension does not affect the world-volume directions (see
Section \ref{secmeas}). Note that while the relation \eqref{ens}
follows in full generality from our definition of non-extremal
branes on a circle, the relation \eqref{rar} is proven here only
for the class of near-extremal branes obtained in
Section \ref{secnelim} via U-duality and the near-extremal limit.

\subsection{Ansatz for near-extremal branes on a circle}
\label{bhsans}

For the neutral Kaluza-Klein black holes with local $SO(d-1)$
symmetry we can write the metric in the ansatz \eqref{ansatz},
as reviewed in Section \ref{review}.
We can therefore use the map \eqref{gensolnh}-\eqref{Hhat}
to near-extremal $p$-branes on a circle.
The resulting near-extremal $p$-brane solution is ($D=d+p+1$)
\begin{equation}
\label{nearsol1}
ds ^2
= \hat{H}^{-\frac{d-2}{D-2}} \left( - f dt^2
+ \sum_{i=1}^p (du^i)^2 + \hat{H} \left[ \frac{A}{f} dR^2
+ \frac{A}{K^{d-2}} dv^2 + K R^2 d\Omega_{d-2}^2 \right] \right) \ ,
\end{equation}
\begin{equation}
 e^{2\phi} = \hat{H}^a
\spa
A_{(p+1)} = \hat{H}^{-1} dt \wedge du^1 \wedge \cdots \wedge du^p \ ,
\end{equation}
\begin{equation}
\label{nearsol3}
f = 1 - \frac{R_0^{d-3}}{R^{d-3}}  \spa
\hat{H} = \frac{\hat h_d}{R^{d-3}} \spa \hat h_d \equiv \frac{
l^2 (2\pi)^{d-5} }{(d-3) \Omega_{d-2}} \ .
\end{equation}
Note that it can also be obtained directly from \eqref{pans1}-\eqref{pans3}
in the near-extremal limit \eqref{nelimit}.

The ansatz \eqref{nearsol1}-\eqref{nearsol3} for near-extremal branes
on a circle was in fact already obtained in \cite{Harmark:2002tr}.
Here we see that it origins from the ansatz \eqref{ansatz}
for neutral Kaluza-Klein black holes with $SO(d-1)$ symmetry by
first doing the boost/U-duality transformation of Section \ref{secUdual},
and then the near-extremal limit \eqref{nelimit}. This gives
a physical understanding of the consistency of the ansatz
\eqref{nearsol1}-\eqref{nearsol3}.

We can now find the energy $\epsilon$, relative tension $r$,
temperature $\hmt$, entropy $\hms$ and relative
world-volume tension $r_u$ directly from the
ansatz \eqref{nearsol1}-\eqref{nearsol3}.
Defining $\chi$ and $A_h$ as in \eqref{defchi} and \eqref{defAh},
we find \cite{Harmark:2002tr}
\begin{equation}
\label{neart1}
\epsilon = \frac{(d-3)\Omega_{d-2}}{(2\pi)^{d-3}} R_0^{d-3}
\left[ \frac{d-1}{2(d-3)} - \chi \right]\spa
r =2 \frac{1-(d-2)(d-3)\chi}{d-1 - 2 (d-3)  \chi} \ ,
\end{equation}
\begin{equation}
\label{neart2}
\hat{\mt} = \frac{(d-3)^{3/2} \sqrt{\Omega_{d-2}}}{2 (2\pi)^{(d-3)/2}}
\frac{1}{\sqrt{A_h} }
R_0^{\frac{d-5}{2}}  \spa
 \hat{\ms} = \frac{2 \sqrt{\Omega_{d-2}}}{(2\pi)^{(d-3)/2}\sqrt{d-3} }
\sqrt{A_h} R_0^{\frac{d-1}{2}} \ ,
\end{equation}
\begin{equation}
\label{neart3}
r_u = \frac{5-d - 2 (d-3) \chi}{d-1 - 2(d-3)\chi} \ .
\end{equation}
One can easily check that \eqref{neart1}-\eqref{neart3}
are consistent with the mapping relations \eqref{nemap}, \eqref{nets}
and \eqref{ra},
and Eqs.~\eqref{ct1}-\eqref{ct2} for neutral Kaluza-Klein black holes
described by the ansatz \eqref{ansatz}.

It is important
to notice that there are two ways of computing $\epsilon$ and $r$.
One can apply \eqref{energ} and \eqref{tens} directly to find $E$
and $\hat{\CT}$. This is done in Appendix \ref{appnearex}.
Alternatively, one can use \eqref{nont1} and \eqref{nont3} to
find $E$ and $\hat{\CT}$ via \eqref{rel1}.
That these two ways of finding the energy and tension
give the same result is important since it shows the
consistency of the near-extremal limit, and in particular
the validity of the relations \eqref{rel1}.

\section{General consequences of near-extremal map}
\label{s:gencon}

Before examining some specific applications of the
map from neutral Kaluza-Klein black holes to near-extremal
branes on a circle found in Section \ref{secnelim},
and the induced map of the phase diagrams \eqref{nemap},
we first discuss here some further
general properties and consequences.

\subsection{General features of the $(\epsilon,r)$ phase diagram}
\label{s:genphase}

We begin by discussing some general features of the $(\epsilon,r)$
phase diagram that one can derive from the map of the phase
diagrams \eqref{nemap}.

We first notice that if we use the bounds \eqref{neutbound}
on $n$ for neutral Kaluza-Klein black holes, we get from the
map \eqref{nemap} the bounds
\begin{equation}
\label{nebound}
0 \leq r \leq d-2 \ ,
\end{equation}
for the relative tension $r$ of near-extremal branes on a circle.
Clearly the bounds \eqref{nebound} are derived here for the class
of near-extremal branes that is generated by the U-duality map and
near-extremal limit. We believe, however, that \eqref{nebound}
is generally valid, and it would be interesting to
show this from the general definitions of energy and tension
in Section \ref{phasenearex}.
Note that the upper bound on $r$
also makes physical sense in view of the near-extremal Smarr
formula \eqref{smarrneex} that we find below.

The results of Section \ref{getnee} mean that we can map all
the phases of neutral Kaluza-Klein black holes to phases of
near-extremal branes on a circle.
In this paper we map all three known phases of Kaluza-Klein
black holes with $n \leq 1/(d-2)$ to phases of near-extremal
branes on a circle.
Since we have from \eqref{nemap} that $n=1/(d-2)$ maps
to $r=2/(d-1)$, we see that all near-extremal branes on a circle
obtained here have $0 \leq r \leq 2/(d-1)$.
We deal with the known phases with
$n > 1/(d-2)$, which consist of solutions that have Kaluza-Klein
bubbles present, in a future publication \cite{Harmark:2004bb}.
Notice that these phases have $2/(d-1) < r \leq d-2$.

As reviewed in Section \ref{review}, the known phases of Kaluza-Klein
black holes with $n \leq 1/(d-2)$ consist of three phases:
The uniform black string branch, the non-uniform black string branch
and the black hole on cylinder branch.
These are the three known phases with a local $SO(d-1)$ symmetry,
and we can thus in all three cases use the ansatz \eqref{ansatz}
for the neutral solution, mapping to the near-extremal
ansatz \eqref{nearsol1}-\eqref{nearsol3}.
{}From the map in Section \ref{getnee}, we now get the
following three phases of near-extremal branes on a circle:
\begin{itemize}
\item {\sl The uniform phase.}
This phase is a near-extremal brane uniformly
smeared on the transverse circle. It comes from the uniform
black string, so it is given by the ansatz \eqref{nearsol1}-\eqref{nearsol3}
with $A(R,v)=K(R,v)=1$. For completeness, we write the solution here:
\begin{equation}
\label{uninear1}
ds ^2
= \hat{H}^{-\frac{d-2}{D-2}} \left( - f dt^2
+ \sum_{i=1}^p (du^i)^2 + \hat{H} \left[ \frac{1}{f} dR^2
+ dv^2 + R^2 d\Omega_{d-2}^2 \right] \right) \ ,
\end{equation}
\begin{equation}
e^{2\phi} = \hat{H}^a
\spa
A_{(p+1)} = \hat{H}^{-1} dt \wedge du^1 \wedge \cdots \wedge du^p \ ,
\end{equation}
\begin{equation}
\label{uninear3}
f = 1 - \frac{R_0^{d-3}}{R^{d-3}}  \spa
\hat{H} = \frac{\hat h_d}{R^{d-3}} \spa \hat h_d \equiv \frac{
l^2 (2\pi)^{d-5} }{(d-3) \Omega_{d-2}} \ .
\end{equation}
Using $n=1/(d-2)$ in \eqref{nemap} and \eqref{ra} gives
the relative tension $r = 2/(d-1)$ and
the relative world-volume tension $r_u = - (d-5)/(d-1)$.
The thermodynamics for the uniform phase is%
\footnote{$\hmf$ is the rescaled free energy defined in \eqref{defhmf}.}
\begin{equation}
\label{unise}
\hms_u (\epsilon) = \frac{4\pi}{\sqrt{d-3}} (\Omega_{d-2})^{-\frac{1}{d-3}}
\left( \frac{2 \, \epsilon }{d-1} \right)^{\frac{d-1}{2(d-3)}} \ ,
\end{equation}
\begin{equation}
\label{unifr}
\hmf_u (\hmt) = - \frac{d-5}{2} (\Omega_{d-2})^{-\frac{2}{d-5}}
\left( \frac{4 \pi \, \hmt }{(d-3)^{3/2}} \right)^{\frac{2(d-3)}{d-5}} \ .
\end{equation}
Note that we discuss a possible classical instability of
the uniform phase in Section \ref{secnune}.
\item {\sl The non-uniform phase.}
This phase is a configuration of near-extremal branes that are
non-uniformly distributed around a circle.
This phase is obtained by applying the near-extremal map of
Section \ref{secnelim} to the non-uniform black string branch
reviewed in Section \ref{review}, and in more detail in Section \ref{revnune}.
We consider the non-uniform phase in detail in Section \ref{secnune}.
\item {\sl The localized phase.} This phase is a near-extremal
brane localized on a transverse circle. It is obtained by applying
the map to the black hole on cylinder branch.
Since the black hole on cylinder branch starts in $(\mu,n)=(0,0)$,
we get from the map \eqref{nemap} that the
localized phase starts in $(\epsilon,r)=(0,0)$.
That the tension along the transverse circle is zero
is expected since the brane becomes completely localized on the circle
in the limit $\epsilon \rightarrow 0$.
Note that from \eqref{rar} this corresponds to the
relative world-volume tension $r_u = - (d-4)/d$, in agreement
with the result when we do not have any transverse circle
(see e.g. \cite{Harmark:2004ch}).
We consider the localized phase in detail in Section \ref{secloc}.
\end{itemize}

\subsubsection*{Copies of near-extremal branes on a circle}

Since we know that for any Kaluza-Klein black hole in the ansatz \eqref{ansatz}
we have an infinite number of copies \cite{Horowitz:2002dc,Harmark:2003eg},
it also follows that we have an infinite number of copies of
near-extremal branes in the ansatz \eqref{nearsol1}-\eqref{nearsol3}.
Using the transformation \eqref{neutcopies} and the map \eqref{mapp}
we can easily determine the corresponding expressions for
the physical quantities of the near-extremal copies
\begin{equation}
\label{necopies}
\epsilon' = k^{-(d-3)} \epsilon
\spa
r' = r
\spa
\hat{\mt}' = k^{-\frac{d-5}{2}} \hat{\mt}
\spa
\hat{\ms}' = k^{-\frac{d-1}{2}} \hat{\ms}  \ ,
\end{equation}
in terms of $(\epsilon,r,\hat\mt,\hat\ms)$ of the original
near-extremal solution, with $k$ being a positive integer.

\subsubsection*{Mapping of curves}

In specific applications, it is useful to convert
 the near-extremal map \eqref{nemap} to a map
of curves from the neutral $(\mu,n)$ phase diagram to the non-extremal
$(\epsilon,r)$ diagram.  Given curves $\mu (n)$, $\ms (n)$ of Kaluza-Klein
black holes we get
\begin{equation}
\epsilon (r) = \frac{ d }{2(d-1) -r} \mu \Big(\frac{dr}{2(d-1)-r} \Big)
\spa
\hms (r) = \left( \frac{2(d-1) -r }{
 2(d-2-r)  } \right)^{1/2}
 \frac{ \ms \Big(\frac{dr}{2(d-1)-r} \Big)}{\mu^{1/2}
 \Big(\frac{dr}{2(d-1)-r} \Big)} \ ,
\end{equation}
and $\hmt$ can then be found from the near-extremal Smarr
formula \eqref{smarrneex}.

\subsection{Thermodynamics}
\label{s:genthne}

We consider here the some general features of the
thermodynamics for near-extremal branes on a circle, following
from the maps \eqref{nemap} and \eqref{nets}.

We first consider the Smarr formula.
Using the Smarr formula \eqref{neutsmarr} of the neutral case and
the maps \eqref{nemap} and \eqref{nets}
we obtain the near-extremal Smarr formula
\begin{equation}
\label{smarrneex}
\hat{\mt} \hat{\ms} = 2 \frac{d-2-r}{d} \epsilon \ .
\end{equation}
This formula holds for all near-extremal branes generated via
U-duality and the near-extremal limit.
Using this in the first law \eqref{firstne}, one finds that
\begin{equation}
\label{frer0}
\frac{\delta \log \hms}{\delta \log \epsilon} = \frac{d}{2(d-2-r)} \ .
\end{equation}
A consequence of this is that given a curve
$r(\epsilon)$ in the near-extremal phase diagram, we can
calculate the entropy function $\hms (\epsilon)$ by integration
and hence obtain the entire thermodynamics.

It is important to realize that if we take two neutral Kaluza-Klein
black hole solutions with same mass $\mu$ and use the map \eqref{nemap}
 the two solutions do in general
not have the same energy $\epsilon$. This means that
one cannot directly translate comparison of entropies
between branches from the neutral case to the
near-extremal case. Thus, if one has two Kaluza-Klein black
hole branches A and B with branch A having higher entropy than B,
then after the map \eqref{nemap}, this might be reversed so
that branch B has the higher entropy for a given energy $\epsilon$.
However, thanks to the following Intersection Rule for
near-extremal branes on a circle, some features of comparison
between entropies of different branches still hold.

Consider two branches A and B that intersect
in the same solution with energy $\epsilon_* \neq 0$.
Using the first law of thermodynamics \eqref{firstne}
and subsequently the Smarr formula \eqref{smarrneex} we have
\begin{equation}
\frac{\ms_A (\epsilon)}{\ms_B (\epsilon)}
= 1 + \int_{\epsilon_*}^\epsilon d\epsilon'
\frac{\hmt_B \hms_B - \hmt_A \hms_A}{\hmt_A \hmt_B \hms_B^2}
= 1 + \int_{\epsilon_*}^\epsilon d\epsilon'
\frac{2\epsilon' }{d \, \hmt_A \hmt_B \hms_B^2}(r_A-r_B) \ .
\end{equation}
{}From this we get
\begin{itemize}
\item {\sl The Intersection Rule.}
For two branches A and B that intersect
in the same solution with energy $\epsilon_* \neq 0$
we have the following rule:
If we have $r_A(\epsilon') > r_B(\epsilon')$
for all $\epsilon'$ with $\epsilon_* < \epsilon' < \epsilon$, we have that
$\hms_A (\epsilon) > \hms_B (\epsilon)$.
On the other hand, if we have $r_A(\epsilon') > r_B(\epsilon')$
for all $\epsilon'$ with $\epsilon < \epsilon' < \epsilon_*$,
we have instead that
$\hms_A (\epsilon) < \hms_B (\epsilon)$.
\end{itemize}
Note that this is completely analogous to the Intersection Rule
for neutral Kaluza-Klein black holes found in \cite{Harmark:2003dg}.

We now turn to the canonical ensemble and consider the free energy.
The Helmholtz free energy is defined as
\begin{equation}
\label{defhmf}
\hmf = \epsilon - \hmt \hms \spa \delta \hmf = -\hms \, \delta \hmt \ .
\end{equation}
Using the Smarr formula \eqref{smarrneex} one finds
\begin{equation}
\label{frer}
\hmf   = \frac{4-d+2r}{d} \epsilon \ .
\end{equation}
In parallel with \eqref{frer0} we then have
\begin{equation}
\label{intf}
\frac{\delta \log \hmf}{\delta \log \hmt} =
2\frac{d-2-r }{d-4-2r} \ .
\end{equation}
This shows that if we know the curve $r(\hmt)$ the free energy
$\hmf (\hmt)$ can be directly calculated by integration.

\subsection{World-volume tension, pressure and free energy}

We consider in this section
the consequences of \eqref{ra} and \eqref{rar}
for the world-volume tension and relative tension that we defined
in \eqref{tensl} and \eqref{epsrdef}.

We first note that for near-extremal branes the
world-volume tension is generally negative (see also below).
This is contrary to the non-extremal case, where the world-volume
tension is always positive. It is therefore more natural to
introduce the pressure, which is related to  minus the tension and
hence generally positive. Another reason to introduce the pressure
is that this quantity has direct physical relevance in the field
theory  dual to the near-extremal brane.

In terms of the relative world-volume tension $r_u$ the dimensionless pressure
is%
\footnote{The dimensionful relation is $P \equiv - \Lu \CTu/V_p =
 - r_u E/V_p$.}
\begin{equation}
\label{pres}
\mathfrak{p} = - r_u \epsilon \ .
\end{equation}
Using the relation \eqref{rar} between world-volume and transverse
tension, and comparing with the free energy in \eqref{frer} we
observe
\begin{equation}
\label{pf}
\mathfrak{p} = -\hmf \ ,
\end{equation}
so that world-volume pressure is minus the free energy. Note that
the same result was found for non-extremal branes in \eqref{tauf}.
The result \eqref{pf} can be equivalently stated by saying
that the Gibbs free energy $G=E-TS + PV $ vanishes
for {\it all} near-extremal $p$-branes on a transverse circle.
For the standard near-extremal $p$-branes this was shown in
\cite{Harmark:2004ch}.
It would be interesting to understand the general property  we have
found here from the dual field theories.

We now further examine the consequences of the
expression for pressure in \eqref{pres}.
For \eqref{pres} to make physically sense from the dual
field theory point of view, we should be in a regime where the
world-volume tension is negative and hence the pressure positive.
{}From the mapping relation \eqref{ra} this means that the
original Kaluza-Klein black hole solution should have
\begin{equation}
\label{nne}
n \leq \frac{d-4}{3} \ ,
\end{equation}
or, in terms of the relative tension,
\begin{equation}
r \leq \frac{d-4}{2} \ .
\end{equation}
Note that when the bound is saturated the free energy (and pressure)
is zero.
Taken together with the bounds \eqref{neutbound} on $n$, we find that
the way we can satisfy \eqref{nne} is highly dependent on the dimension
$d$ and hence on the dimension  of the brane:
\begin{itemize}
\item For $d=4$ the bound \eqref{nne} is only satisfied for $n=0$,
which corresponds to a near-extremal brane localized on the transverse
circle (in the decompactification limit). This case thus corresponds to the
near-extremal NS5-brane in type II string theory. Indeed, the
thermodynamics of the near-extremal NS5-brane is known to be degenerate
with vanishing free energy.
\item For $d=5$ the bound \eqref{nne} becomes $n \leq 1/3$, so we need to take Kaluza-Klein black holes
with tensions not exceeding that of the uniform black string. So
only the Kaluza-Klein black holes with $SO(d-1)$ symmetry are allowed. The limiting
case $n=1/3$ corresponds to the M5-brane uniformly smeared on the transverse
circle, and has the same thermodynamics as the NS5-brane. The lower limit
$n=0$ is that of the standard near-extremal M5-brane with negative
free energy.
\item For $d \geq 6$ it is not difficult to see that the entire
black hole/string region $0 \leq n \leq 1/(d-2)$ is included, as
well as a subset of the region  $1/(d-2) \leq n \leq d-2$ containing
Kaluza-Klein bubbles, namely $1/(d-2) < n \leq (d-4)/3$. This case includes
e.g. phases of near-extremal D$p$-branes with $0 \leq p \leq 4$, the
F-string and the near-extremal M2-brane.
\end{itemize}

Note that
at present we only know explicit solutions in the Kaluza-Klein bubble
region $1/(d-2) < n \leq d-2$ for the cases $d=4$ and 5.
This part of the phase diagram was recently considered in detail in
\cite{Elvang:2004iz}. However, the considerations above show that
it would be very interesting to obtain such configurations for $d\geq 6$.
Near-extremal solutions with black holes and Kaluza-Klein
bubbles will be discussed in
a forthcoming work \cite{Harmark:2004bb}.

\section{Near-extremal branes localized on the circle \label{secloc}}

In this section we use the analytical results of Ref.~\cite{Harmark:2003yz}
for small black holes on the cylinder together with the U-duality
mapping of Sections \ref{getnoe} and \ref{getnee} to
obtain the first correction to the solution
 and thermodynamics of non- and near-extremal branes
localized on a circle. We refer to this as the {\sl localized phase}
of non- and near-extremal branes.

\subsection{Review of small black holes on cylinders}
\label{smallBH}

We begin by reviewing the metric for small black holes on
cylinders, obtained in Ref.~\cite{Harmark:2003yz}.%
\footnote{See also \cite{Gorbonos:2004uc} for analytic
work on small black holes on cylinders.}
This metric was obtained in the ansatz
\begin{equation}
\label{newansatz} ds^2 = - f dt^2 + \frac{L^2}{(2\pi)^2} \left[
\frac{\tilde{A}}{f} d\rht^2 +
\frac{\tilde{A}}{\tilde{K}^{d-2}} \rht^2 d\tht^2 + \tilde{K}
\rht^2 \sin^2 \tht d\Omega_{d-2}^2 \right] \spa f = 1 -
\frac{\rho_0^{d-2}}{\rht^{d-2}}  \ ,
\end{equation}
 obtained from the original ansatz \eqref{ansatz} via the
coordinate transformation
\begin{equation}
\label{coordnew}
R^{d-3} = k_d \rht^{d-2} \spa v = \pi - \frac{d-2}{d-3} k_d^{-1}
\int_{x=0}^{\tht} dx (\sin x)^{d-2} \spa k_d \equiv
\frac{1}{2\pi} \frac{d-2}{d-3} \frac{\Omega_{d-1}}{\Omega_{d-2}}
\ .
\end{equation}
We thus have the relation $\rho_0^{d-2} = k_d^{-1} R_0^{d-3}$
for the horizon radius.
The new coordinates $(\rht,\tht)$ are more suitable than
the $(R,v)$ coordinates since the solution should approach the
black hole metric on $\R^d$ as the mass becomes smaller. Indeed,
 the $(\rht,\tht)$ coordinates become like spherical coordinates as $\rho_0
\rightarrow 0$.

The flat space limit $\CM^d \times
S^1$ of the metric in the $(\rht,\tht)$ coordinate system is reviewed
in Appendix \ref{appcoord}. This takes the form of \eqref{newansatz}
 with $\rho_0 =0$ ($f=1$) and the functions $\tilde A=\tilde A_0(\rht,\tht)$,
 $\tilde K=\tilde K_0(\rht,\tht)$ given  in \eqref{AK0}.
By considering the Newtonian limit of the Einstein equations,
the leading correction to the metric for small black holes on
cylinders is for $\rht \gg \rho_0$ given by
\begin{equation}
\label{AKfirst}
\tilde{A} = \tilde{A}_0 -
\frac{\rht}{2(d-2)} \frac{\rho_0^{d-2}}{\rht^{d-2}}
\partial_{\rht} \tilde{A}_0
\spa
\tilde{K} = \tilde{K}_0 -
\frac{\rht}{2(d-2)} \frac{\rho_0^{d-2}}{\rht^{d-2}}
\partial_{\rht} \tilde{K}_0 \ .
\end{equation}
which holds to first order in $\rho_0^{d-2}$ when $\rho_0 \ll 1$.
Using the $\rht \ll 1$ expansion of $\tilde{A}_0$ and
$\tilde{K}_0$ reviewed in  \eqref{tilA0}-\eqref{tilK0} this becomes
\begin{equation}
\label{cortA}
\tilde{A} = 1+
\frac{2(d-1)\zeta(d-2)}{(d-2) (2\pi)^{d-2}}
\left[ 2\rht^{d-2} - \rho_0^{d-2} \right] + \CO ( \rht^d ) \ ,
\end{equation}
\begin{equation}
\label{cortK}
\tilde{K} = 1+
\frac{2\zeta(d-2)}{(d-2) (2\pi)^{d-2}}
\left[ 2\rht^{d-2} - \rho_0^{d-2} \right] + \CO ( \rht^d ) \ .
\end{equation}
which describe the metric for $\rho_0 \ll \rht \ll 1$.

The metric for small black holes on  cylinders for $\rho_0 \leq \rht \ll 1$
was then found by solving the vacuum Einstein equations and using the
fact that in this range $\tilde{A}, \tilde{K}$ are independent of $\tht$.
The result is given  by
\begin{equation}
\label{AKw}
\tilde{A}^{- \frac{d-2}{2(d-1)}}
= \tilde{K}^{- \frac{d-2}{2}} =
\frac{1-w^2}{w} \frac{\rht^{d-2}}{\rho_0^{d-2}} + w \ ,
\end{equation}
with $w$ a constant. This constant was subsequently fixed
by comparing  \eqref{AKw} to \eqref{cortA}-\eqref{cortK} yielding
\begin{equation}
\label{thew}
w = 1 + \frac{\zeta(d-2)}{(2\pi)^{d-2}} \rho_0^{d-2}
+ \CO (\rho_0^{2(d-2)} ) \ .
\end{equation}
We recall that the metric for larger $\rht$ is given
by \eqref{AKfirst} in the ansatz \eqref{newansatz}.

The result may then be summarized as follows:
For $\rho_0 \leq \rht \ll 1$
the metric of a small black hole on a cylinder $\R^{d-1} \times S^1$ is
given by \cite{Harmark:2003yz}
\begin{equation}
\label{met1}
ds^2 = - f dt^2 + f^{-1} G^{-\frac{2(d-1)}{d-2}} d\rht^2
+ G^{-\frac{2}{d-2}} \rht^2
\left( d\tht^2 + \sin^2 \tht \,  d\Omega_{d-2}^2 \right)
 \ ,
\end{equation}
\begin{equation}
\label{met2}
f = 1 - \frac{\rho_0^{d-2}}{\rht^{d-2}}
\spa
G(\rht) = \frac{1-w^2}{w} \frac{\rht^{d-2}}{\rho_0^{d-2}} + w
\spa
w = 1 + \frac{\zeta(d-2)}{(2\pi)^{d-2}} \rho_0^{d-2}
+ \CO ( \rho_0^{2(d-2)} ) \ ,
\end{equation}
to first order in $\rho_0^{d-2}$.
Since $M \propto \rho_0^{d-2}$ this means that we have the complete metric
for small black holes on cylinders
to first order in the mass.
Notice also that $w=1$ in the metric above yields the
$(d+1)$-dimensional Schwarzschild black hole metric, so that the small
black hole metric correctly asymptotes to that metric in the limit
$\rho_0 \rightarrow 0$.

The corrected thermodynamics can then be computed from \eqref{met1},
and we refer to Ref.~\cite{Harmark:2003yz}  
for the expressions of
$(\mu,n,\mt,\ms)$ in terms of $\rho_0$, including the first correction.
The results are nicely summarized by the simple relation
\begin{equation}
\label{nofM}
n (\mu)= \frac{(d-2)\zeta(d-2)}{2(d-1)\Omega_{d-1}} \, \mu + \CO ( \mu^2 ) \ .
\end{equation}
This shows that to first order, the relative tension
increases linearly with the (rescaled) mass, and, in particular,
gives an analytic expression for the slope, which we define by
\begin{equation}
\label{lambdadef}
\lambda_d \equiv \frac{(d-2)\zeta(d-2)}{2(d-1)\Omega_{d-1}} \ .
\end{equation}
See Fig.~\ref{fig_neut} for
a plot of \eqref{nofM} in the phase diagram for $d=5$.

{}From \eqref{nofM} and the first law of thermodynamics \eqref{neutsmarr}
we then find \cite{Harmark:2003yz}
\begin{equation}
\label{logSlogM2}
\frac{\delta \log \ms}{\delta \log \mu} = \frac{d-1}{d-2}
\left( 1 +  \frac{\zeta(d-2)}{2(d-1)\Omega_{d-1}} \, \mu
+ \CO ( \mu^2 ) \right) \ .
\end{equation}
This relation can be integrated to give
\begin{equation}
\label{cors} \ms (\mu) = C_1^{(d)} \mu^{\frac{d-1}{d-2}} \left( 1
+ \frac{\zeta(d-2)}{2(d-2)\Omega_{d-1}} \,  \mu
 + \CO ( \mu^2 )\right) \spa
C_1^{(d)} \equiv 4\pi (\Omega_{d-1})^{-\frac{1}{d-2}} (d-1)^{-\frac{d-1}{d-2}} \ ,
\end{equation}
where the constant of integration is fixed by the physical requirement
that in the limit of vanishing mass we should recover the thermodynamics
of a Schwarzschild black hole in $(d+1)$-dimensional
Minkowski space.

\subsection{Localized phase of non- and near-extremal branes}
\label{s:smallne}

We can now use the general U-duality results
\eqref{gen1}-\eqref{gen3} and the ansatz \eqref{newansatz} to
obtain the metric for a non-extremal $p$-brane solution in
$D=d+p+1$ dimensions, localized on a circle. The non-extremal
solution then takes the form
\begin{equation}
\label{pans1bh}
ds^2 = H^{-\frac{d-2}{D-2}} \left( - f dt^2 + \sum_{i=1}^p (du^i)^2
+ H  \frac{L^2}{(2\pi)^2}\left[ \frac{\tilde A}{f} d \rht^2
+ \frac{\tilde A}{\tilde K^{d-2}} d \tht^2 + \tilde K \rht^2
d\Omega_{d-2}^2 \right] \right) \ ,
\end{equation}
\begin{equation}
e^{2\phi} = H^a \spa
A_{(p+1)} = \coth \alpha \, (H^{-1} -1) dt \wedge du^1 \wedge \cdots \wedge
du^p \ ,
\end{equation}
\begin{equation}
\label{pans3bh}
f = 1 - \frac{\rho_0^{d-2}}{\rht^{d-2}} \spa
H = 1 + \sinh^2 \alpha \frac{\rho_0^{d-2}}{\rht^{d-2}} \ ,
\end{equation}
with $\tilde{A}, \;\tilde{K}$ given by the expressions in
\eqref{AKfirst} and \eqref{AKw}.

The mass $\bar \mu$ and relative tension $\bar n$ now follow
 immediately by substituting the relative tension \eqref{nofM}
 in the general map \eqref{mapp}.
Moreover, we can eliminate $\mu$ and $n$ completely and
obtain for a given $q$ the curve
$\bar n(\bar \mu;q)$ in the non-extremal $(\bar \mu,\bar n)$ phase
diagram. After some straightforward algebra, the result is
\begin{equation}
\label{mapploc2}
\bar n (\bar \mu;q) =\frac{2(d-1)}{d} \lambda_d (\bar \mu -q)
+  \Ord \Big( (\bar \mu -q)^2\Big) \ ,
\end{equation}
where $\lambda_d$ is defined in \eqref{lambdadef}.
Here we have expanded to first order in $\bar \mu -q$, keeping in mind
that the neutral relation \eqref{nofM} is the linear approximation
in terms of the neutral mass $\mu$. Eq.~\eqref{mapploc2} thus
shows that this branch is the linear approximation
around the extremal point $(\bar \mu,\bar n) = (q,0)$.
 Thus in the $(\bar \mu,\bar n)$ phase diagram, this phase
 of non-extremal branes localized on a circle starts in the
 extremal point and goes upwards with a slope that can be read off
 from \eqref{mapploc2}. The corresponding entropy and temperature
 of the branch can also be computed from the neutral
 thermodynamics \eqref{cors} and the mapping relations \eqref{mapp2}.

We recall that an essential ingredient in finding the neutral
black hole on cylinder branch was the (consistent) assumption
that in the limit of vanishing mass the black hole behaves as a point-like
object. We see here that
this assumption translates into the property that a non-extremal
$p$-brane localized on a circle becomes point-like in the extremal
limit, i.e. for small temperatures.

It is important to notice that a non-extremal $p$-brane on a circle in
the localized phase has an event horizon
with topology $\R^p \times S^{d-1}$.

\subsubsection*{Near-extremal branes localized on a circle}

The most interesting application is to near-extremal branes.
Using \eqref{newansatz} in the general expression
\eqref{nearsol1}-\eqref{nearsol3} for the
near-extremal branes on a circle in the ansatz, we find
\begin{equation}
\label{nearsol1bh}
ds^2 = \hat{H}^{-\frac{d-2}{D-2}}
\left( - f dt^2 + \sum_{i=1}^p (du^i)^2
+ \hat H  \left[ \frac{\tilde A}{f} d \rht^2
+ \frac{\tilde A}{\tilde K^{d-2}} d \tht^2 + \tilde K \rht^2
d\Omega_{d-2}^2 \right] \right) \ ,
\end{equation}
\begin{equation}
e^{2\phi} = \hat{H}^a
\spa
A_{(p+1)} = \hat{H}^{-1} dt \wedge du^1 \wedge \cdots \wedge du^p \ ,
\end{equation}
\begin{equation}
\label{nearsol3bh}
f = 1 - \left( \frac{\rho_0}{\rht} \right)^{d-2} \spa
\hat{H} =  \frac{\hat h_d}{k_d \rht^{d-2}} \spa \hat h_d \equiv \frac{
l^2 (2\pi)^{d-5} }{(d-3) \Omega_{d-2}} \ ,
\end{equation}
with $k_d$ defined in \eqref{coordnew}.

More explicitly, we can substitute the result
 \eqref{met1}  in \eqref{nearsol1bh}, giving
\begin{eqnarray}
\label{nearsol2}
ds^2 &=&  \hat{H}^{-\frac{d-2}{D-2}} \left(
- f dt^2  + \sum_{i=1}^p (du^i)^2  \right. \nn
\\
 & & \left. + \hat H  \left[ f^{-1} G^{-\frac{2(d-1)}{d-2}} d\rht^2
+ G^{-\frac{2}{d-2}} \rht^2
\left( d\tht^2 + \sin^2 \tht \,  d\Omega_{d-2}^2 \right) \right]  \right) \ ,
\end{eqnarray}
where the functions $f$ and $G$ are given in \eqref{met2}.
This background thus describes a near-extremal $p$-brane
localized on a transverse circle.

The energy $\epsilon$ and relative tension $r$ of this near-extremal
solution follow
 immediately by substituting the relative tension \eqref{nofM}
 in the general map \eqref{nemap}. Again, we can eliminate
 $\mu$ and $n$ completely,  so that we obtain
\begin{equation}
\label{rofeps}
r (\epsilon) = \frac{4(d-1)^2}{d^2} \lambda_d \epsilon + \Ord (\epsilon^2) \ ,
\end{equation}
where $\lambda_d$ is defined in \eqref{lambdadef}. Here, we have
kept only the linear term in $\epsilon$, in accordance with the
fact that the neutral solution contains only the first order correction
in $\mu$.

This equation is an important result as it determines
the first order correction to the relative tension for near-extremal branes
localized on a transverse circle. As for the neutral black hole
localized on a circle \cite{Harmark:2003yz},
we observe that $r \rightarrow 0$ for $\epsilon \rightarrow 0$.
This means that as the energy above extremality goes to zero
the near-extremal brane becomes more and more like a
near-extremal $p$-brane in asymptotically flat space.

The relation \eqref{rofeps} enables the determination of the
entire thermodynamics, using for example \eqref{frer0}. Alternatively,
one can use the neutral thermodynamics \eqref{cors}
and the near-extremal map \eqref{nemap}.
In particular, using \eqref{rofeps} in \eqref{frer0} we find
the corrected entropy function
\begin{equation}
\label{entne}
\hms (\epsilon) = \hat{C}_1^{(d)} \epsilon^{\frac{d}{2(d-2)}}
\left( 1 + \frac{(d-1) \zeta (d-2)}{d(d-2) \Omega_{d-1}} \epsilon
 +\Ord (\epsilon^2) \right) \spa
 \hat{C}_1^{(d)} \equiv \frac{ 4 \pi ( \Omega_{d-1})^{-\frac{1}{d-2}}}{\sqrt{d-2}}
 \left( \frac{2}{d} \right)^{\frac{d}{2(d-2)}} \ .
\end{equation}
Here the constant of integration $\hat{C}_1^{(d)}$ is fixed by
 the physical requirement that in
the limit of vanishing energy we should recover the thermodynamics
of the $p$-brane in asymptotically flat space.

{}From \eqref{entne} one finds using the first law of thermodynamics
\eqref{firstne} that
\begin{equation}
\label{entropyne}
\hmt  (\epsilon) = \frac{2(d-2)}{d \hat{C}_1^{(d)}} \epsilon^{\frac{d-4}{2(d-2)}}
\left( 1 - \frac{ (d-1)(3d-4) \zeta(d-2)}{d^2 (d-2) \Omega_{d-1}}
\epsilon + \Ord (\epsilon^2) \right) \ .
\end{equation}
{}From this and \eqref{rofeps} the free energy can be computed,
either by integrating \eqref{intf}, or by inverting and directly
using \eqref{frer}. We find that the free energy of the localized
phase of near-extremal $p$-branes is to first order given by%
\footnote{Note that since $\hat{K}_1^{(d)}\propto d-4$
the leading term in the free energy is zero for $d=4$, i.e. the D5- or
NS5-brane, as it should be. The first correction term on the other
hand is finite in this case.}
\begin{equation}
\label{freene}
\hmf  (\hmt) = - \hat{K}_1^{(d)} \hmt^{\frac{2(d-2)}{d-4}}
\left( 1 +  \frac{2(d-1) \zeta (d-2) \hat{K}_1^{(d)}}{(d-4)^2 \Omega_{d-1}}
 \hmt^{\frac{2(d-2)}{d-4}}
+\Ord \Big( (\hmt^{\frac{2(d-2)}{d-4}} )^2 \Big)  \right) \ ,
\end{equation}
\begin{equation}
\label{K1def}
\hat{K}_1^{(d)} \equiv \frac{d-4}{2} (\Omega_{d-1})^{-\frac{2}{d-4}} \left(
\frac{ 4\pi}{(d-2)^{3/2}} \right)^{\frac{2(d-2)}{d-4}} \ .
\end{equation}
This expression for the free energy including the first correction
at low temperatures for the localized phase of near-extremal branes
 is one of the central results of the paper.

Since the near-extremal branes localized on a circle,
fall into the ansatz
\eqref{nearsol1}-\eqref{nearsol3}, it follows from the results of
Section \ref{s:genphase} that we have an infinite number of copies
for near-extremal brane localized on a circle. The corresponding
thermodynamics of these copies can be easily found using the
mapping relation \eqref{necopies} and the thermodynamics obtained above.

As an important application of the results above, we compute in
Sections \ref{s:LST} and \ref{secqft} the resulting expressions for the corrected free energy
for the field theories that are dual to respectively the
M5-brane, D-branes and M2-brane on a transverse circle.
In particular, for the case of the M5-brane the result is
relevant for thermal (2,0) Little String Theory, for
D$(p-1)$-branes the above results are relevant for the
localized phase of thermal $(p+1)$-dimensional supersymmetric Yang-Mills
theory
 on $\R^{p-1} \times S^1$ and for the M2-brane we find new results in
uncompactified thermal $(2+1)$-dimensional supersymmetric Yang-Mills
theory on $\R^{2}$.

\section{New non-uniform phase for near-extremal branes on a circle
\label{secnune}}

Another interesting branch of neutral and static solutions
on $\CM^d \times S^1$
to map to near-extremal solutions is the non-uniform string branch.
As reviewed in Section \ref{review}, it was found in
\cite{Gregory:1988nb,Gubser:2001ac,Wiseman:2002zc,Sorkin:2004qq}
that a neutral and static
non-uniform
string branch will emerge out of
the neutral and static uniform string branch at
the Gregory-Laflamme mass $\mu = \mu_{\rm GL}$.
Thus, there exist in particular non-uniform string branches
on $\CM^d \times S^1$ with $4 \leq d \leq 9$.
Using the maps of Section \ref{getnoe} and \ref{getnee},
this means that we get new non-extremal and near-extremal $p$-brane solutions
on a circle that have horizons connected
around the circle but that are not translationally invariant
along the circle. We refer below to a $p$-brane
configuration in such a new type of phase as being in the
{\sl non-uniform phase}.

This section is structured as follows. In Section \ref{revnune}
we review certain important facts about the non-uniform string branch
for $4 \leq d \leq 9$. In Section \ref{secnunon} we consider the
map to non-extremal branes on a circle, and in Section \ref{secnunear}
to near-extremal branes on a circle.
We treat the case $d=5$ in detail in Section \ref{secM5}.

\subsection{Review of non-uniform black string branch}
\label{revnune}

We give in this section further details on the non-uniform string
branch, focusing on the behavior near the Gregory-Laflamme point
$\mu=\mu_{\rm GL}$.

As reviewed in Section \ref{review},
the non-uniform string branch
starts at $\mu = \mu_{\rm GL}$ with $n=1/(d-2)$ in
the uniform string branch and continues with decreasing $n$.
We can therefore write $n(\mu)$ for the non-uniform string branch
near the Gregory-Laflamme point as
\begin{equation}
\label{nuni}
n (\mu) = \frac{1}{d-2} - \gamma (\mu - \mu_{\rm GL} )
+ \CO ((\mu - \mu_{\rm GL})^2 )
\spa 0 \leq  \mu -\mu_{\rm GL} \ll 1 \ .
\end{equation}
The approximate behavior of the branch near the Gregory-Laflamme
point $\mu=\mu_{\rm GL}$
is studied for $d=4$ in \cite{Gubser:2001ac}, for $d=5$
in \cite{Wiseman:2002zc} and for general $d$ in \cite{Sorkin:2004qq}.
These computations give the values for the constant $\gamma$ in
\eqref{nuni} as listed in Table \ref{tabnonuni}
for $4 \leq d \leq 9$.%
\footnote{$\gamma$ in Table \ref{tabnonuni} is found
in terms of $\eta_1$ and $\sigma_2$ given in Figure 2 in \cite{Sorkin:2004qq}
by the formula
\begin{equation}
\gamma = - \frac{2(d-1)(d-3)^2 }{(d-2)^2}
\frac{\sigma_2}{(\eta_1)^2} \frac{1}{\mu_{\rm GL}} \ .
\end{equation}
$\eta_1$ and $\sigma_2$ are also found in \cite{Gubser:2001ac,Wiseman:2002zc}
for $d=4,5$.}
We see that since $\gamma$ is positive we have
that the non-uniform branch has decreasing $n$ and increasing $\mu$
for $4 \leq d \leq 9$.%
\footnote{Interestingly, in \cite{Sorkin:2004qq} it was found that
$\gamma$ is negative for $d > 12$.}
\begin{table}[ht]
\begin{center}
\begin{tabular}{|c||c|c|c|c|c|c|}
\hline
$d$ & $4$ & $5$ & $6$ & $7$ & $8$ & $9$
\\ \hline
$\mu_{\rm GL}$ & $3.52$ & $2.31$ & $1.74$ & $1.19$ & $0.79$ & $0.55$ \\
\hline
$\gamma$  & $0.14$ & $0.17$ & $0.21$ & $0.31$ & $0.47$ & $0.74$ \\
\hline
\end{tabular}
\caption{The critical masses $\mu_{\rm GL}$ for the
Gregory-Laflamme instability and the constant $\gamma$ determining
the behavior of the non-uniform branch for $0 \leq \mu - \mu_{\rm GL} \ll 1$.
\label{tabnonuni}}
\end{center}
\end{table}

Using \eqref{neutfirst2} with \eqref{nuni}, we get the following
expansion of the entropy around $\mu = \mu_{\rm GL}$
\begin{equation}
\label{sofmu}
\frac{\ms(\mu)}{\ms_{\rm GL} } =
1 + \frac{d-2}{d-3} \frac{\mu-\mu_{\rm GL}}{\mu_{\rm GL}}
+ \frac{d-2}{(d-3)^2} \left( 1- \frac{d-2}{d-1} \gamma \mu_{\rm GL} \right)
 \frac{(\mu-\mu_{\rm GL})^2}{2 \, \mu_{\rm GL}^2}
+ \CO ( (\mu-\mu_{\rm GL})^3 ) \ ,
\end{equation}
with $\ms_{\rm GL}$ being the entropy for the uniform black string
at the Gregory-Laflamme point, given by
\begin{equation}
\ms_{\rm GL} = 4\pi (\Omega_{d-2})^{-\frac{1}{d-3}} (d-2)^{-\frac{d-2}{d-3}}
\mu_{\rm GL}^{\frac{d-2}{d-3}} \ .
\end{equation}
If we compare the entropy of the non-uniform branch to
the entropy $\ms_{\rm u} (\mu)$
of the uniform branch, we get
\begin{equation}
\frac{\ms(\mu)}{\ms_{\rm u}(\mu)}
= 1 - \frac{(d-2)^2}{2(d-1)(d-3)^2} \frac{\gamma}{ \mu_{\rm GL}}
(\mu-\mu_{\rm GL})^2 + \CO ( (\mu-\mu_{\rm GL})^3 )
\end{equation}
We see from this that the difference in entropy appears
only at the second order in $\mu-\mu_{\rm GL}$.
Note that the entropy for the non-uniform branch
is less than the entropy of the uniform branch, for a given
mass $\mu$, as also mentioned in Section \ref{review}.

In the following two sections
we use the formula \eqref{nuni} for the non-uniform string
branch together with the mapping of solutions and the  phase diagrams,
to find the solutions and phase diagrams for non-extremal and near-extremal branes
on a circle.

\subsection{Non-uniform phase of non-extremal branes on a circle}
\label{secnunon}

We first remind the reader that the neutral uniform string branch is
mapped to what we can call the {\sl uniform phase} of the
non-extremal $p$-brane on a circle. I.e. it is really a $(p+1)$-brane
since it is uniformly distributed on the transverse circle.
As already explained in Section \ref{phasnonex}
the uniform phase has $\bar{n} = 1/(d-2)$.
In terms of solutions we have that the uniform string
branch can be written in the ansatz \eqref{ansatz} with $A=K=1$.
We have therefore that the uniform phase of a non-extremal
brane on a circle can be written in the ansatz \eqref{pans1}-\eqref{pans3}
with $A(R,v)=K(R,v)=1$.

We now turn to the neutral non-uniform string branch.
As reviewed above, solutions for $4 \leq d \leq 9$
were found in \cite{Gubser:2001ac,Wiseman:2002zc,Sorkin:2004qq}
near the Gregory-Laflamme mass, i.e. for  $|\mu-\mu_{\rm GL}| \ll 1$.
These solutions can be written in the ansatz \eqref{ansatz} since,
as already explained in Section \ref{review}, any solution with $SO(d-1)$
symmetry can be written in that ansatz.

Take now a neutral non-uniform string solution for a particular $d$.
Using the boost/U-duality transformation of Section \ref{secUdual}
we can clearly transform this solution to a corresponding
solution for a non-extremal $p$-brane on a circle (with $D=d+p+1$).
More specifically, we can write this solution in the
ansatz \eqref{pans1}-\eqref{pans3} using the same functions $A(R,v)$ and
$K(R,v)$ as for the neutral solution.

Thus, we have shown that we have a new branch of solutions
for non-extremal branes on a circle that are non-uniformly
distributed on the circle. We denote this branch as the {\sl non-uniform
phase}. Moreover, we have shown that
it can be written in the ansatz \eqref{pans1}-\eqref{pans3}.
We now study the properties of this new branch of solutions.

The physics of the neutral non-uniform string branch
near $\mu = \mu_{\rm GL}$
is captured by the formula \eqref{nuni}.
Using the map \eqref{mapp} from the neutral case
to the non-extremal case we see that the new branch
emerges out of the uniform phase (with $\bar{n} = 1/(d-2)$)
at the critical mass
\begin{eqnarray}
\label{e:muc}
\bar{\mu}_{\rm c} &=& \mu_{\rm GL} + \nu (\mu_{\rm GL},1/(d-2),q) q
\nn \\
&=& q + \frac{(d-1) }{2(d-2)} \mu_{\rm GL} + \frac{b_c^2}{1+\sqrt{1+b_c^2}} q
\spa b_c \equiv \frac{(d-3)\mu_{\rm GL}}{2(d-2)q} \ .
\end{eqnarray}
Note that $\bar{\mu}_c - q \simeq \mu_{\rm GL}$ for $q \ll 1$
and that $\bar{\mu}_c - q \simeq \mu_{\rm GL} (d-1)/(2(d-2))$
for $q \gg 1$, thus the value of $\bar{\mu}_c$ is decreasing
as the charge $q$ increases.
We can furthermore use the map \eqref{mapp} on Eq.~\eqref{nuni}.
This gives
\begin{equation}
\label{e:nunonex}
\bar{n} (\bar \mu;q) = \frac{1}{d-2} - \bar{\gamma}(q) (\bar{\mu} - \bar{\mu}_c)
\spa 0 \leq \bar{\mu} - \bar{\mu}_c  \ll 1 \ ,
\end{equation}
\begin{equation}
\bar{\gamma}(q) = \gamma \left[
\frac{d-1}{2(d-2)} - \frac{\gamma \mu_{\rm GL}}{2(d-1)}
+\frac{b_c}{\sqrt{1+b_c^2}}
\frac{(d-1)(d-3)+(d-2)\gamma \mu_{\rm GL}}{2(d-1)(d-2)} \right]^{-1} \ .
\end{equation}
{}From Eq.~\eqref{e:nunonex} one can see the behavior of the non-uniform
phase of non-extremal branes on a circle for masses slightly higher than
$\bar{\mu} = \bar{\mu}_c$.
Note that $\bar{\gamma}(q) \rightarrow \gamma$
for $q\rightarrow 0$, as expected.
We see from \eqref{e:nunonex} that non-uniform phase starts in
$(\bar{\mu},\bar{n})=(\bar{\mu}_c,1/(d-2))$ and continues
with increasing $\bar{\mu}$ and decreasing $\bar{n}$
with a slope given by $\bar{\gamma}(q)$.
One can now straightforwardly use Eq.~\eqref{e:nunonex} along
with the first law of thermodynamics \eqref{nonfirst}
and the Smarr formula \eqref{nonexsmarr}
to get the thermodynamics of the non-uniform phase
for $0 \leq \bar{\mu} - \bar{\mu}_c \ll 1$.
Using this one can derive that
the uniform phase has higher entropy than the non-uniform
phase for same mass $\bar{\mu}$, just as in the neutral case.

Finally, we note that there is a natural physical
interpretation of the mass $\bar{\mu}_c$ given in Eq.~\eqref{e:muc}
as being a Gregory-Laflamme critical mass of
non-extremal branes uniformly distributed on a circle
(i.e. the uniform phase). This should be understood in the sense that
for $\bar{\mu} < \bar{\mu}_c$ the uniform non-extremal branch,
as given by the solution \eqref{uni1}-\eqref{uni3}, is classically
unstable, while it is classically stable for $\bar{\mu} > \bar{\mu}_c$.
This seems natural from many points of view: $\bar{\mu}_c$
is the mass in which the non-uniform non-extremal branch
starts, and this points to having a marginal deformation for the
uniform non-extremal branch in that mass, which precisely is what one expects
for a Gregory-Laflamme critical mass.
Moreover, one can show that the
entropy of the localized phase is higher than that of the uniform phase
for a given mass $\bar{\mu} -q \ll 1$, which suggests a decay of the
uniform phase just as in the neutral case.
We note also that it seems that one can
make a precise argument in favor of $\bar{\mu}_c$
being a Gregory-Laflamme critical mass by taking the
unstable mode for the neutral uniform black string
 and use the boost and U-duality transformation
of Section \ref{secUdual} to transform it into an unstable mode
for non-extremal branes on a circle. This would be interesting
to pursue further.

\subsection{Non-uniform phase of near-extremal branes on a circle}
\label{secnunear}

As explained in Section \ref{s:genphase}, the uniform black string phase
of neutral Kaluza-Klein black holes is mapped
onto the uniform phase of near-extremal branes on a circle, obtained
as the near-extremal limit of non-extremal branes smeared on a circle.
The uniform phase of near-extremal branes on a circle is described
by \eqref{uninear1}-\eqref{uninear3},%
\footnote{As noted in Section \ref{s:genphase} this is the solution
found by putting $A(R,v)=K(R,v)=1$ in the ansatz for near-extremal
branes \eqref{nearsol1}-\eqref{nearsol3}.}
and it has relative tension $r = 2/(d-1)$.

If we instead consider the map of Section \ref{secnelim} from the neutral
to the near-extremal case applied to the non-uniform black string phase of
neutral Kaluza-Klein black holes, we clearly have that this phase
is mapped to a non-uniform phase of near-extremal branes on a circle.
A near-extremal $p$-brane on a circle in this
non-uniform phase is characterized by having an event horizon
with topology $\R^p \times S^{d-2} \times S^1$, and by
not having translational invariance along the transverse circle.
We remark that since we can write the non-uniform black string
branch in the ansatz \eqref{ansatz}, it follows that
the non-uniform phase of near-extremal branes on a circle
can be written in the ansatz \eqref{nearsol1}-\eqref{nearsol3}.

We can now use the map \eqref{nemap} to obtain the $(\epsilon,r)$
phase diagram for the non-uniform phase of near-extremal branes on
a circle.
We first note that the point $(\mu_{\rm GL},1/(d-2))$ in
which the non-uniform black string branch emanates from the
uniform black string branch, is mapped to the point
$(\epsilon_c,2/(d-1))$ in the $(\epsilon,r)$
phase diagram with critical energy
\begin{equation}
\epsilon_c = \frac{d-1}{2(d-2)} \mu_{\rm GL} \ .
\end{equation}
We then use the map \eqref{nemap} on the first part of the
non-uniform black string branch, which is described by \eqref{nuni}.
This gives
\begin{equation}
\label{nearr}
r (\epsilon) = \frac{2}{d-1} - \hat{\gamma} ( \epsilon - \epsilon_c )
+ \CO ( ( \epsilon - \epsilon_c )^2 )
\spa 0 \leq \epsilon-\epsilon_c \ll 1 \ ,
\end{equation}
with $\hat{\gamma}$ given by
\begin{equation}
\hat{\gamma} = \frac{4d(d-2)^3}{(d-1)^2} \frac{\gamma}{(d-1)^2-(d-2)\gamma\mu_{\rm GL}} \ .
\end{equation}
In Table \ref{tabnear}
we have listed the explicit values of the two key numbers
$\epsilon_c$ and $\hat{\gamma}$ which characterize the behavior
of the non-uniform phase of near-extremal branes on a circle
for $0 \leq \epsilon - \epsilon_c \ll 1$.

\begin{table}[ht]
\begin{center}
\begin{tabular}{|c||c|c|c|c|c|c|}
\hline
$d$ & $4$ & $5$ & $6$ & $7$ & $8$ & $9$
\\ \hline \hline
$\epsilon_c$ & $2.64$ & $1.54$ & $1.09$ & $0.71$ & $0.46$ & $0.31$ \\
\hline
$\hat{\gamma}$  & $0.25$ & $0.39$ & $0.55$ & $0.88$ & $1.42$ & $2.33$ \\
\hline
\hline
$\hmt_c$  & $0.75$ & $1$ & $1.07$ & $1.05$ & $0.97$ & $0.89$ \\
\hline
$\hms_c$  & $2.33$ & $1.54$ & $1.22$ & $0.91$ & $0.68$ & $0.53$ \\
\hline
$c$  & $-433$ & $6.46$ & $3.06$ & $1.77$ & $1.14$ & $0.79$ \\
\hline
\end{tabular}
\caption{In this table we list the critical energy $\epsilon_c$
 and the constant $\hat{\gamma}$ determining
the non-uniform phase of near-extremal branes on a circle
for $0 \leq \epsilon - \epsilon_c \ll 1$.
In addition we list the critical temperature $\hmt_c$ and the
critical entropy $\hms_c$, as well as the heat capacity $c$ of the non-uniform
phase at the critical point.
\label{tabnear}}
\end{center}
\end{table}

{}From \eqref{nearr} we see that the non-uniform phase
of near-extremal branes on a circle starts out
at $(\epsilon,r) = (\epsilon_c,2/(d-1))$ and then
continues with increasing $\epsilon$ and decreasing $r$,
since $\hat{\gamma}$ is positive.

In Section \ref{secM5} we treat the case $d=5$ in detail, since
we can use the maps \eqref{nemap} and \eqref{nets} on Wiseman's
numerical data for the non-uniform black string on $\CM^5 \times S^1$
to get an $(\epsilon,r)$ phase diagram in the near-extremal case.
The $(\epsilon,r)$ diagram for $d=5$ is depicted in Figure \ref{fig_ne}.

As in the non-extremal case, the natural interpretation
of the critical energy $\epsilon_c$ is that of
a Gregory-Laflamme type critical energy for near-extremal branes
smeared on a circle, i.e. the uniform phase.%
\footnote{In particular, it is easy to see by comparing \eqref{unise}
and \eqref{entne}
that the uniform phase has lower entropy than the localized phase
for $\epsilon \ll 1$.}
Assuming this is the right interpretation of $\bar{\mu}_c$ for non-extremal
branes smeared on a circle, that feature should still be true after
the near-extremal limit.
Thus, it seems evident that the solution describing
near-extremal branes smeared on a circle is classically unstable
for small energies $\epsilon < \epsilon_c$, and classically stable
for large energies $\epsilon > \epsilon_c$.
The fact that the non-uniform phase emanates from
the uniform phase in $(\epsilon_c,2/(d-1))$ then
corresponds to the existence of a
marginal deformation of the near-extremal
uniform phase for $\epsilon = \epsilon_c$.

\subsubsection*{Thermodynamics of non-uniform phase}

We consider first the thermodynamics in the microcanonical
ensemble, i.e. for fixed energy $\epsilon$.
Using \eqref{nearr} in \eqref{frer0} we find the expansion
of the entropy for the non-uniform phase expanded around $\epsilon
= \epsilon_c$ to be
\begin{equation}
\label{sofeps}
\frac{\hms ( \epsilon ) }{\hms_c}
= 1 + \frac{d-1}{2(d-3)} \frac{\epsilon - \epsilon_c}{\epsilon_c}
- \left[ \frac{(d-1)(d-5)}{4(d-3)^2}
+ \frac{(d-1)^2 \hat{\gamma} \epsilon_c}{2d(d-3)^2}
 \right]  \frac{(\epsilon - \epsilon_c)^2}{2 \, \epsilon_c^2}
+ \CO ( (\epsilon - \epsilon_c)^3 ) \ ,
\end{equation}
with $\hms_c$ the critical entropy
\begin{equation}
\label{scdef}
\hms_c = \frac{4\pi}{\sqrt{d-3}} (\Omega_{d-2})^{-\frac{1}{d-3}}
\left( \frac{2 \, \epsilon_c}{d-1} \right)^{\frac{d-1}{2(d-3)}} \ .
\end{equation}
The numerical values of $\hms_c$ are listed in Table \ref{tabnear}.
Comparing the entropy $\hms (\epsilon)$ of the non-uniform phase
to the entropy $\hms_u (\epsilon)$ in \eqref{unise} of the uniform
phase we get
\begin{equation}
\label{ssu}
\frac{\hms ( \epsilon ) }{\hms_u ( \epsilon ) }
= 1 - \frac{(d-1)^2 }{4d(d-3)^2} \frac{ \hat{\gamma}}{ \epsilon_c}
(\epsilon - \epsilon_c)^2
+ \CO ( (\epsilon - \epsilon_c)^3 ) \ .
\end{equation}
We see from this that the entropy of the non-uniform phase
is less than the entropy of the uniform phase for a given
energy $\epsilon$. This can also be seen directly
from \eqref{nearr} using the Intersection Rule for near-extremal
branes on a circle derived in Section \ref{s:genthne}.
Notice that the entropy of the non-uniform phase deviates from
that of the uniform phase only to second order.

We now turn to the canonical ensemble, i.e. the thermodynamics
with fixed temperature $\hmt$.
The critical temperature $\hmt_c$
at which the non-uniform phase emanates
from the uniform phase is given by
\begin{equation}
\label{tcdef}
\hmt_c = \frac{(d-3)^{\frac{3}{2}}}{4\pi}
(\Omega_{d-2})^{\frac{1}{d-3}}
\left( \frac{2 \, \epsilon_c}{d-1} \right)^{\frac{d-5}{2(d-3)}} \ .
\end{equation}
The numerical values of $\hmt_c$ are listed in Table \ref{tabnear}.
{}From \eqref{sofeps}
we find the free energy $\hmf = \epsilon - \hmt \hms$ of the non-uniform phase
for $0 \leq \hmt - \hmt_c \ll 1$ to be
\begin{equation}
\label{freeen}
\hmf ( \hmt ) = - \frac{d-5}{d-1} \epsilon_c - \hms_c ( \hmt - \hmt_c )
- \frac{c}{2\hmt_c} ( \hmt - \hmt_c )^2
+ \CO (( \hmt - \hmt_c )^3) \ ,
\end{equation}
with
\begin{equation}
\label{heatcap}
c = \frac{d (d-1) \hms_c}{
d(d-5) + 2 (d-1) \hat{\gamma} \epsilon_c } \ ,
\end{equation}
where
$c = \hmt \delta \hms / \delta \hmt$ is the heat capacity
of the non-uniform phase for $\hmt = \hmt_c$ (see Table
\ref{tabnear} for numerical values).%
\footnote{$c$ is the heat capacity for constant volume $V_p$
of the world-volume of the $p$-brane.}
Note that the heat capacity $c$ is positive for $d\geq 5$.
The free energy $\hmf_u ( \hmt )$
for the uniform phase around $\hmt = \hmt_c$
is also given by \eqref{freeen}, but with the heat capacity
being $c_u = (d-1)\hms_c / (d-5)$.%
\footnote{However, for $d=5$ we have $\hmf = 0$ and $\hmt$ is
constant. See Section \ref{secM5} for more details.}
Comparing the free energy $\hmf ( \hmt )$
for the non-uniform phase with the free energy $\hmf_u ( \hmt )$
for the uniform phase, we see that
$\hmf ( \hmt ) > \hmf_u ( \hmt )$ for $d > 5$.
This means that for $d > 5$ the uniform phase is
thermodynamically preferred over the non-uniform phase, since
the free energy of the uniform phase is less than the free energy
of the non-uniform phase.

In Section \ref{secM5} we go into more detail with the case $d=5$.
This case is of particular interest since the thermodynamical
behavior is quite different than for $d > 5$,
and furthermore since we have a greater knowledge of the
phase diagram for $d=5$.

\subsubsection*{Copies of the non-uniform phase}

As reviewed in Section \ref{review}, the neutral non-uniform
black string branch has copies, with the physical parameters
given by \eqref{neutcopies} \cite{Horowitz:2002dc,Harmark:2003eg}.
This means that the non-uniform phase
of near-extremal branes on a circle also has copies,
which can be written in the ansatz \eqref{nearsol1}-\eqref{nearsol3}.
The physical parameters $\epsilon$, $r$, $\hmt$ and $\hms$ can
be found using the relation \eqref{necopies} in Section \ref{s:genphase}.

{}From \eqref{necopies} we see that the energy at which
the $k$'th copy of the non-uniform starts
is $k^{-(d-3)} \epsilon_c $. Then from \eqref{nearr} we
see that the $k$'th copy for $0 \leq \epsilon - k^{-(d-3)} \epsilon_c
\ll 1$ has
\begin{equation}
r (\epsilon) = \frac{2}{d-1} - \hat{\gamma} \left( k^{d-3} \epsilon - \epsilon_c
\right)
+ \CO ( ( \epsilon - k^{-(d-3)} \epsilon_c )^2 ) \ .
\end{equation}
This determines the behavior of the $k$'th copy of the
non-uniform phase of near-extremal branes on a circle close
to the point $(\epsilon,r)=(k^{-(d-3)} \epsilon_c,2/(d-1))$.
One now easily finds the entropy as function of energy and
the free energy as function of temperature
for the $k$'th copy.

In the right part of Figure \ref{fig_ne} we depicted the $(\epsilon,r)$
phase diagram including the copies for the case $d=5$.

\subsubsection*{Possible violation of the Gubser-Mitra conjecture}

In \cite{Gubser:2000ec,Gubser:2000mm} Gubser and Mitra made
the following general conjecture about black branes:
\begin{itemize}
\item
A black brane with a non-compact translational symmetry is
free of dynamical instabilities if and only if it is
locally thermodynamically stable.
\end{itemize}
This conjecture claims a connection between the thermodynamics of
a black brane and the classical stability of the solution under
small fluctuations. A more precise argument for this conjecture
was given in \cite{Reall:2001ag} and further considered in
 \cite{Gregory:2001bd}.

However, above we have argued that the near-extremal
$p$-brane smeared on a circle (i.e. the uniform phase)
is classically unstable for $\epsilon < \epsilon_c$.
On the other hand, we have that the heat capacity
for the near-extremal
$p$-brane smeared on a circle in general is given by
$c_u = (d-1)\hms_c / (d-5)$ which means that it is
positive and finite for $d > 5$.
We see that the near-extremal
$p$-brane smeared on a circle gives an example of a black brane
with a translational symmetry that is classically unstable
but locally thermodynamically stable.

Now, consider a near-extremal $p$-brane with $d>5$ smeared
on a non-compact direction. This will inherit the classical
instability of the near-extremal brane smeared on a circle
(for $\epsilon < \epsilon_c$). But it will also inherit
the local thermodynamical stability. This thus seems to directly
violate the Gubser-Mitra conjecture stated above.

We note here that these arguments rest on the assumption
that the non- and near-extremal $p$-branes smeared on a circle
have a Gregory-Laflamme instability. It would be interesting
to check if the predicted unstable mode in these solutions
really exists. This would enable one to prove the existence of
a counter-example to the Gubser-Mitra conjecture.

\section{Non-uniform phase of M5-branes on a circle}
\label{secM5}

In this section we consider M5-branes on a circle%
\footnote{To be more precise,
the M5-branes are coincident and the circle is transverse
to the M5-branes, precisely as specified in Section \ref{phasnonex}.},
corresponding to the case $d=5$ for our general class of solutions.
By M/IIA S-duality M5-branes on a circle are dual to coincident NS5-branes
in type IIA string theory. By learning about
the thermal phases of M5-branes on a circle we can thus learn
about the thermal phases of NS5-branes in type IIA string theory.
We consider in the following only near-extremal M5-branes
on a circle.
However, one could easily use the results
above to examine also the non-extremal case.

As we shall see, there are at least two reasons why the $d=5$ case, i.e.
the case of M5-branes on a circle, is of particular
interest.
One reason is that for $d=5$ we can map the numerical data
of Wiseman for the non-uniform black string on $\CM^5 \times S^1$
to the non-uniform phase with $d=5$.
Another reason is that the behavior of the thermodynamics
of the uniform phase with $d=5$ is of a rather singular
nature, thus making the
behavior of the thermodynamics of the non-uniform phase
particularly interesting to study.

Moreover, it is important to remark that the interesting
consequences for the thermodynamics of the near-extremal M5-branes on
a circle will again have interesting interpretations
for the thermal behavior of $(2,0)$ Little String Theory.
This we discuss in detail in Section \ref{s:LST}.

\subsection{Phase diagram for near-extremal M5-branes on a circle}

We find in this section the $(\epsilon,r)$ phase diagram
for near-extremal M5-branes on a circle (with $r \leq 1/2$).

\begin{figure}[ht]
\centerline{\epsfig{file=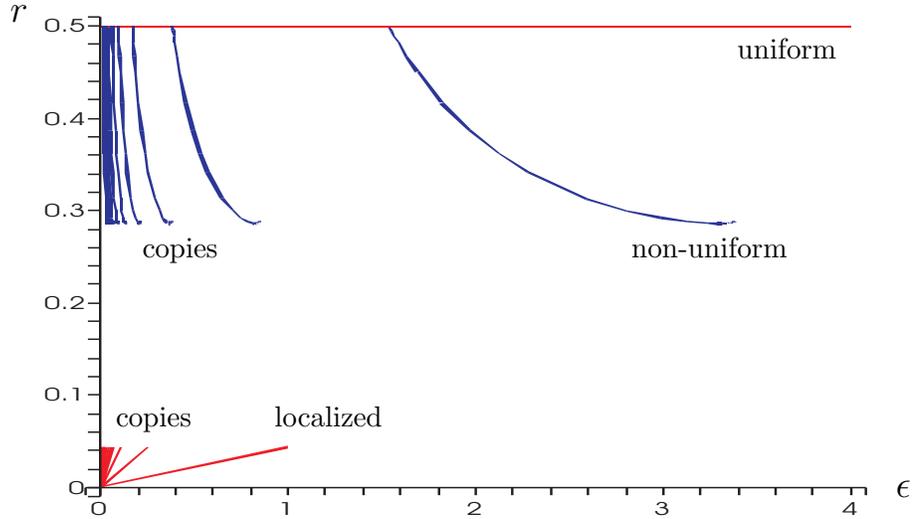,width=13 cm,height=8cm} }
 \caption{Phase diagram for near-extremal M5-branes on a circle.}
\label{fig_ne}
\begin{picture}(0,0)(0,0)
\put(370,65){\Large $\epsilon$}
\put(35,245){\Large $r$}
\put(75,92){copies}
\put(85,156){copies}
\put(135,92){localized}
\put(270,156){non-uniform}
\put(310,231){uniform}
\end{picture}
\end{figure}

We consider first the uniform phase, as reviewed in Section \ref{s:genphase}.
This phase correspond
to near-extremal M5-branes smeared on a circle, as also explained above.%
\footnote{Note that all of the following results hold
for any near-extremal $p$-brane on a circle with
$d=5$. Thus, we could equivalently consider the D4-brane
instead. However, the D4-brane is in any case trivially related to the
M5-brane by U-duality.}
This is dual
to the standard solution for
near-extremal NS5-branes in type IIA string theory.
The uniform phase has $r=1/2$,
whereas the energy $\epsilon$ can take all positive values.
We depicted the uniform phase in the $(\epsilon,r)$ phase
diagram for near-extremal M5-branes on a circle in
Figure \ref{fig_ne}.

For the non-uniform phase we first note that \eqref{nearr}
gives us the starting point and starting slope
of the non-uniform phase in the $(\epsilon,r)$ phase diagram.
However, for $d=5$ we have in addition the numerical data
of Wiseman \cite{Wiseman:2002zc} (see Figure \ref{fig_neut}
for the $(\mu,n)$ phase diagram for these data).
Using the map \eqref{nemap} we can now map the data of Wiseman
to data for the $d=5$ non-uniform phase in the $(\epsilon,r)$
phase diagram. This is depicted in Figure \ref{fig_ne}.
We see that the non-uniform phase starts out
in $\epsilon_c \simeq 1.54$, and then continues with
decreasing $r$ and increasing $\epsilon$ until the
endpoint which is $(\epsilon,r) = (3.39,0.29)$.%
\footnote{Note that the curve has increasing $r$ near
the endpoint. It is not clear if this is
an actual physical feature of the non-uniform
phase or if it is due to an inaccuracy in the data of Wiseman
in \cite{Wiseman:2002zc}.
See also below for a comment regarding the heat capacity.}

In addition to the uniform and non-uniform phase,
we also have the localized phase, corresponding to near-extremal
M5-branes localized on a circle. This phase is considered in detail in
Section \ref{s:smallne}. {}From \eqref{rofeps} we see that
$r \simeq 0.044 \, \epsilon$ corresponds to the leading slope
of the curve for the localized phase. This is also depicted
in Figure \ref{fig_ne}.

We also have
the copies of the localized and non-uniform phase,
as explained above in Section \ref{s:genphase} and \ref{secnunear}.
These are also depicted in Figure \ref{fig_ne}.

To draw all phases in the $(\epsilon,r)$ phase diagram for near-extremal
M5-branes on a circle one would have to also consider
solutions with Kaluza-Klein bubbles, i.e. with $r > 1/2$. This
we consider in a forthcoming publication \cite{Harmark:2004bb}.

\subsection{Thermodynamics of non-uniform phase}

\subsubsection*{Uniform phase}

Before considering the thermodynamics of the non-uniform
phase, we first review the thermodynamics of the uniform
phase. The thermodynamics is
\begin{equation}
\label{thuni}
\hmt = 1 \spa
\hmf = 0 \spa
\hms_u (\epsilon)  = \epsilon \ .
\end{equation}
Thus, the temperature is constant, the free energy is zero
and the entropy grows linearly with energy.
As we mention in Section \ref{s:LST}
this is a signature
of Hagedorn behavior of the thermodynamics with $\hmt = 1$
being the Hagedorn temperature.

We see that the thermodynamics \eqref{thuni} of the uniform phase
is of a rather singular
nature. In the microcanonical ensemble we can put in an arbitrary
amount of energy into the system without affecting the temperature.
In the canonical ensemble we see that the temperature is fixed, so
the heat capacity is infinite.
One should therefore understand the thermodynamics \eqref{thuni}
as being a special limit of the thermodynamics of near-extremal
M5-branes on a circle. I.e. the hope is to find corrections
to \eqref{thuni} or other phases
of near-extremal M5-branes on a circle that describe the
thermodynamics for temperatures $\hmt \neq 1$.
For $\hmt \ll 1$ we have the localized phase, as considered
in Section \ref{s:smallne} (see also
Section \ref{s:LST}), but for $\hmt > 1$ we need new input
to determine the physics. This will be provided below.

\subsubsection*{Thermodynamics of non-uniform phase
in microcanonical ensemble}

\begin{figure}[ht]
\centerline{\epsfig{file=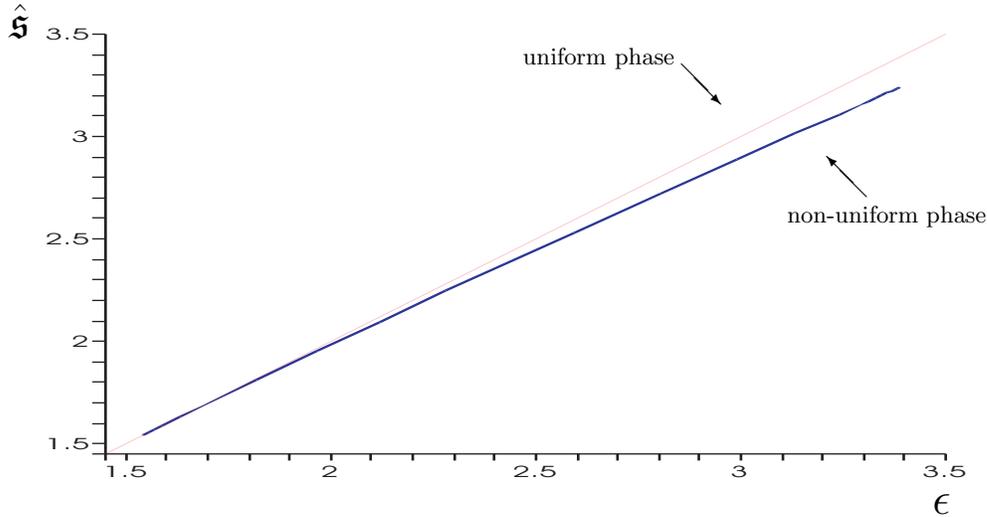,width=14cm,height=7cm} }
\caption{The entropy $\hms(\epsilon)$ as a function of energy
for near-extremal M5-branes on a circle.}
\label{fig_ESdiag}
\begin{picture}(0,0)(0,0)
\put(210,210){\footnotesize uniform phase}
\put(270,210){\vector(1,-1){15}}
\put(310,150){\footnotesize non-uniform phase}
\put(340,160){\vector(-1,1){15}}
\put(365,40){\LARGE $\epsilon$}
\put(16,220){\LARGE $\hms$}
\end{picture}
\end{figure}

We first consider the microcanonical ensemble, i.e.
the thermodynamics for fixed energy.
We use now the map \eqref{nets} to find the
temperature $\hmt$ and entropy $\hms$ of the non-uniform
phase of near-extremal M5-branes on a circle, using the
data of Wiseman \cite{Wiseman:2002zc} for the temperature and entropy
of the non-uniform black string branch.
Using this we plot the non-uniform phase in the
$(\epsilon,\hms)$ diagram as depicted in Figure \ref{fig_ESdiag}.
In this figure we also plotted the uniform phase using
the thermodynamics \eqref{thuni}, giving a straight line
in the $(\epsilon,\hms)$ diagram.

As we clearly see in Figure \ref{fig_ESdiag}
the non-uniform phase emanates from the uniform
phase at the energy $\epsilon = \epsilon_c$, and has
lower entropy than the uniform phase for a given
energy $\epsilon$. Therefore, in the microcanonical
ensemble the uniform phase is the dominant phase since it
has higher entropy.

If we consider the behavior of the entropy near the energy
$\epsilon = \epsilon_c$, we see from \eqref{ssu} that
\begin{equation}
\frac{\hms (\epsilon)}{\hms_u (\epsilon)} =
1 - 0.05 \cdot ( \epsilon- \epsilon_c)^2 \spa \epsilon_c = 1.54 \ .
\end{equation}
for $0 \leq \epsilon-\epsilon_c \ll 1$.
This is in accordance with the numerical data plotted in
Figure \ref{fig_ESdiag}.

\subsubsection*{Thermodynamics of non-uniform phase
in canonical ensemble}

\begin{figure}[ht]
\centerline{\epsfig{file=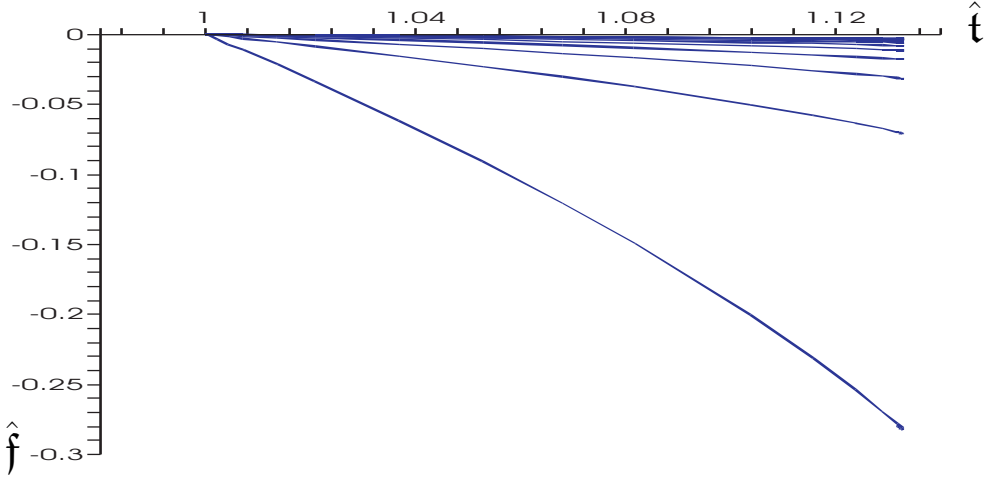,width=14cm,height=7cm} }
\caption{The free energy $\hmf(\hmt)$ as a function of temperature
for the non-uniform phase of near-extremal M5-branes on a circle,
and for the copies of the non-uniform phase.}
\label{fig_TFdiag}
\begin{picture}(0,0)(0,0)
\put(380,235){\LARGE $\hmt$}
\put(16,75){\LARGE $\hmf$}
\end{picture}
\end{figure}

We now turn to the canonical ensemble, i.e. thermodynamics for
fixed temperature.
As stated above, we can use the map \eqref{nets} to
find the temperature $\hmt$ and entropy $\hms$ for the non-uniform phase,
from the data of Wiseman on non-uniform black strings \cite{Wiseman:2002zc}.
Using in addition \eqref{nemap} to get the energy $\epsilon$, we can
also find the free energy $\hmf = \epsilon - \hmt \hms$.
Therefore, we find the free energy as function of temperature
$\hmf(\hmt)$ for the non-uniform phase. This is plotted
in Figure \ref{fig_TFdiag}.
Note that also the copies of the non-uniform phase
are plotted in Figure \ref{fig_TFdiag}, using that for the
$k$'th copy $\hmf' = k^{-2} \hmf$ and $\hmt' = \hmt$.

We see in Figure \ref{fig_TFdiag} that the free energy
of the non-uniform phase is negative.
This is also
the case for the copies, although the free energy for these
is higher. Therefore, the system prefers to
be in the non-uniform phase rather than in one of its
{\it copy-phases}, since the non-uniform phase has the lowest
free energy.

\begin{figure}[ht]
\centerline{\epsfig{file=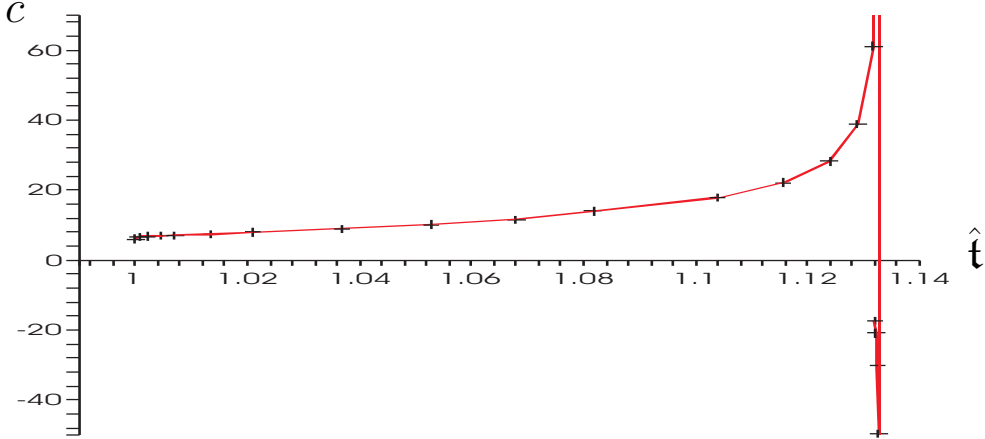,width=14cm,height=7cm} }
\caption{The heat capacity $c$ as function of the temperature $\hmt$
for the non-uniform phase of near-extremal M5-branes on a circle.}
\label{fig_heatcap}
\begin{picture}(0,0)(0,0)
\put(388,142){\LARGE $\hmt$}
\put(24,235){\LARGE $c$}
\end{picture}
\end{figure}

We can also consider the heat capacity $c = \hmt \delta \hms/\delta \hmt$
of the system. We have plotted this in Figure \ref{fig_heatcap}
for the non-uniform phase as a function of the temperature
$\hmt$.

We see from Figure \ref{fig_heatcap} that the heat capacity is
positive, at least for $\hmt < 1.13$.%
\footnote{The divergence of the heat capacity around $\hmt = 1.13$
should, if true, be a signal of a phase transition. However,
it seems doubtful that the data used to compute the heat capacity
is sufficiently accurate to determine that such
a behavior occurs.}
This means that the non-uniform phase is in fact a stable phase
for near-extremal M5-branes on a circle, with temperatures
$\hmt > 1$.
This is highly interesting since so far the only known stable
phase for M5-branes on a circle has been the low temperature
phase for $\hmt \ll 1$ where the M5-brane localizes on the circle.
Here we see that there is a new stable phase with temperatures
above the critical temperature $\hmt = 1$.

In conclusion, it seems that the non-uniform phase for
near-extremal M5-branes on a circle plays a crucial role
for the thermal phase structure of near-extremal
M5-branes on a circle in the canonical ensemble, since
it provides a new phase with temperatures $\hmt > 1$.
This will be further commented on in Section \ref{s:LST},
where we also relate it to the dual non-gravitational
theory living on the NS5-brane.

Finally, we note that if
we consider the free energy for $0 \leq \hmt - 1 \ll 1$
we see from \eqref{freeen} that
\begin{equation}
\label{hmfex}
\hmf(\hmt) = - 1.54 \cdot (\hmt - 1) - 3.23 \cdot (\hmt - 1)^2
\end{equation}
The numerical values in the expression can be found
using Table \ref{tabnear}.

\section{New results for $(2,0)$ LST from M5-branes on a circle}
\label{s:LST}

In Section \ref{s:smallne} and Section
\ref{secM5} we discussed the localized and non-uniform phases
of near-extremal M5-branes on a circle.
As we explain below,
we have a duality between near-extremal M5-branes on a circle
and  $(2,0)$ Little String Theory.
In the following we use this to give predictions
for the thermal behavior
of $(2,0)$ Little String Theory, using the results
on near-extremal M5-branes on a circle.

\subsubsection*{M5-branes on a circle and $(2,0)$ LST}

The $(2,0)$ Little String Theory (LST) is a $5+1$ dimensional
non-gravitational string theory that is conjectured to live on
the world-volume of the type IIA NS5-brane.
The precise definition of $(2,0)$ LST (of type $A_{N-1}$) is that it is the
$g_s \rightarrow 0$ limit of $N$ coincident NS5-branes
with $l_s$ being fixed \cite{Seiberg:1997zk,Berkooz:1997cq,Dijkgraaf:1998ku}.
Here $g_s$ is the string coupling and $l_s$ is the string length.

$N$ coincident type IIA NS5-branes are S-dual to
$N$ coincident M5-branes on a (transverse) circle,
with the circumference of the circle $L$ related
to the eleventh dimensional Planck length $l_p$ and the
string coupling $g_s$ and string length $l_s$ as
$l_s^2 = 2\pi l_p^3 / L$ and $g_s^2 = L^3/(2\pi l_p)^3$.
Therefore, we can alternatively define $(2,0)$ LST
of type $A_{N-1}$ as being the
$l_p \rightarrow 0$ limit of $N$ coincident M5-branes on a circle,
with the circumference
$L$ of the circle going to zero such that $L/l_p^3$ is fixed.

As conjectured in \cite{Maldacena:1997cg,Aharony:1998ub},
the near horizon limit of $N$ coincident M5-branes located
at a point on a (transverse) circle
is the supergravity dual of $(2,0)$ LST for large $N$.
The low energy phase in the supergravity side
corresponds to $AdS_7 \times S^4$.
This is dual to the 5+1 dimensional superconformal
$(2,0)$ theory which is the low energy phase of $(2,0)$ LST.
The high energy phase is the near-horizon limit of an M5-brane smeared
on the circle, which is the S-dual supergravity solution to the
type IIA NS5-brane.
As explained in \cite{Harmark:2002tr} the supergravity description
of $(2,0)$ LST does not break down in the transition between the
low- and high energy phases since the curvature of the
eleven dimensional supergravity solution is always small.

If we turn on a temperature, the following duality is conjectured
to be true \cite{Maldacena:1997cg,Aharony:1998ub} (see also
\cite{Harmark:2002tr})
\begin{itemize}
\item Thermal $(2,0)$ LST of type $A_{N-1}$
is dual to $N$ coincident near-extremal M5-branes
on a circle, for $N \gg 1$.
\end{itemize}
As for the zero-temperature duality, one expects the smeared
M5-brane solution, which is S-dual to near-extremal
type IIA NS5-branes, to be valid at high energies.
The near-extremal NS5-brane has a constant temperature
and the entropy is proportional to the energy (see \eqref{thuni}).
This fits well with the conjecture that this solution is dual
to a non-gravitational string theory
\cite{Maldacena:1996ya,Maldacena:1997cg}.

However, the fact that the temperature is constant (or, equivalently
that the free energy is zero) for the near-extremal NS5-brane
means that any correction to the thermodynamics, no matter how
small, will contribute to the leading order behavior of the thermodynamics.
Thus, to understand the thermodynamics of near-extremal NS5-branes,
it seems crucial to study the corrections to the solution.

In \cite{Harmark:2000hw,Berkooz:2000mz,Kutasov:2000jp} string corrections
to the supergravity description were considered, with the result
that the leading correction gives rise to a negative specific heat
of the NS5-brane, and that the temperature of the near-extremal
NS5-brane is larger than $\hat T_{\rm hg}$ \cite{Kutasov:2000jp}.

In \cite{Harmark:2002tr} it was advocated that one instead should
look for a non-uniform phase of near-extremal M5-branes on a circle,
in order to find a dual to the thermal $(2,0)$ LST.
The expectation of \cite{Harmark:2002tr} was that such a non-uniform
phase would have positive specific heat, and could therefore
provide a dual of thermal $(2,0)$ LST in a stable phase.

In the following we shall see that the new non-uniform phase
of near-extremal M5-branes on a circle found in Section \ref{secM5}
provides a stable phase of $(2,0)$ LST in the canonical ensemble
for a certain range of temperatures.
This confirms in part the ideas of \cite{Harmark:2002tr} since we find
a new dual of thermal $(2,0)$ LST from a non-uniform phase.

\subsubsection*{Interpretation of non-uniform phase
for $(2,0)$ LST}

We now make the statement of the duality between $(2,0)$ LST
of type $A_{N-1}$ and the near-extremal limit of $N$
coincident M5-branes on a circle more precise.
We have defined the near-extremal limit by \eqref{nelimit}.
By employing the IIA/M S-duality, one gets that the
fixed quantities $g$ and $l$ in \eqref{nelimit} are given as
\begin{equation}
\label{M5gl}
g = \frac{(2\pi)^5 l_s^6}{V_5}
\spa
l = 2\pi l_s \sqrt{N} \ .
\end{equation}
where $l_s$ is the string length in type IIA string theory, that
is kept fixed in the limit defining $(2,0)$ LST.
One can now use \eqref{M5gl} together with
\eqref{epsrdef} and \eqref{tsneex} as a
dictionary between dimensionless and
dimensionful quantities.
Considering the thermodynamics of the uniform phase
\eqref{thuni}, we see that in dimensionful quantities we have
\begin{equation}
\hat{T} = \hat T_{\rm hg} \equiv \frac{1}{2\pi l_s \sqrt{N}} \spa
\hat{F} = 0 \spa
\hat{S} = \frac{1}{\hat T_{\rm hg}} E \ .
\end{equation}
We see that this corresponds to the thermodynamics of a
system near a Hagedorn temperature $\hat T_{\rm hg}$,
since the entropy is proportional
to the energy \cite{Maldacena:1996ya,Maldacena:1997cg}.

We can now interpret the non-uniform phase of near-extremal
M5-branes on a circle found in Section \ref{secM5}.
If we consider the canonical ensemble, we found in
Section \ref{secM5} that the heat capacity is positive
(see Figure \ref{fig_heatcap}),
the free energy is negative (see Figure \ref{fig_TFdiag})
and the temperature can vary
in the range $[ \hat T_{\rm hg} , 1.13 \hat T_{\rm hg}]$ (using
that $\hmt = \hat{T} / \hat T_{\rm hg}$).
This means that this is a stable phase of near-extremal
M5-branes on a circle, and by the duality conjecture
this should describe thermal $(2,0)$ LST in that temperature
range.
We have thus found a stable phase of $(2,0)$ LST in the canonical
ensemble for
temperature above the Hagedorn temperature $\hat T_{\rm hg}$.

Notice that the copies of the non-uniform phase,
as depicted in Figure \ref{fig_TFdiag} have positive heat capacity
and they exist in the same temperature range,
so they also correspond to stable phases of thermal $(2,0)$ LST.
However, the free energy of these phases are higher than
that of the non-uniform phase, so the non-uniform phase
dominates over its copies.

Translating the dimensionless expression \eqref{hmfex}
for the free energy into dimensionful quantities,
we see that the free energy for temperatures
near the Hagedorn temperature, i.e.
$0 \leq \hat T- \hat T_{\rm hg} \ll \hat T_{\rm hg}$, is
\begin{equation}
\hat{F} (\hat{T} ) = - 2 \pi V_5 N^3 \hat T_{\rm hg}^6 \left[
1.54 \cdot \left( \frac{\hat T}{\hat T_{\rm hg}} - 1 \right)
+ 3.23 \cdot \left( \frac{\hat T}{\hat T_{\rm hg}} - 1 \right)^2 \right] \ .
\end{equation}
It would interesting if one could reproduce this expression from
the $(2,0)$ LST side.

Turning to the microcanonical ensemble, we see from Figure
\ref{fig_ESdiag} that the uniform phase dominates over
the non-uniform phase, since the uniform phase has
higher entropy.
Thus, for a fixed energy, the temperature is expected from
this to decrease down to the Hagedorn temperature. It would
be interesting to understand what the physics behind this is.
Perhaps the resolution of this could be that the localized
phase also exists in the same energy range as the non-uniform phase,
but with higher entropy than the uniform phase.

\subsubsection*{Prediction for thermodynamics at
low temperatures in $(2,0)$ LST}

In addition to the new non-uniform phase that we interpreted above,
we also have the new results for the localized phase
found in Section \ref{s:smallne}.

The localized phase of near-extremal M5-branes on a circle corresponds
to the low temperature/low energy phase of thermal $(2,0)$ LST.
This is consistent with the fact that the localized phase
dominates over the uniform phase, both in the canonical and
microcanonical ensembles, for low temperature/low energy.
Note that the infrared fixed point of $(2,0)$ LST
is the superconformal $(2,0)$ theory, thus by considering
the localized phase of near-extremal M5-branes for
low temperatures (or energies) we find the corrections
to the thermodynamics of the superconformal $(2,0)$ theory, as one moves away
from the infrared fixed point.

The corrected free energy of the localized
near-extremal M5-brane  is computed from \eqref{freene} with $d=5$
using the translation dictionary \eqref{M5gl} to go  to
dimensionful quantities.
In the canonical ensemble we then find the
following expression for the corrected thermodynamics of the
superconformal $(2,0)$ theory
\begin{equation}
\hat{F}(\hat{T}) = - k_{\rm M5} V_5 N^{3} \hat T^6 \left[ 1 +
 \frac{3 \zeta (3)k_{\rm M5}}{2 \pi^3}
 \left(\frac{\hat T}{\hat T_{\rm hg}}\right)^6 + {\cal{O}}\left( \left(\frac{\hat T}{\hat T_{\rm hg}}
 \right)^{12}
 \right) \right]
\spa
k_{\rm M5} =    \frac{2^6 \pi^3}{3^7} \ .
\end{equation}
This shows that the dimensionless expansion parameter is
$\delta = \hat T/\hat T_{\rm hg}$, as expected on physical grounds.

Note that one can obtain the free energy for the $k$'th copy
of the localized phase from
\begin{equation}
\hat{F}_{(k)}(\hat{T}) = k^{-2} \hat{F} (\hat{T}) \ ,
\end{equation}
for $k=1,2,...$. We see that the copy-phases have higher free
energy than the localized phase, so the localized phase dominates
over the copies.

We list here only the results for the canonical ensemble, both since
this is the most interesting for comparison with calculations
on the $(2,0)$ LST side, but also since one in any case
easily can transform the result for the canonical ensemble
to that of the microcanonical ensemble.

\section{Predictions for supersymmetric Yang-Mills theories  \label{secqft}}

In this section
we apply the new results on near-extremal D-branes and the M2-brane
on a transverse circle to obtain non-trivial predictions for the thermodynamics
of the corresponding dual non-gravitational theories that live on these branes.

We first treat the case of near-extremal D0, D1, D2 and D3-branes
on a circle. Then we discuss the case of
 the near-extremal M2-brane on a circle.

\subsection{$(p+1)$-dimensional Yang-Mills theory on a circle}
\label{s:SYMcirc}

If we take the near-extremal limit of a D$p$-brane it
is well known that the bulk dynamics decouples
from the supersymmetric Yang-Mills (SYM) theory
living on the brane.%
\footnote{With
the exception of the D6-brane, as discussed for example in
\cite{Itzhaki:1998dd}.}
The near-extremal D$p$-brane is therefore
conjectured to be a dual description of
the strongly coupled large $N$ limit of ($p+1$)-dimensional
SYM with 16 supercharges
(on the manifold $\R^{p}$ for the spatial directions)
\cite{Maldacena:1997re,Itzhaki:1998dd}.%
\footnote{That it describes the strongly coupled large $N$ limit
regime of the SYM theory is because one needs to ensure that
the supergravity
description is valid, i.e. that the geometry
is weakly curved and that the effective string coupling is small
\cite{Itzhaki:1998dd}.}

To determine what the dual field theories are for near-extremal D-branes
on a circle we use T-duality.
A near-extremal D$(p-1)$-brane on a transverse circle is
T-dual to a D$p$-brane wrapped on a circle.
It follows from this that the field theory dual of
a near-extremal D$(p-1)$-brane on a transverse
circle is that of a near-extremal D$p$-brane with a circle
along the world-volume.
We see from this that $N$ coincident near-extremal D$(p-1)$-branes
on a transverse circle should be dual to
$(p+1)$-dimensional SYM on $\R^{p-1}\times S^1$
with gauge group $SU(N)$ and
16 supercharges.%
\footnote{See also
\cite{Susskind:1997dr,Barbon:1998cr,Li:1998jy,%
Martinec:1998ja,Martinec:1999bf,Fidkowski:2004fc} and
references therein.}

In particular, the phases of near-extremal D2-branes on a transverse
circle are dual to those of ${\cal{N}}=4$ SYM theory in 3+1 dimensions
on $\R^2 \times S^1$ . Including the thermal time direction
the field theory lives on $\R^2 \times T^2$, where we define
the torus here as $T^2 \equiv S^1 \times S^1_T$.
Two other  interesting examples are  near-extremal D1 and D0-branes on a
transverse circle which are respectively  dual to 2+1 dimensional
SYM on $\R \times S^1$ (i.e. including the thermal time direction
on $ \R \times T^2$)
and 1+1 dimensional SYM theory on
$S^1$ (i.e. including the thermal time direction on the torus $T^2$).
In this section we restrict ourselves to the D0, D1, D2 and D3-branes
on a circle.%
\footnote{However, one could easily extend the results of this section
to include the F-strings and the remaining D-branes.}

We can thus use the results of Sections \ref{secloc} and \ref{secnune}
to obtain predictions for the thermodynamics of the dual gauge theories
described above. In particular, the localized phase
should correspond to the low temperature/low energy regime,
while the non-uniform phase corresponds to a new phase, emerging
from the uniform  phase, which describes the
 high temperature/high energy regime.

In order to make precise predictions for the SYM theory we should
first understand better the dictionary between near-extremal
D$(p-1)$-branes on a (transverse) circle and the SYM theory.
To this end, define $L$ to be the circumference of the circle
transverse to the D$(p-1)$-branes, $g_s$ to be the string coupling
and $l_s$ the string length. In the T-dual theory, with the T-duality
along the circle direction, we define instead $\hat{L}$ to be
the circumference of the T-dual circle and $\hat{g}_s$ to be
the T-dual string coupling.
We have then that $L \hat{L} = (2\pi l_s)^2$ and
$g_s \hat{L} = 2\pi l_s \hat{g}_s$.
Using this along with
the quantization condition for $N$ D$p$-branes in the T-dual
theory, the near-extremal limit \eqref{nelimit}
and the definition of $l$ in \eqref{nearsole2}, we get that
$g$ and $l$ are given as%
\footnote{We use the notation and conventions of
Ref.~\cite{Harmark:1999xt}.}
\begin{equation}
\label{dictdp}
{\rm D}(p-1) \;\,\mbox{on circle}\; \, : \qquad g =
\frac{1}{(2\pi)^3 N^2 } \frac{\hat L^p}{V_{p-1}}
\left( \lambda \hat L^{3-p} \right)^2
\spa
l  = \frac{\hat L }{\sqrt{2\pi}}\sqrt{ \lambda \hat L^{3-p} } \ .
\end{equation}
Here we wrote $g$ and $l$ in terms of the SYM theory variables, i.e.
$N$ is the rank of the $SU(N)$ gauge group, $\lambda = g_{\rm YM}^2
N$, $g_{\rm YM}^2 = (2\pi)^{p-2} \hat{g}_s l_s^{p-3}$ and $\hat L$ is
the circumference of the field theory circle $S^1$.

Note in particular that from $g$, $l$ and $\hat T$ we can form
two independent dimensionless parameters, $l/g$ and $l \hat T$.
These two parameters will play a physical role in the field theory
expressions below.

We can now reinstate the dimensions in the thermodynamics,
so that it can be written in terms of SYM variables.
For this we recall the definition
of $\epsilon$ in \eqref{epsrdef}, of $\hmt$, $\hms$ in \eqref{tsneex}
and of $\hmf$ in \eqref{defhmf}. It follows that
given the thermodynamic functions $\hms (\epsilon)$ and $\hmf (\hmt)$ that
we computed before, we can compute the entropy $\hat S$ and free energy
$\hat F$ of the dual field theory with
\begin{equation}
\label{physsf}
\hat S (E) = \frac{l}{g} \hms ( g E) \spa
 \hat{F} (\hat T) = \frac{1}{g} \hmf ( l \hat T) \ ,
\end{equation}
where $E$ and $\hat T$ are the energy and temperature in the field theory.

We now present the new results for
($p+1$)-dimensional SYM theory on $\R^{p-1} \times S^1$
that follow from the localized and non-uniform
phases of  near-extremal D$(p-1)$-branes.
We remind the reader that these phases are directly
linked via U-duality and the near-extremal limit to the
localized black hole phase and non-uniform
black string phase of pure gravity on $\CM^{10-p} \times S^1$.
In the quantitative analysis below we focus on the canonical
ensemble.

\subsubsection*{Uniform phase: High temperature phase of SYM on a circle}

We begin by considering the uniform phase
of near-extremal D$(p-1)$-branes on a circle.
This corresponds to the high temperature phase of
($p+1$)-dimensional SYM theory on $\R^{p-1} \times S^1$.
This is easily understood on physical grounds since
high temperatures correspond to short distances, and this means
that one does not see the compact direction in  $\R^{p-1} \times S^1$,
i.e. the circumference $\hat{L}$ should only appear as a trivial
proportionality factor in the free energy.
This is precisely what we get for the uniform phase since it
corresponds to a uniformly smeared near-extremal D$(p-1)$-brane,
which has the same thermodynamics as a near-extremal D$p$-brane.
It is not difficult to check this explicitly
using  the thermodynamics \eqref{unifr} and the translation dictionary
\eqref{physsf}. The result for the free energy in the uniform phase
is then
\begin{equation}
\label{unidp}
\hat{F}^{\rm u}_{{\rm SYM}(p+1)} (\hat T) =
\hat{F}^{\rm u}_{{\rm D}(p-1)}(\hat T)
= - k_{p}V_{p-1} \hat L  N^2 \lambda^{- \frac{3-p}{5-p}}
\hat T^{2 \frac{7-p}{5-p}} \ ,
\end{equation}
where the coefficients $k_p$ are given in Table \ref{tabcoef}.
The expression \eqref{unidp} is indeed the familiar result for the free
energy of the D$p$-brane theory, with $V_p = V_{p-1} \hat L$.

\subsubsection*{Localized phase: Low temperature phase of SYM on a circle}

We now turn to the localized phase, corresponding to the
low temperature phase of ($p+1$)-dimensional SYM theory
on $\R^{p-1} \times S^1$.
Low temperatures correspond to large distances in the SYM theory,
so in this regime the presence of the circle enters in a non-trivial
manner. If one consider free ($p+1$)-dimensional SYM theory
on $\R^{p-1} \times S^1$, it is clear that one has a tower of
Kaluza-Klein states for the circle, and that the lowest Kaluza-Klein
mode corresponds to free $p$-dimensional SYM theory
on $\R^{p-1}$.
Similarly, we shall see below that for the temperature $\hat{T}$ going
to zero, the leading order thermodynamics for the localized phase
of D$(p-1)$-branes on a circle corresponds to the
thermodynamics of a near-extremal D$(p-1)$-brane, which is dual
to strongly coupled $p$-dimensional SYM theory on $\R^{p-1}$.
The corrections to the leading order thermodynamics for the localized phase
found in Section \ref{secloc} then correspond to
including corrections from the Kaluza-Klein modes
on the circle.

Using the expressions for the corrected
free energy in  \eqref{freene} along with the translation dictionary
\eqref{physsf}, \eqref{dictdp} we find for the localized phase
of thermal
($p+1$)-dimensional SYM theory on $\R^{p-1} \times S^1$ the free energy
\begin{eqnarray}
\label{locdp}
 & & \hat{F}^{\rm loc}_{{\rm SYM}(p+1)} (\hat T) =
\hat{F}^{\rm loc}_{{\rm D}(p-1)}(\hat T) \\ \nn
&& \simeq - k_{p-1}V_{p-1} N^2 \left( \frac{\lambda}{\hat L} \right)^{- \frac{4-p}{6-p}}
\hat T^{2 \frac{8-p}{6-p}} \left\{ 1
+ \frac{2 (9-p) \zeta (8-p) k_{p-1} }{(6-p)^2 (2\pi)^3 \Omega_{9-p}}
\left[ \hat L \hat T  \sqrt{ \lambda\hat L^{3-p}}
\right]^{2\frac{8-p}{6-p}}  \right\} \ .
\end{eqnarray}
Here the coefficients $k_p$ are listed in Table \ref{tabcoef}.%
\footnote{These coefficients are related to
$\hat K_1^{(d)}$ defined in \eqref{K1def} via the
relation $\hat K_1^{(9-p)} (2\pi)^{2\frac{4-p}{5-p}} = k_p$.}

\begin{table}
\begin{center}
\begin{tabular}{|c||c|c|c|c|c|}
\hline
$p$ & 0 & 1 & 2 & 3 & 4 \\ \hline
$k_p$ & $(2^{21} 3^2 5^7 7^{-19} \pi^{14})^{1/5} $
& $2^4 3^{-4} \pi^{5/2}  $ & $ (2^{13} 3^5 5^{-13} \pi^8)^{1/3}$ &
$2^{-3} \pi^2 $ & $2^5 3^{-7} \pi^2$ \\ \hline
\end{tabular}
\end{center}
\caption{Coefficients for the free energy of D$p$-branes.
\label{tabcoef} }
\end{table}

Note first of all that the leading term in \eqref{locdp}
is in perfect agreement with the expected result for the strong
coupling limit of the $p$-dimensional SYM theory. In particular,
we observe the correct 't Hooft coupling $\lambda/\hat L$ that
follows from compactifying the $(p+1)$-dimensional theory.
Moreover, \eqref{locdp} gives a quantitative prediction for
the first correction term in terms of the dimensionless
parameter
\begin{equation}
\delta = \hat L \hat T \sqrt{\lambda \hat L^{3-p}} \ .
\end{equation}
The next corrections will be of order $(\delta^{2\frac{8-p}{6-p}})^2$.
It would be highly interesting if one could find a way to
reproduce the corrections computed above from the field theory side.

The explicit expressions of \eqref{locdp} for the cases $p=1,2,3$
and 4 are
\begin{eqnarray}
\label{locdp1}
& & \hat{F}^{\rm loc}_{{\rm SYM}(1+1)} (\hat T) =
\hat{F}^{\rm loc}_{{\rm D}0}(\hat T) \\ \nn
& &  = - k_{0} N^2 \left(
\frac{\lambda}{\hat L} \right)^{- 3/5} \hat T^{14/5} \left\{ 1 +
\frac{21 \zeta (7) k_{0} }{80\pi^7} ( \lambda^{1/2} \hat L^2 \hat
T )^{14/5}  +   {\cal{O}}  \left( ( \lambda^{1/2} \hat L^2 \hat T
)^{28/5} \right) \right\} \ ,
\end{eqnarray}
\begin{eqnarray}
\label{locdp2}
& & \hat{F}^{\rm loc}_{{\rm SYM}(2+1)} (\hat T) =
\hat{F}^{\rm loc}_{{\rm D}1}(\hat T) \\ \nn
& &  = - k_{1} V_1 N^2 \left(
\frac{\lambda}{\hat L} \right)^{- 1/2} \hat T^{3} \left\{ 1 +
\frac{ k_{1} }{2880\pi} \left(  \lambda^{1/2} \hat L^{3/2} \hat T
\right)^{3}  + {\cal{O}}  \left( ( \lambda^{1/2} \hat L^{3/2} \hat T)^{6}
\right) \right\} \ ,
\end{eqnarray}
\begin{eqnarray}
\label{locdp3}
& & \hat{F}^{\rm loc}_{{\rm SYM}(3+1)} (\hat T) =
\hat{F}^{\rm loc}_{{\rm D}2}(\hat T) \\ \nn
 &  & = - k_{2} V_2 N^2 \left(
\frac{\lambda}{\hat L} \right)^{- 1/3} \hat T^{10/3} \left\{ 1 +
\frac{5 \zeta(5) k_{2} }{32\pi^6} \left(  \lambda ^{1/2} \hat L
\hat T \right)^{10/3}  +  {\cal{O}}  \left( ( \lambda^{1/2} \hat L \hat T )^{20/3}
\right) \right\} \ ,
\end{eqnarray}
\begin{eqnarray}
\label{locdp4}
& & \hat{F}^{\rm loc}_{{\rm SYM}(4+1)} (\hat T) =
\hat{F}^{\rm loc}_{{\rm D}3}(\hat T) \\ \nn
 &  & = - k_{3}V_3 N^2  \hat T^{4} \left\{ 1 +
\frac{ k_{3} }{288 \pi^2} \left(  \lambda ^{1/2} \hat L^{1/2}
\hat T \right)^{4}  +  {\cal{O}}  \left( ( \lambda^{1/2} \hat L^{1/2} \hat T )^{8}
\right) \right\} \ ,
\end{eqnarray}
where we recall that in each case $\lambda$ is the 't Hooft
coupling of the ($p+1$)-dimensional SYM theory.

We note that the corresponding results in the microcanonical
ensemble are easily obtained using the corrected entropy
\eqref{entropyne} of the localized near-extremal branes
and  \eqref{dictdp}, \eqref{physsf}.

\subsubsection*{Non-uniform phase: New phase in SYM on a circle}

Turning to the non-uniform phase of near-extremal $D(p-1)$-branes
on a circle, we have that
the results of Section \ref{secnune}
for the non-uniform phase
give us
i) a prediction of the existence of a new phase of ($p+1$)-dimensional
SYM theory on $\R^{p-1} \times S^1$ at intermediate temperatures, and
ii) the first correction to the thermodynamics around the point
where this phase connects to the uniform phase.

In particular, using the expressions for the corrected
free energy in  \eqref{freeen} along with the translation dictionary
\eqref{physsf}, \eqref{dictdp} we find for the non-uniform phase
of thermal
($p+1$)-dimensional SYM theory on $\R^{p-1} \times S^1$ the free energy
\begin{eqnarray}
\label{nudp}
& & F^{\rm nu}_{{\rm SYM}(p+1)} (\hat T) =
F^{\rm nu}_{{\rm D}(p-1)}(\hat T) \\ \nn
& & \simeq
-(2\pi)^3 V_{p-1} N^2 \lambda \hat{L}^{3-2p}
 \left[
\frac{5-p}{9-p} \epsilon_c
+ \hmt_c \hms_c \left( \frac{\hat T}{\hat T_c} -1\right)
+ \frac{\hmt_c c}{2} \left( \frac{\hat T}{\hat T_c} -1\right)^2  \right] \ .
\end{eqnarray}
Here the critical temperature is given by
\begin{equation}
\hat T_c =\frac{\hmt_c}{\hat L} \sqrt{\frac{2\pi}{\lambda  \hat L^{3-p}}}  \ .
\end{equation}
and the values for $\hmt_c$,
$\epsilon_c$, $\hms_c$ and $c$ can be read off from Table \ref{tabnear}
using $d=10-p$. Note that the first term in the expression \eqref{nudp}
corresponds to  $\hat F^{\rm u}_{{\rm SYM}(p+1)}(\hat T_c)$, i.e. the
free energy \eqref{unidp} of the uniform phase evaluated at the
critical temperature, while the second term is
$-\hat S^{\rm u}_{{\rm SYM}(p+1)}(\hat T_c) (\hat T - \hat T_c)$, and
hence involves the entropy of the uniform phase evaluated
at the critical temperature. The third term contains the departure
of the free energy of the uniform phase in this non-uniform phase.

As for the localized phase,  the corresponding results
for the non-uniform phase in the microcanonical
ensemble are not difficult to obtain using the corrected entropy
\eqref{ssu} of the localized near-extremal branes
and \eqref{physsf}, \eqref{dictdp}.

It would be very interesting to understand the new results
\eqref{locdp} and  \eqref{nudp}
for thermal $(p+1)$-dimensional SYM  theory
on $\R^{p-1} \times S^1$ from the gauge theory side.

\subsubsection*{Copies of the localized and non-uniform phase}

Since both the localized and non-uniform phase of near-extremal
branes have copies, we also have copies of the thermal
$(p+1)$-dimensional SYM phases obtained above.
Using \eqref{necopies} the free energy of the $k$'th copy is given by
\begin{equation}
(\hat F_{{\rm SYM}(p+1)})_{(k)} (\hat T) =k^{p-7} \hat F_{{\rm SYM}(p+1)}
(k^{(5-p)/2} \hat T) \spa k = 1,2,3,... \ ,
\end{equation}
in terms of either $\hat F_{{\rm SYM}(p+1)}^{\rm loc}$ given in
\eqref{locdp} or $\hat F_{{\rm SYM}(p+1)}^{\rm nu}$ given in \eqref{nudp}

\subsection{New results in (2+1)-dimensional Yang-Mills theory}

In this section we consider near-extremal M2-branes on a circle.
By IIA/M S-duality, this is dual to a near-extremal D2-brane.
Therefore,
it is conjectured that $N$ coincident
near-extremal M2-branes on a circle are dual to
(2+1)-dimensional SYM theory with 16 supercharges
with gauge group $SU(N)$
(on the non-compact space $\R^2$), $N \gg 1$ \cite{Maldacena:1997re,Itzhaki:1998dd}.
We see that this is analogous to the near-extremal
M5-brane on a circle and the duality to $(2,0)$ LST
in Section \ref{s:LST}.

In more detail, we take the circle transverse to the M2-brane
to have circumference $L = 2\pi g_s l_s$, and using the
quantization condition on the M2-branes together with
\eqref{nelimit} and \eqref{nearsole2} we get
\begin{equation}
\label{m2gl}
g = \frac{(2\pi)^2}{V_2} \frac{N^3}{\lambda^3}
\spa
l = 2\pi \frac{N^{3/2}}{\lambda} \ ,
\end{equation}
where $\lambda = \gym^2 N = g_s l_s^{-1} N$.
This provides the dictionary for translating the
results of the near-extremal M2-branes on a circle to the SYM theory.

As we shall see,
the localized and non-uniform phase of the near-extremal M2-brane on a circle
provide us with new information for thermal
(2+1)-dimensional SYM theory on $\R^2$.
Note that the results below are written in the canonical ensemble.

\subsubsection*{Uniform phase: High temperature phase}

The uniform phase of near-extremal M2-branes on a circle corresponds
to the high temperature/high energy limit of thermal (2+1)-dimensional SYM,
i.e. the field theory dual of near-extremal D2-branes.
We can check this explicitly using the thermodynamics \eqref{unifr}
together with \eqref{physsf} and \eqref{m2gl}.
The result for the free energy in the uniform phase is
\begin{equation}
\label{unim2}
\hat F^{\rm UV}_{\rm SYM(2+1)} (\hat T) =
\hat F^{\rm u}_{{\rm M}2}(\hat T)
= - k_{2}V_{2}  N^2 \lambda^{- 1/3} \hat T^{10/3} \ ,
\end{equation}
where the coefficient $k_2$ is given in Table \ref{tabcoef}.
This is indeed the familiar result for the free energy
of the near-extremal D2-brane theory.

\subsubsection*{Localized phase: Low temperature phase}

The localized phase of near-extremal M2-branes on a circle corresponds
to the low temperature/low energy limit of thermal (2+1)-dimensional SYM.
The infrared fixed point of (2+1)-dimensional SYM is a superconformal
field theory with $SO(8)$ R-symmetry, which for large $N$ is dual
to the near-extremal M2-brane solution.
The corrections obtained for the localized phase
thus compute the corrections to the M2-brane theory as we move away
from the infrared fixed point.

In particular, using $d=8$ in \eqref{freene} together with
\eqref{physsf} and \eqref{m2gl} we then obtain the corrected
free energy of the localized M2-brane phase as
\begin{eqnarray}
\label{locm2}
&& \hat{F}_{{\rm SYM}(2+1)}^{\rm IR} (\hat T) =
\hat{F}_{\rm M2}^{\rm loc} (\hat T)
\\ \nn &&
=
- k_{\rm M2} V_2 N^{3/2} \hat T^3 \left[ 1 +
\frac{ \pi^4  k_{\rm M2} }{90}
\left( \frac{N^{3/2} \hat T}{\lambda} \right)^3 +
{\cal{O}} \left(
\left( \frac{N^{3/2} \hat T}{\lambda} \right)^6 \right)
\right]
\spa
k_{\rm M2} = \frac{2^{7/2}\pi^2}{3^4} \ .
\end{eqnarray}
Note that the dimensionless expansion parameter is
$\delta = N^{3/2} \hat T/\lambda$ where $\lambda $ is the 't Hooft coupling
of the (2+1)-dimensional SYM theory.

\subsubsection*{Non-uniform phase: New phase in 2+1 dimensional SYM}

For the non-uniform phase of the near-extremal M2-brane on a circle,
the results of Section \ref{secnune} give us i)
a prediction of the existence of a new non-uniform phase of
uncompactified ($2+1$)-dimensional SYM theory on $\R^{2}$ at
intermediate temperatures and ii) the first
correction to the thermodynamics around the point where this phase
connects to the uniform phase.

In particular, using the expressions for the corrected
free energy in  \eqref{freeen} along with the translation dictionary
\eqref{physsf}, \eqref{m2gl} we find for the non-uniform phase
of thermal ($2+1$)-dimensional SYM theory on $\R^{2}$ the free energy
\begin{equation}
\label{num2}
\hat F^{\rm UV'}_{{\rm SYM}(2+1)} (\hat T) =
\hat F^{\rm nu}_{{\rm M}2}(\hat T)
\simeq - \frac{V_2}{(2\pi)^2} \frac{\lambda^3}{N^{3}}
\left[ \frac{3}{7} \epsilon_c
+ \hmt_c \hms_c \left( \frac{\hat T}{\hat T_c} -1\right)
+ \frac{\hmt_c c}{2}
\left( \frac{\hat T}{\hat T_c} -1\right)^2 \right]
\end{equation}
Here the critical temperature is given by
\begin{equation}
\hat T_c =   \frac{\lambda}{2\pi N^{3/2}} \hmt_c \spa \hmt_c = 0.97 \ ,
\end{equation}
and one should substitute $\epsilon_c = 0.46$, $\hms_c = 0.68$ and
$c = 1.14$ which are read off for $d=8$ from Table \ref{tabnear}.

As for the case of SYM on a circle, it would be
 interesting to understand the new results \eqref{locm2} and  \eqref{num2}
 for uncompactified (2+1)-dimensional SYM theory from the gauge theory side.

\subsubsection*{Copies of the localized and non-uniform phase}

Since both the localized and non-uniform phase of near-extremal
branes have copies, we also have copies of the thermal uncompactified
($2+1$)-dimensional SYM phases obtained above.
Using \eqref{necopies} the free energy of the $k$'th copy is given by
\begin{equation}
(\hat F_{{\rm SYM}(2+1)})_{(k)} (\hat T) =\frac{1}{k^{5}} \hat F_{{\rm SYM}(p+1)}
(k^{3/2} \hat T) \spa k = 1,2,3,... \ ,
\end{equation}
in terms of either $\hat F_{{\rm SYM}(p+1)}^{\rm IR}$ given in
\eqref{locm2} or $\hat F_{{\rm SYM}(p+1)}^{\rm UV'}$ given in \eqref{num2}

\section{Discussion and conclusions}
\label{s:concl}

We conclude the paper with the following remarks and open problems:
\newline

\noindent {\sl Comparison to weakly coupled SYM theory on a circle:}
In Section \ref{s:SYMcirc} we saw that we can give quantitative
predictions for the thermodynamics of $(p+1)$-dimensional
SYM theory on $\R^{p-1} \times S^1$, using the results on Kaluza-Klein
black holes.
One of the central ingredients is that the free energy is
expressible as a function of the variable
$\hmt \sim \sqrt{\lambda \hat{L}^{3-p}} \hat{L} \hat{T}$.
Here $\lambda \hat{L}^{3-p}$ should be large in order
for the supergravity description to hold, which means
the SYM theory is in the strongly coupled regime.
Thus, the scale in which we shift between
the low temperature and high temperature regimes
is around $\hmt \sim 1$, i.e. when $\hat{L} \hat{T}
\sim 1/\sqrt{\lambda \hat{L}^{3-p}}$. We thus see that this
scale is lower than for the free field theory, where it is
instead at $\hat{L} \hat{T} \sim 1$.
Also, for the localized phase we see that the expansion
of the free energy for low temperatures is in powers
of $\hmt$.
Therefore, it is not clear if the predictions of
Section \ref{s:SYMcirc} can be connected to thermodynamics
of weakly coupled field theory in a quantitative manner.
Perhaps a double-scaling limit could be engineered to make
such a quantitative connection.
However, it would in any case be interesting to examine
free $(p+1)$-dimensional SYM on $\R^{p-1} \times S^1$
to see if there should be at least a qualitative similarity
of the thermodynamics.%
\footnote{See \cite{Aharony:2003sx} for recent work on large-$N$
SYM theories on compact spaces and the
relation to the strong coupling limit.}
In particular, it would be interesting
if one could find a similar new non-uniform phase as the
one we predicted should be there in the strongly coupled
regime.
\newline

\noindent {\sl $(2,0)$ LST and near-extremal M5-branes on a circle: }
We have seen in Sections \ref{secM5} and \ref{s:LST}
that the non-uniform phase of near-extremal M5-branes on
a circle predicts a new stable phase for the dual $(2,0)$ LST
in the canonical ensemble, with temperatures above the Hagedorn
temperature.
This would be interesting to understand from the Little String
Theory side.
\newline

\noindent {\sl Gregory-Laflamme type instability and Gubser-Mitra
conjecture:} We have shown that one non-trivial consequence of the
map from Kaluza-Klein black holes to non- and near-extremal branes
is that there exists a non-uniform phase for both these classes of
branes, emanating from the uniform phase. In particular, the
Gregory-Laflamme mass $\mu_{\rm GL}$ of the uniform black string
branch is mapped onto a critical mass $\bar \mu_{c}$ of the
uniform non-extremal branch and a critical energy $\epsilon_{c}$
of the uniform near-extremal branch. This, together with the fact
that the entropy of the localized phase is higher than that of the
uniform phase for small energies (above extremality),  strongly
suggests that non-/near-extremal branes on a circle have a
critical mass/energy below which the uniform phase is classically
unstable. It seems possible that one could use the boost and
U-duality transformation of Section \ref{secUdual} to show that
the unstable mode for the neutral uniform black string transforms
into an unstable mode of the non-/near-extremal brane on a circle.
It would be interesting to further examine this.

A related issue is the possibility of a counter-example to the
Gubser-Mitra conjecture
\cite{Gubser:2000ec,Gubser:2000mm,Reall:2001ag}, stating that a
black brane with a non-compact translational symmetry is free of
dynamical instabilities if and only if it is locally
thermodynamically stable. The uniform near-extremal branch has a
translational symmetry and (for $d > 5$) a positive heat capacity,
so if indeed near-extremal branes are classically unstable for
energies below $\epsilon_c$, this would be in contradiction to the
conjecture.
\newline

\noindent {\sl On conjectures of Horowitz and Maeda:}
Another interesting consequence of the results of this paper
is obtained if we revisit the results and conjectures of
Horowitz and Maeda in \cite{Horowitz:2001cz,Horowitz:2002ym}.
In \cite{Horowitz:2001cz} it was conjectured that there
exists a non-uniform black string branch for arbitrarily small
masses, with entropy higher than that of the uniform black
string branch.
This conjecture was based on the result that it would
take infinite ``time'' (i.e. affine parameter on the horizon)
to change the topology of the horizon
in a classical evolution.
Furthermore, in \cite{Horowitz:2002ym} it was conjectured that
for the D3, M2 and M5-branes on a (transverse) circle,
new non-uniform phases exist with arbitrarily small energy
and with entropy higher than that of the uniform phase.
The conjecture was based on a construction of suitable
initial data and on the conjecture of \cite{Horowitz:2001cz}.

But if we assume that the conjecture of \cite{Horowitz:2001cz}
is correct, we see that using the map from
static and neutral
Kaluza-Klein black holes to near-extremal branes on a circle,
we get immediately that the conjectured
small mass non-uniform black string
branch implies the existence of a low-energy non-uniform
phase of near-extremal branes on a circle,
with entropy higher than that of the uniform phase. This includes in particular
the D3, M2 and M5-branes, but in addition also the other $1/2$
BPS branes of string theory and M-theory.

However, it is important to emphasize that it is unclear whether
the conjecture of \cite{Horowitz:2001cz} is correct. The
above point is therefore, so far, more of a theoretical nature.
\newline

\noindent {\sl Non- and near-extremal bubble-black hole
sequences:} Recently, it was found \cite{Elvang:2004iz} that in
the region $1/(d-2) <  n \leq d-2$ of the $(\mu,n)$ phase
diagram of Kaluza-Klein black holes  involve solutions
that combine event horizons and static Kaluza-Klein bubbles.
We leave the application of the map to these bubble-black
hole sequences to a future publication \cite{Harmark:2004bb}.
\newline

\noindent {\sl New phases:} More generally, we emphasize
that any new phase that one may find in the $(\mu,n)$ phase
diagram, analytically or numerically, will immediately
generate via the map new non- and near-extremal phases.
Moreover, any such new phase would immediately have
consequences for the non-gravitational theories
dual to the near-extremal phases.
Finally, it should be clear that any non-trivial phase structure
and  phase transition for Kaluza-Klein black holes, such as the conjectured
topology-changing black string/hole transition,%
\footnote{See
\cite{Horowitz:2001cz,Gubser:2001ac,Harmark:2002tr,Horowitz:2002dc,Kol:2002xz,Wiseman:2002zc,Wiseman:2002ti,Harmark:2003fz,Kol:2003ja,Choptuik:2003qd,Harmark:2003dg,Kol:2003if,Harmark:2003eg,Sorkin:2003ka,Kudoh:2003ki,Harmark:2003yz,Sorkin:2004qq,Gorbonos:2004uc,Elvang:2004iz,Elvang:2004ny,Harmark:2004bb}
for a fairly complete list of references on this.}
will have an interesting counterpart in the dual non-gravitational theories.

\section*{Acknowledgments}

We thank H.~Elvang, G.~Horowitz, B. Jelstrup and S.~Ross  for illuminating
discussions.

\begin{appendix}

\section{Energy and tension for near-extremal branes \label{appnearex}}

In this appendix we use the general expressions \eqref{energ},
\eqref{tens} to compute the energy and tension for the class of
near-extremal branes on a circle that fall into the ansatz
\eqref{nearsol1}-\eqref{nearsol3}.
As remarked at the end of Section \ref{bhsans}, this is an important check on
the consistency of the near-extremal
limit. Moreover, since these branes are not asymptotically flat,
it also provides a
useful illustration of the general formula for tension in
non-asymptotically flat spaces given in Ref.~\cite{Harmark:2004ch}.
Following the discussion of Section \ref{phasenearex},
the reference space that we use for the near-extremal
 $p$-brane is the near-horizon limit of the extremal $p$-brane on
 a circle, the asymptotics of which is given in
 \eqref{nearsole}-\eqref{nearsole2}.

 To compute the energy and tension, we first recall a useful
 expression for the extrinsic curvature $\CK^{(D-2)}$ entering the
 computation of these quantities. We have
\begin{equation}
\label{excur}
\CK^{(D-2)} = \frac{1}{\sqrt{g_{RR}}} \partial_R \log \sqrt{h} \ .
\end{equation}
Here $h = \sqrt{ \det h^{D-2}}$ with $h^{D-2}$ the metric obtained
from $g$ by omitting $t,R$ when computing $E$ and omitting
$v,r$ when computing $\hat \CT$. For clarity we denote the corresponding
expressions by $\CK_t$ and $\CK_v$ respectively.

For the computation below we use that in the metric
\eqref{nearsol1}-\eqref{nearsol3} we have asymptotically
\begin{equation}
g_{tt} \simeq \hat H^{-\frac{d-2}{D-2}}  \left(-1 + \frac{\hat c_t}{R^{d-3}} \right)
\spa
g_{uu} \simeq \hat H^{-\frac{d-2}{D-2}} \ ,
\end{equation}
\begin{equation}
g_{RR} \simeq \hat H^{\frac{p+1}{D-2}} \left(1 + \frac{\hat c_R}{R^{d-3}} \right)
\spa
g_{vv} \simeq \hat H^{\frac{p+1}{D-2}} \left( 1 + \frac{\hat c_v}{R^{d-3}} \right)
\spa
g_{\Omega \Omega} \simeq \hat H^{\frac{p+1}{D-2}} R^2
\left(1 + \frac{\hat c_\Omega}{R^{d-3}} \right) \ ,
\end{equation}
with
\begin{equation}
\label{cval}
\hat c_t = R_0^{d-3} \spa \hat c_R = (1-  \chi) R_0^{d-3} \spa
\hat c_v = (d-3)  \chi R_0^{d-3} \spa \hat c_\Omega =-  \chi R_0^{d-3} \ ,
\end{equation}
where we used \eqref{defchi}.
It is then not difficult to use \eqref{excur}
and compute
\begin{equation}
\CK_t \simeq \left[ \hat H^{ \frac{p+1}{D-2}} (R_{\rm m})
\left( 1 + \frac{\hat c_R}{R_{\rm m}^{d-3}} \right)
\right]^{-1/2} \frac{1}{R_{\rm m}} \left[ \frac{d-1}{2} -
\frac{d-3}{2R_{\rm m}^{d-3}}
(\hat c_v + \hat c_\Omega) \right] \ ,
\end{equation}
\begin{equation}
\CK_v = \left[ \hat H^{ \frac{p+1}{D-2}} (R_{\rm m})
\left( 1 + \frac{\hat c_R}{R_{\rm m}^{d-3}} \right)
\right]^{-1/2} \frac{1}{R_{\rm m}} \left[ d-2 - \frac{d-3}{2R_{\rm m}^{d-3}}
(- \hat c_t + \hat c_\Omega) \right] \ ,
\end{equation}
where $R_m$ is sent to infinity.
Next we compute the extrinsic curvature of the reference space in both
cases, giving
\begin{equation}
\CK_t^{(0)} = \left[ \hat H^{ \frac{p+1}{D-2}} (R_{\rm eff}) \right]^{-1/2}
\frac{d-1}{2 \, R_{\rm eff}} \spa
\CK_v^{(0)} = \left[ \hat H^{ \frac{p+1}{D-2}} (R_{\rm eff}) \right]^{-1/2}
\frac{ d-2}{ R_{\rm eff}} \ ,
\end{equation}
where we used that asymptotically $\hat r \simeq R$ and $\hat z \simeq
v$ \cite{Harmark:2002tr}.
The relation between $R_{\rm m}$ and $R_{\rm eff}$ is obtained
by imposing that the radius of the $S^{d-2}$ in the brane
space time is equal to that of the radius of the $S^{d-2}$ in
the reference space. This gives
\begin{equation}
\left[ \hat H^{ \frac{p+1}{D-2}} (R_{\rm eff}) \right]^{1/2}
 R_{\rm eff} =
 \left[ \hat H^{ \frac{p+1}{D-2}} (R_{\rm m}) \left( 1 + \frac{\hat c_\Omega}{R_{\rm m}^{d-3}} \right)
\right]^{1/2} R_{\rm m} \ ,
\end{equation}
and a little algebra then shows
\begin{equation}
\CK_t - \CK_t^{(0)} \simeq \frac{1}{2 \, R_{\rm m}^{d-2} }
\left[ \hat H^{ \frac{p+1}{D-2}} (R_{\rm m}) \right]^{-1/2}
\left[ \frac{d-1}{2} (\hat c_\Omega-\hat c_R) - (d-3)( \hat c_v + (d-2)\hat c_\Omega)
 \right] \ ,
\end{equation}
\begin{equation}
\CK_v -  \CK_v^{(0)} \simeq
\frac{1}{2 \, R_{\rm m}^{d-2} }
\left[ \hat H^{ \frac{p+1}{D-2}} (R_{\rm m}) \right]^{-1/2}
\left[ (d-2) (\hat c_\Omega-\hat c_R) - (d-3)(- \hat c_t + (d-2)
\hat c_\Omega)  \right] \ .
\end{equation}

Finally, we need for each case the lapse functions and integration
measures: $N_t= (g_{tt}^{(0)})^{1/2} = \hat H^{-\frac{d-2}{2(D-2)}}$,
$\sqrt{g_{D-2}} = \sqrt{\hat H} R^{d-2}$ for the energy
and $N_v = (g_{vv}^{(0)})^{1/2} = \hat H^{\frac{p+1}{2(D-2)}}$,
$\sqrt{g_{D-2}} =  R^{d-2}$ for the tension. Using all this in \eqref{energ}, \eqref{tens}
we have the final results
\begin{equation}
E =
\frac{\Omega_{d-2}}{(2\pi)^{d-3} g }
\left[ \frac{d-1}{2 } (\hat c_R-\hat c_\Omega) + (d-3)(\hat c_v + (d-2)\hat c_\Omega )
\right] \ ,
\end{equation}
\begin{equation}
\hat \CT = \frac{\Omega_{d-2}}{(2\pi)^{d-2} g }
\left[ (d-2) (\hat c_R-\hat c_\Omega) + (d-3)( - \hat c_t + (d-2)\hat c_\Omega)  \right] \ ,
\end{equation}
for the energy and tension for near-extremal $p$-branes on a circle.
Substituting the values \eqref{cval} and using the definitions
\eqref{epsrdef} we then find the results given in
\eqref{neart1} for $\epsilon$ and $r$.

We also quote the result for the tension in the world volume of
the brane
\begin{equation}
\Lu \hCTu = \frac{\Omega_{d-2}}{(2\pi)^{d-3} g }
\left[ \frac{d-1}{2} (\hat c_R-\hat c_\Omega) + (d-3)(- \hat c_t + \hat c_v+ (d-2)\hat c_\Omega)  \right] \ ,
\end{equation}
which easily follows from a similar computation as the one given above.
Substituting the values \eqref{cval} and using the definitions
\eqref{epsrdef} this gives  the expression for $r_u$ given in
\eqref{neart3}.

\section{Flat space in $(\rht,\tht)$ coordinates \label{appcoord}}

In this appendix we review the flat space metric of $\CM^d \times
S^1$ in the $(\rht,\tht)$ coordinates which was used in
\cite{Harmark:2003yz} to write down the metric of small black
holes on the cylinder.

To write down the metric we need the function \cite{Harmark:2002tr}
\begin{equation}
\label{defF}
F(r,z) = \sum_{k=-\infty}^{\infty}
\frac{1}{(r^2 + (z-2\pi k)^2)^{\frac{d-2}{2}}} \ ,
\end{equation}
that enters the Newtonian potential for a black hole on the cylinder.
This definition employs cylindrical coordinates $(r,z)$, which
are related to spherical coordinates $(\rho,\theta)$ via
\begin{equation}
r = \rho \sin \theta
\spa
z = \rho \cos \theta \ .
\end{equation}
In terms of the latter coordinates the function in \eqref{defF} is
denoted by $F(\rho,\theta)$, which for $\rho \ll 1$ can be
expanded as
\begin{equation}
\label{Fexp}
F(\rho,\theta) = \frac{1}{\rho^{d-2}}
+ \frac{2\zeta(d-2)}{(2\pi)^{d-2}}
+ \frac{\zeta(d)}{(2\pi)^{d}} (d-2) \left[ d \cos^2 \theta - 1 \right] \rho^2
+ \mathcal{O} (\rho^4) \ .
\end{equation}

The function $F(\rho,\theta)$ enters the coordinate
transformation to the final $(\rht,\tht)$ coordinates, which are
given by%
\footnote{We note that these expressions equivalently follow from
the coordinate change \eqref{coordnew} and the relation between $(r,z)$
and $(R,v)$ as written in \cite{Harmark:2002tr}.}
\begin{equation}
\rht^{d-2} = \frac{1}{F(\rho,\theta)} \ ,
\end{equation}
\begin{equation}
\label{deftht}
(\sin \tht)^{d-2} \partial_\rho \tht =
\frac{\rho^{d-3} }{d-2} (\sin \theta)^{d-2} \partial_\theta F
\spa
(\sin \tht)^{d-2} \partial_\theta \tht =
- \frac{\rho^{d-1} }{d-2} (\sin \theta)^{d-2} \partial_\rho F \ .
\end{equation}
Using the expansion \eqref{Fexp} this can be expanded for
$\rho \ll 1$ as
\begin{equation}
\label{expthr}
\rho = \rht \left( 1 + \frac{2\zeta(d-2)}{(d-2) (2\pi)^{d-2}} \rht^{d-2}
+ \CO ( \rht^d ) \right) \ ,
\end{equation}
\begin{equation}
\label{exptht}
\sin^2 \theta = \sin^2 \tht \left( 1 + \frac{4 \zeta(d)}{(2\pi)^d}
\cos^2 \tht \rht^d + \CO (\rht^{d+2} ) \right) \ .
\end{equation}

Finally, the flat space metric of $\CM^d \times
S^1$  in $(\rht,\tht)$ coordinates takes the form \cite{Harmark:2003yz}
\begin{equation}
\label{newansatzflat}
ds^2 = - dt^2 + \tilde{A}_0 d\rht^2
+ \frac{\tilde{A}_0}{\tilde{K}_0^{d-2}} \rht^2 d\tht^2
+ \tilde{K}_0 \rht^2 \sin^2 \tht d\Omega_{d-2}^2 \ ,
\end{equation}
where
\begin{equation}
\label{AK0}
\tilde{K}_0 = \frac{\rho^2 \sin^2 \theta}{\rht^2 \sin^2 \tht}
\spa
\tilde{A}_0 = \left[ (\partial_\rho \rht)^2
+ \rht^2 \tilde{K}_0^{-(d-2)} (\partial_\rho \tht)^2 \right]^{-1} \ .
\end{equation}
In particular, one may use  the expansions \eqref{expthr}, \eqref{exptht}
to obtain the $\rho \ll 1$ expressions
\begin{equation}
\label{tilA0}
\tilde{A}_0 (\rht,\tht)
= 1 + \frac{4(d-1)\zeta(d-2)}{(d-2) (2\pi)^{d-2}} \rht^{d-2}
+ \CO (\rht^d) \ ,
\end{equation}
\begin{equation}
\label{tilK0}
\tilde{K}_0 (\rht,\tht)
= 1 + \frac{4\zeta(d-2)}{(d-2) (2\pi)^{d-2}} \rht^{d-2}
+ \CO (\rht^d) \ .
\end{equation}

\end{appendix}

\addcontentsline{toc}{section}{References}


\providecommand{\href}[2]{#2}\begingroup\raggedright\endgroup

\end{document}